\documentclass[prd,,aps,showpacs,nofootinbib,amsmath,amssymb]{revtex4}
\usepackage{amssymb,epsfig}
\usepackage{mathrsfs}
\usepackage[usenames,dvipsnames]{color}
\usepackage[bookmarks]{hyperref}
\usepackage[usenames,dvipsnames]{color}

\newtheorem{theorem}{Theorem}

\newcommand{\proof}{\noindent {\bf Proof. }}

\newcommand{\qed}{\hfill $\fbox{\hspace{0.3mm}}$ \vspace{.3cm}} 

\newcommand{\Real}{\mathbb{R}}

\newcommand{\re}{\mbox{Re}}

\newcommand{\ScriPlus}{{\mathscr{I}^+}}

\newcommand{\hz}{\mbox{\em \r{h}\hspace{0.3mm}}} 
\newcommand{\Gammaz}{\Gamma\hspace{-0.25cm}{}^{\mbox{\r{~}}}{}\hspace{-0.12cm}}

\newcommand{\norm}[1]{\left\lVert#1\right\rVert}

\begin{document}

\title{Evolution of scalar fields surrounding black holes on compactified constant mean curvature hypersurfaces}

\author{Manuel D. Morales and Olivier Sarbach}
\affiliation{Instituto de F\'isica y Matem\'aticas,
Universidad Michoacana de San Nicol\'as de Hidalgo,
Edificio C-3, Ciudad Universitaria, 58040 Morelia, Michoac\'an, Mexico}

\begin{abstract}
Motivated by the goal for high accuracy modeling of gravitational radiation emitted by isolated systems, recently, there has been renewed interest in the numerical solution of the hyperboloidal initial value problem for Einstein's field equations in which the outer boundary of the numerical grid is placed at null infinity. In this article, we numerically implement the tetrad-based approach presented in [J.M. Bardeen, O. Sarbach, and L.T. Buchman, Phys. Rev. D 83, 104045 (2011)] for a spherically symmetric, minimally coupled, self-gravitating scalar field. When this field is massless, the evolution system reduces to a regular, first-order symmetric hyperbolic system of equations for the conformally rescaled scalar field which is coupled to a set of singular elliptic constraints for the metric coefficients. We show how to solve this system based on a numerical finite-difference approximation, obtaining stable numerical evolutions for initial black hole configurations which are surrounded by a spherical shell of scalar field, part of which disperses to infinity and part of which is accreted by the black hole. As a non-trivial test, we study the tail decay of the scalar field along different curves, including one along the marginally trapped tube, one describing the world line of a timelike observer at a finite radius outside the horizon, and one corresponding to a generator of null infinity. Our results are in perfect agreement with the usual power-law decay discussed in previous work. This article also contains a detailed analysis for the asymptotic behavior and regularity of the lapse, conformal factor, extrinsic curvature and the Misner-Sharp mass function along constant mean curvature slices.
\end{abstract}

\date{\today}


\maketitle

\section{Introduction}

The recent discovery of gravitational waves from a binary black hole merger by the Laser Interferometer Gravitational-wave Observatory (LIGO) in September 2015 is truly a milestone in the history of science~\cite{bA06}. From an observational point of view, this discovery opens a new window into our universe~\cite{kKwNwP15}, with fascinating implications for astrophysics and new possibilities for testing general relativity and alternative theories of gravity (see~\cite{eB16} and references therein). It provides a strong impetus to improve the computer modeling of gravitational waves, considering the indispensable role that numerical relativity has played and will continue to play in the understanding of black hole mergers~\cite{fP05,jB06,mC06}.

A crucial issue in the modeling of gravitational waves is the study of isolated systems~\cite{rG77}. In practice, we want to focus on certain physical systems, describing their physical features without the influence of their environment. Isolated systems do not exist in the real world, they are an idealization, of course. Nevertheless, when gravitational effects of the environment are non significant in comparison with those produced by the physical system in which we are interested, the behavior of the latter can be approximated as being produced by an isolated system. In particular, if we model physical systems in which strong gravity is involved (as in binary black hole mergers, supernova core collapses, etc.), we want to calculate numerically the radiation emitted by the system in order to identify and analyze the experimental data provided by gravitational wave detectors, such as LIGO~\cite{LIGO}, VIRGO~\cite{VIRGO} or KAGRA~\cite{KAGRA}.

Here we need to take into account an important aspect, namely that isolated systems are mathematically described by asymptotically flat spacetimes (see Refs.~\cite{HawkingEllis-Book,Wald-Book} and references therein) which, by definition, are infinite in extension. So the crucial question is: How can we numerically model such infinite systems? The standard procedure in numerical relativity to handle this issue has been to consider a Cauchy evolution based on a foliation by spacelike hypersurfaces approaching spacelike infinity, introducing an artificial timelike boundary far enough from the strong field region which truncates the spacetime domain. This procedure requires the specification of suitable ``absorbing" boundary conditions which, ideally, should reproduce the same solution one would obtain from a Cauchy evolution on the infinite domain.

However, this pragmatic approach comes with several difficulties. First, the boundary conditions need to be specified in such a way that the resulting initial-boundary value problem (IBVP) is well posed. Due to gauge freedom and constraint modes propagating with non-trivial speeds, this problem turns out to be much more difficult than other typical IBVPs in physics. Despite of these difficulties, a solution of this problem has been given in recent years, at least for certain formulations of Einstein's field equations (see~\cite{oSmT12} for a recent review). Second, truncating the physical domain with an artificial boundary almost always introduces undesirable reflections when the emitted radiation reaches the boundary. Although classes of absorbing boundary conditions have been introduced which are exact for gravitational linearized perturbations on Minkowski spacetime~\cite{lBoS06,lBoS07} or the Schwarzschild spacetime~\cite{sL05}, completely eliminating the spurious reflections remains a formidable task in the full nonlinear theory. Third, and more importantly, if we are interested in computing the physically relevant quantities associated with the radiated field, such as the radiated energy, the only place where these quantities are well-defined is at future null infinity. Therefore, the only way to unambiguously numerically model them is to include future null infinity (henceforth denoted by $\ScriPlus$) in the numerical domain.

To achieve this goal, the ideas of Penrose~\cite{rP64} about conformal infinity have proven fundamental, and provide the basis of many theoretical and numerical approaches that have been developed today, including the one adopted in this work. Friedrich, in his pioneering work~\cite{hF83}, used conformal compactification to provide a well-posed Cauchy formulation of Einstein's field equations on hyperboloidal spacelike hypersurfaces approaching null infinity. Such hyperboloidal surfaces have the advantage of behaving like conventional time slices in the strong field region while approaching outgoing null surfaces asymptotically, and hence it is expected that they are well suited for describing the radiation emitted by an isolated system. One of the key features of Friedrich's formulation consists of a symmetric hyperbolic evolution system involving the tetrad and connection fields as well as components of the Weyl curvature tensor which is manifestly regular at $\ScriPlus$. Later, H\"ubner~\cite{pH95,ph96,pH01b}, Frauendiener~\cite{jF98,jF98b,jF00}, and Husa~\cite{sH02,sH03} numerically implemented Friedrich's scheme for different scenarios. H\"ubner applied this formalism to study numerically the global structure of spacetimes describing the spherical collapse of a self-gravitating scalar field, while Frauendiener, H\"ubner and Husa studied vacuum, asymptotically flat spacetimes containing gravitational radiation with special emphasis on wave extraction or decay of curvature invariants. For a more detailed account on these works we refer to reader to~\cite{jF04}.

Although the above numerical implementations constituted an important achievement, they contain difficulties associated with the constraint equations. Unlike the evolution equations, the constraints in Friedrich's formulation involve (apparently) singular terms at future null infinity, requiring a special treatment in the construction of the initial data. The existence of hyperboloidal Cauchy data in the vacuum case has been studied in~\cite{lApChF92,lApC94}. However, one would like to know whether such hyperboloidal data may arise from standard, asymptotically Euclidean data. Here the analytic work of Corvino~\cite{jC00} has been key, since it allows a gluing of a bounded domain of a time-symmetric, asymptotically flat initial data set satisfying the vacuum Einstein equations to a static slice of the (exact) Schwarzschild metric, which is known in explicit form. This avoids the necessity of dealing with the singular terms, since in a region near $\ScriPlus$ the initial data can be described by a suitable hyperboloidal slice of the Schwarzschild solution. For the non-time-symmetric case, this construction was later generalized by Corvino and Schoen~\cite{jCrS06}. In this case, the initial data can be glued to a suitable slice of the Kerr metric. However, these gluing techniques are not very explicit and numerical implementations of such methods have only started recently~\cite{gDoR16}. Another, maybe more serious, issue, which significantly hindered new developments in the numerical modeling of self-gravitating physical systems based on Friedrich's formalism, was the rapid growth of constraint violations with time, triggered by numerical error~\cite{sH02,jF04}. In fact, similar problems were also present in other symmetric hyperbolic formulations of Einstein's equations used in numerical relativity at that time, see for example~\cite{oB99,mA00,lKmSsT01}.

Considering the aforementioned, it is not surprising  that many works began taking a few steps back, analyzing in detail the theory and numerical stability of test fields propagating on a fixed background foliated by spacelike hyperboloidal hypersurfaces, see for instance~\cite{gCcGdH05,aZ08a,aZ08b}. A distinguished class of such hypersurfaces are those having positive and constant mean curvature (CMC) since this property is shared by the hyperboloids in Minkowski spacetime. For explicit expressions of such CMC foliations for the Schwarzschild spacetime, the numerical construction of CMC binary black hole initial data and axisymmetric CMC foliations of the Kerr spacetime, see Refs.~\cite{eMnM03,lBhPjB09,dSpRmA14}.

More recently, Moncrief and Rinne proposed a new formulation of Einstein's equations~\cite{vMoR09} which is based on the Arnowitt-Deser-Misner $3+1$ decomposition and which uses a CMC foliation and spatial harmonic coordinates. This approach has the advantage (compared to Friedrich's) of being much closer to the traditional schemes used in numerical relativity, such as the popular Baumgarte-Shapiro-Shibata-Nakamura (BSSN) formulation~\cite{mStN95,tBsS98} used in binary black hole collisions. However, compared to Friedrich's formulation, there are two extra complications in the proposal by Moncrief and Rinne: first, the resulting equations constitute a hyperbolic-elliptic system, due to the gauge conditions and choice for the conformal factor. Second, the evolution equations are not manifestly regular at $\ScriPlus$, but contain apparently singular terms which require Taylor expansions of the fields to evaluate them. Despite these complications, Rinne was able to successfully implement this formulation in an axisymmetric vacuum code~\cite{oR10}, obtaining long-term stable and convergent evolutions of a Schwarzschild black hole perturbed by a gravitational wave. More recently, Rinne and Moncrief used their approach to develop a spherically symmetric code including matter fields, and to study the collapse and tail decay of self-gravitating scalar and Yang-Mills fields~\cite{oRvM13,oRvM14}.

A related but different line of work has been initiated recently by Va\~n\'o-Vi\~nuales, Husa and Hilditch~\cite{aVsHdH15} who implemented an unconstrained evolution scheme based on hyperboloidal foliations and generalized versions of the BSSN equations in the spherically symmetric case. They couple Einstein's field equations to a massless scalar field and study the evolution from regular initial data. See also~\cite{aV15} for more details and the evolution from black hole initial data. The advantage of their formulation is that it does not require solving any elliptic equations during the evolution. However, in order to achieve stability, they need a rather sophisticated evolution equation for the lapse and the addition of damping terms to the right-hand side (RHS) of the evolution equations with an {\it ad hoc} choice of parameters to deal with the formally singular terms at $\ScriPlus$.

Finally, we mention that a completely alternative method for reading off the radiation at $\ScriPlus$ which is emitted by an isolated system can be achieved by matching at a timelike surface a standard Cauchy code with a characteristic code which extends the simulation to null infinity, see~\cite{jW12} for a review. Recently, a slightly simplified version of this approach, called Cauchy-characteristic extraction, in which data on a timelike tube obtained from a stand-alone Cauchy code is propagated out to null infinity using a characteristic code has been successfully applied to several neutron star and black hole systems, see~\cite{nBlR16} for a recent review. The advantages and disadvantages of the Cauchy-characteristic extraction approach compared to the hyperboloidal one have been summarized in~\cite{jBoSlB11}.

Having briefly reviewed previous work in the field, we now describe the approach adopted in the present article for numerically modeling asymptotically flat spacetimes. In spirit, our approach is quite similar to the one by Moncrief and Rinne described above; the main difference is that we use tetrad fields rather than metric variables. Our method is based on the tetrad formalism of numerical relativity on conformally compactified CMC hypersurfaces developed in~\cite{jBoSlB11}, and a main motivation for this work is to provide a first numerical test for the viability of this evolution scheme. Unlike the traditional metric-based formulations of the Einstein equations in which the components of the metric and other tensor fields are expanded in terms of a coordinate basis, here we decompose them in terms of an orthonormal frame ${\bf e}_0,{\bf e}_1,{\bf e}_2,{\bf e}_3$. As in~\cite{jBoSlB11} we adopt the hypersurface-orthogonal gauge in which the timelike leg ${\bf e}_0$ of this frame is orthogonal to the CMC hypersurfaces. The remaining rotational degrees of freedom in the choice for the spacelike legs ${\bf e}_a$, $a=1,2,3$, is fixed (up to a global rotation) by imposing to the 3D Nester gauge condition~\cite{jN89,jN91}. From a mathematical point of view, the use of tetrad fields (instead of metric ones) has some attractive properties. First, the frame components of tensor fields behave as scalars under coordinate transformations, and further the raising and lowering of indices becomes trivial, since the frame components of the metric are the same as the ones of the Minkowski metric in inertial coordinates. Second, while the Levi-Civita connection in the metric formulation leads to $40$ independent Christoffel symbols, in the tetrad formulation the connection gives rise to only $24$ connection coefficients. Finally, their $3+1$ decomposition has clear geometric interpretations. These properties lead to evolution and constraint equations which are rather elegant; they are described in detail in~\cite{jBoSlB11}. The resulting evolution scheme consists of a hyperbolic-elliptic system of equations. The CMC slicing condition, the Hamiltonian constraint and the preservation of the Nester gauge yield an elliptic system of equations for the conformal lapse, the conformal factor and some of the connection coefficients. As in the scheme by Moncrief and Rinne these equations are formally singular at $\ScriPlus$, where the conformal factor vanishes, and hence they require the imposition of appropriate regularity conditions. Using the constraints, one can derive formal expansions for all the relevant quantities near $\ScriPlus$ from which the singular terms can be evaluated. In general, these expansions are polyhomogeneous, that is, they contain $\log$ terms (see~\cite{lApChF92,lApC94,pCmMdS95} and the discussion in Appendix A of~\cite{jBoSlB11}). See also~\cite{jBlB12} for a recent discussion and explicit formulas for the Bondi-Sachs energy and momentum in terms of the coefficients of the asymptotic expansions.

In this article, we numerically implement the formulation put forward in~\cite{jBoSlB11} for a simple, yet non-trivial scenario, namely, the propagation of a minimally coupled, self-gravitating scalar field configuration surrounding a black hole. After presenting a brief summary in Sec.~\ref{Sec:BSB} of the hyperbolic-elliptic system derived of Ref.~\cite{jBoSlB11}, in Sec.~\ref{Sec:Scalar} we generalize this system to include a (minimally coupled) scalar field $\Phi$ with arbitrary potential $V(\Phi)$ without symmetry assumptions. Using the Einstein equations, we show that the equations of motion for the scalar field can be cast as a first-order symmetric hyperbolic system for the rescaled field $\tilde{\phi} = \Phi/\Omega$, which is manifestly regular at $\ScriPlus$, provided the potential $V(\Phi)$ falls off sufficiently fast as $\Phi\to 0$. Next, in Sec.~\ref{Sec:SphSym} we reduce the equations to spherical symmetry, where as it turns out, there is a preferred choice for the spatial triad fields which automatically satisfies the 3D Nester gauge condition. Furthermore, spatially harmonic coordinates can be chosen such that, with the choice for the conformal factor in~\cite{jBoSlB11}, the conformal metric is the Euclidean metric written in spherical coordinates. After giving a summary of all the evolution and constraint equations in this conformally flat gauge, in Sec.~\ref{Sec:SphSym} we also provide a discussion of useful geometric quantities, such as the in- and outgoing expansions associated with the invariant two-spheres and the Hawking (or Misner-Sharp) mass function. Next, in Sec.~\ref{Sec:Asymptotics} we analyze the asymptotic behavior of the fields in the vicinity of $\ScriPlus$ and derive formal expansions for them. As in the vacuum case without symmetries, these expansions are polyhomogeneous, that is, of the form $f(z) = \sum_{ij}f_{ij} z^i\log^j(z)$, with $f$ the quantity of interest and $z$ the radial proper distance from $\ScriPlus$ along the CMC slices. Even when assuming the vanishing of the Newman-Penrose constant~\cite{eNrP68} (which might be physically justified by excluding incoming radiation at past null infinity) we show that the coefficients in front of the leading $\log$ terms are non-zero whenever outgoing scalar radiation is present at $\ScriPlus$. This is similar to the vacuum case without symmetries, where $\log$ terms appear if and only if gravitational radiation is present at $\ScriPlus$, provided the Penrose regularity condition holds~\cite{jBoSlB11}. The expansions derived in this section play a crucial role for the numerical implementation of the elliptic equations since they provide the means to specify correct boundary conditions near $\ScriPlus$.

The numerical implementation of our system of equations and the results from our simulations are discussed in Sec.~\ref{Sec:NumResults}. We start by setting up initial data on a CMC surface, representing a scalar field distribution surrounding a spherically symmetric black hole. We do this by specifying a Gaussian pulse for the physical field $\Phi$ on this surface and setting the associated canonical momentum to zero. Additionally, we solve the Hamiltonian constraint for the conformal factor $\Omega$, assuming that the inner boundary represents a trapped surface. Then, we numerically evolve the scalar field and the geometric quantities using the hyperbolic-elliptic system derived from the scheme in~\cite{jBoSlB11} and summarized in Sec.~\ref{Sec:SphSym} and perform several tests. We find that it is much more convenient to determine the trace-free part of the conformal extrinsic curvature from the momentum constraint rather than from its evolution equation, as it seems to allow better control of the regularity conditions at $\ScriPlus$. We end Sec.~\ref{Sec:NumResults} with long-term evolutions showing the tail decay of the scalar field along the world lines of different ``observers", including ones at the apparent horizon and at null infinity. In particular, we reproduce the known polynomial tail decays in the literature~\cite{rP72,cGrPjP94,cGrPjP94b,mDiR05}. Conclusions are drawn in Sec.~\ref{Sec:Conclusions}, where we also comment on possible extensions of this work. Finally, some auxiliary, yet important technical results which are relevant to our work are presented in the appendices. In Appendix~\ref{App:ExplicitExpressions} we derive explicit expressions for the fields in Schwarzschild spacetimes. Formal polyhomogeneous expansions for the metric fields at $\ScriPlus$ in the presence of a scalar field are given in Appendix~\ref{App:FormalExpansions}, and in Appendix~\ref{App:LocSolutionsScriPlus} we prove that these expansions are not just formal, but do correspond to genuine local solutions of the constraint equations in the vicinity of $\ScriPlus$.

Throughout this work we use the signature convention $(-,+,+,+,)$ for the metric and units in which the speed of light is one. Greek indices $\mu,\nu,\ldots$ refer to spacetime indices.

\section{Tetrad formulation on compactified CMC hypersurfaces}
\label{Sec:BSB}

In this section, we briefly review the proposal of~\cite{jBoSlB11} for numerically evolving Einstein's field equations on a an asymptotically flat spacetime using compactified CMC hypersurfaces $\Sigma_t$. For ease of reading and for the purpose of fixing the notation, the pertinent results from Ref.~\cite{jBoSlB11} are reproduced here.

We work in the hypersurface-orthogonal gauge in which the timelike leg ${\bf e}_0$ of the  orthonormal frame is aligned with the future-directed normal to $\Sigma_t$ (and hence the spacelike legs ${\bf e}_1, {\bf e}_2, {\bf e}_3$ are tangent to $\Sigma_t$). With respect to local coordinates $t,x^1,x^2,x^3$ adapted to $\Sigma_t$ we thus have
$$
{\bf e}_0 = \frac{1}{\alpha}\left( \frac{\partial}{\partial t} 
 - \beta^i\frac{\partial}{\partial x^i} \right),\qquad
{\bf e}_a = B_a{}^i\frac{\partial}{\partial x^i},\quad a = 1,2,3, 
$$
where here and in the following the letters $a,b,c,d = 1,2,3$ refer to triad indices and $i,j,k$ to spatial coordinate indices. $\alpha$ and $\beta^i$ refer to the lapse and the coordinate components of the shift vector, respectively, and $B_a{}^i$ are the coordinate components of the spatial legs ${\bf e}_a$ of the tetrad fields.

The connection coefficients $\Gamma_{\alpha\beta\gamma} := {\bf g}({\bf e}_\alpha,\nabla_{{\bf e}_\gamma} {\bf e}_\beta) = -\Gamma_{\beta\alpha\gamma}$, $\alpha,\beta,\gamma = 0,1,2,3$, associated with the tetrad field split into the following components (see~\cite{fEhW64}):
$$
K_{ab} := \Gamma_{b0a},\qquad
N_{ab} := \frac{1}{2}\varepsilon_b{}^{cd}\Gamma_{cda}
$$
and
$$
a_b := \Gamma_{b00},\qquad
\omega_b := -\frac{1}{2}\varepsilon_b{}^{cd}\Gamma_{cd0}.
$$
As a consequence of the hypersurface-orthogonal gauge these components have the following nice geometric interpretation: $K_{ab} = K_{ba}$ is the extrinsic curvature of $\Sigma_t$ and is symmetric, $N_{ab}$ (which is not symmetric in general) is the induced connection on $\Sigma_t$, while $a_b$ and $\omega_b$ describe, respectively, the acceleration and the angular velocity  of the triad vectors ${\bf e}_1,{\bf e}_2,{\bf e}_3$ relative to Femi-Walker transport along the normal observers. Furthermore, the acceleration is given by the gradient of the logarithm of the lapse,
$$
a_b = D_b(\log\alpha),
$$
where we denote by $D_b = B_b{}^i\partial_i$ the directional derivative along ${\bf e}_b$.

In the formulation of Ref.~\cite{jBoSlB11} one does not work directly with the fields $\alpha$, $B_a{}^i$, $K_{ab}$, $N_{ab}$ and $\omega_b$ but rather with rescaled fields $\tilde{\alpha}$, $\tilde{B}_a{}^i$, $\tilde{K}_{ab}$, $\tilde{N}_{ab}$ and $\tilde{\omega}_b$ which are obtained by the conformal rescaling ${\bf e}_0 = \Omega\tilde{\bf e}_0$, ${\bf e}_a = \Omega\tilde{\bf e}_a$, where the conformal factor $\Omega$ is zero at $\ScriPlus$ and positive everywhere in the interior domain. This rescaling gives rise to the following conformal transformations:
\begin{equation}
\alpha = \Omega^{-1}\tilde{\alpha},\qquad
\beta^i = \tilde{\beta}^i,\qquad
B_a{}^i = \Omega\tilde{B}_a{}^i
\end{equation}
and
\begin{equation}
K_{ab} = \Omega\tilde{K}_{ab} - \delta_{ab}\tilde{D}_0\Omega,\qquad
N_{ab} = \Omega\tilde{N}_{ab} + \varepsilon_{ab}{}^c\tilde{D}_c\Omega,\qquad
\omega_b = \Omega\tilde{\omega}_b,
\label{Eq:KNomegaConf}
\end{equation}
where here $\tilde{D}_0$ and $\tilde{D}_c$ denote the directional derivatives along $\tilde{\bf e}_0$ and $\tilde{\bf e}_c$, respectively.

The local rotational freedom in the choice of the spatial legs ${\bf e}_a$ is fixed by imposing the 3D Nester gauge~\cite{jN89}, which implies that the trace of $N_{ab}$ vanishes and that its antisymmetric part $n_b := \varepsilon_b{}^{cd} N_{cd}/2$ is a gradient. As already noted by Nester himself~\cite{jN91}, the conformal transformations preserve the Nester gauge and further the conformal factor $\Omega$ might be chosen such that the antisymmetric part of the conformally rescaled variable $\tilde{N}_{ab}$ vanishes completely. This choice lies at the heart of the formulation in Ref.~\cite{jBoSlB11} and fixes $\Omega$  up to a constant rescaling. With these gauge conditions, only the trace-free, symmetric parts $\hat{\tilde K}_{ab}$ and $\hat{\tilde N}_{ab}$ of the conformal extrinsic curvature and spatial connection coefficients are free to evolve, and together with $\tilde{B}_a{}^i$ they obey the first-order symmetric hyperbolic system
\begin{eqnarray}
 \tilde{D}_0\tilde{B}_a{}^i &=&-\hat{\tilde{K}}_a{}^b\tilde{B}_b{}^i -
  \varepsilon_a{}^{cd} \tilde{\omega}_c \tilde{B}_d{}^i -\frac{\tilde{K}}{3}\tilde{B}_a{}^i,
\label{Eq:Ba}\\
\tilde{D}_0 \hat{\tilde{N}}_{ab} + \tilde{D}_c \hat{\tilde{K}}_{d(a}
  \varepsilon_{b)}{}^{cd}
  &=& \left\{ 2 \hat{\tilde{K}}_{(a}{}^c\hat{\tilde{N}}_{b)c} - \frac{\tilde{K}}{3}\hat{\tilde{N}}_{ab}
    + 2 \varepsilon^{cd}{}_{(a} \hat{\tilde{N}}_{b)c} \tilde{\omega}_d 
    + \frac{1}{\tilde{\alpha}} \left[\varepsilon^{cd}{}_{(a} \hat{\tilde{K}}_{b)c} 
      {\tilde{D}_d} {\tilde{\alpha}} - \tilde{D}_{(a}
      \left({\tilde{\alpha}} {\tilde{\omega}}_{b)}\right)
    \right] \right\}^{TF},
  \label{Eq:Nab} \\
  \tilde{D}_0\hat{\tilde{K}}_{ab} - \tilde{D}_c \hat{\tilde{N}}_{d(a} \varepsilon_{b)}{}^{cd} 
  &=& \left\{ - 2 \hat{\tilde{N}}_a{}^c \hat{\tilde{N}}_{bc} - \frac{\tilde{K}}{3}\hat{\tilde{K}}_{ab}
    + 2\varepsilon^{cd}{}_{(a} \hat{\tilde{K}}_{b)c} \tilde{\omega}_d \right.
  \nonumber\\
  &+& \left.  \frac{1}{\tilde{\alpha}} \tilde{\nabla}_a \tilde{\nabla}_b \tilde{\alpha} 
    - \frac{2}{\Omega} \left[\tilde{\nabla}_a \tilde{\nabla}_b \, \Omega 
      + C\hat{\tilde{K}}_{ab} \right] + 8\pi G\tilde{\sigma}_{ab}
  \right\}^{TF},
\label{Eq:Kab}
\end{eqnarray} 
where the super index $TF$ denotes the traceless part and $C := K/3$ is the mean extrinsic curvature. We have also defined $\tilde{\sigma}_{ab} := {\bf T}(\tilde{\bf e}_a,\tilde{\bf e}_b) = \Omega^{-2} {\bf T}({\bf e}_a,{\bf e}_b)$ to be the conformally rescaled stress tensor associated with the stress energy-momentum tensor ${\bf T}$ describing the matter fields in the spacetime.\footnote{Note that in Ref.~\cite{jBoSlB11} the fields $\tilde{\sigma}_{ab}$, $\tilde{\rho}$ and $\tilde{j}_b$ are defined with a factor of $\Omega^{-4}$ instead of $\Omega^{-2}$. Here, we choose the factor $\Omega^{-2}$ because for a scalar field it leads to the correct rescaling, as we will see.} Further, we note that $\tilde{D}_0 \tilde{B}_a{}^i = \tilde{\alpha}^{-1}[\partial_t\tilde{B}_a{}^i - \beta^j\partial_j\tilde{B}_a{}^i + (\partial_j\beta^i)\tilde{B}_a{}^j ]$, is the $i$-th coordinate component of the Lie derivative of the vector field $\tilde{\bf e}_a = \tilde{B}_a{}^i\partial_i$ along the normal vector to the time slices $\Sigma_t$.

While Eqs.~(\ref{Eq:Ba}) and (\ref{Eq:Nab}) are manifestly regular at $\ScriPlus$ since they are independent of $\Omega$, Eq.~(\ref{Eq:Kab}) contains the apparently singular term
$$
S_{ab} := \frac{1}{\Omega}\left( \tilde \nabla_a \tilde \nabla_b \Omega 
    + C\hat{\tilde K}_{ab} \right)^{TF},
$$
which requires the regularity condition
\begin{equation}
\hat{\tilde K}_{ab} = \tilde \kappa _{ab} 
 - \frac{1}{2}\tilde \gamma_{ab}\tilde{\gamma}^{cd}\tilde\kappa_{cd}
\label{Eq:RegCond}
\end{equation}
at $\ScriPlus$, where here $\tilde{\gamma}_{ab} = \delta_{ab} - \tilde{s}_a\tilde{s}_b$ and $\tilde{\kappa}_{ab} := \tilde{\gamma}_a{}^c\tilde{\nabla}_c\tilde{s}_b$ are the first and second fundamental form of the cross sections of $\ScriPlus$ with outward unit normal $\tilde{s}_a$ with respect to the conformal geometry.

In order to close the evolution system described in Eqs.~(\ref{Eq:Ba},\ref{Eq:Nab},\ref{Eq:Kab}), one needs to specify the fields $\tilde{K}$, $\tilde{\omega}_b$, $\tilde{\alpha}$, $\beta^i$ and the conformal factor $\Omega$. As shown in~\cite{jBoSlB11}, $\tilde{K}$ and $\tilde{\omega}_b$ are determined by the requirement of preserving the 3D Nester gauge, which yields the elliptic system
\begin{eqnarray}
  -\tilde{D}^a(\tilde{\alpha}\tilde{\omega}_a) 
  &=& \tilde{\alpha} \hat{\tilde{N}}^{ab}\hat{\tilde{K}}_{ab},
 \label{Eq:DivergenceOmega}\\
 \tilde{D}_a\left( \frac{2}{3}\tilde{\alpha}\tilde{K}\right) 
 + \varepsilon_a{}^{bc}\tilde{D}_b(\tilde{\alpha}\tilde{\omega}_c)
 &=& \tilde{D}^b(\tilde{\alpha}\hat{\tilde{K}}_{ab}).
\label{Eq:GradKTilde}
\end{eqnarray}
The conformal lapse is determined by the requirement to preserve the CMC slicing condition, which gives rise to the following elliptic equation for $\tilde{\alpha}$:
\begin{equation}
  \Omega\tilde{D}^a\tilde{D}_a\tilde{\alpha} -
  3(\tilde{D}^a\Omega)\tilde{D}_a\tilde{\alpha}
  + (\tilde{D}^a\tilde{D}_a\Omega)\tilde{\alpha} - \frac{\Omega}{2}\left(
    \hat{\tilde{N}}^{ab} \hat{\tilde{N}}_{ab} + 3\hat{\tilde{K}}^{ab}
    \hat{\tilde{K}}_{ab} \right) \tilde{\alpha}
  = 4\pi G \Omega \left(3\tilde{\rho} + \tilde{\sigma}^c{}_c \right) 
  \tilde{\alpha},
\label{Eq:CMC}
\end{equation}
where $\tilde{\rho} := {\bf T}(\tilde{\bf e}_0,\tilde{\bf e}_0)$ is the rescaled energy density. The conformal factor $\Omega$, on the other hand, is determined by solving the Hamiltonian constraint which yields
\begin{equation}
  \Omega \tilde{D}^a \tilde{D}_a \Omega = \frac{3}{2} \left[ \left(\tilde{D}^a \Omega \right) 
    \left(\tilde{D}_a \Omega \right) - C^2 \right] + \frac{\Omega^2}{4} 
  \left( \hat{\tilde{K}}^{ab}  \hat{\tilde{K}}_{ab}  +  \hat{\tilde{N}}^{ab}  \hat{\tilde{N}}_{ab} 
  \right)  +  4\pi G \Omega^2 \tilde{\rho}. 
 \label{Eq:Ham}
\end{equation}

The hyperbolic-elliptic evolution system presented in Eqs.~(\ref{Eq:Ba},\ref{Eq:Nab},\ref{Eq:Kab},\ref{Eq:DivergenceOmega}--\ref{Eq:Ham}) is subject to the constraints
\begin{eqnarray} 
\varepsilon_a{}^{bc}\tilde{D}_b\tilde{B}_c{}^k - \hat{\tilde{N}}_a{}^b\tilde{B}_b{}^k &=& 0,
\label{Eq:CurlB}\\
\tilde{D}^a\hat{\tilde{N}}_{ab} &=& 0,
\label{Eq:DivN}\\
\tilde{D}^a\hat{\tilde{K}}_{ab} + \varepsilon_b{}^{cd}\hat{\tilde{K}}_c{}^a
 \hat{\tilde{N}}_{ad} - \frac{2}{\Omega}(\tilde{D}^a\Omega)\hat{\tilde{K}}_{ab}
  &=& -8\pi G\tilde{j}_b,
\label{Eq:Mom}
\end{eqnarray}
with $\tilde{j}_b := -{\bf T}(\tilde{\bf e}_0,\tilde{\bf e}_b)$.

There is also an evolution equation for the conformal factor which follows from taking the trace of the first relation in Eq.~(\ref{Eq:KNomegaConf}),
\begin{equation}
\tilde{D}_0\Omega = -\frac{1}{3}(K - \Omega\tilde{K}).
\end{equation}

Finally, a suitable condition on the spatial coordinates $x^1,x^2,x^3$ needs to be specified in order to convert the aforementioned equations into partial differential equations. Among the different possibilities discussed in~\cite{jBoSlB11}, here we choose spatial harmonic coordinates with respect to the conformal three-metric, such that
\begin{equation}
\tilde{h}^{ij} \left[ \tilde{\Gamma}^k{}_{ij} - \Gammaz^k{}_{ij}  \right] = 0,
\label{Eq:HarmGauge}
\end{equation}
where $\tilde{\Gamma}^k{}_{ij}$ and $\Gammaz^k{}_{ij}$ are, respectively, the Christoffel symbols of the conformal three-metric $\tilde{h}_{ij}$ and of a given reference metric $\hz_{ij}$ on the time slices $\Sigma_t$. The spatial harmonic gauge implies an elliptic system for the shift, see~\cite{jBoSlB11}.

\section{Scalar field matter sources}
\label{Sec:Scalar}

In this section, we couple the gravitational field to a scalar field $\Phi$ whose dynamics are governed by the wave equation
\begin{equation}
\Box\Phi + \frac{\partial V}{\partial\Phi}(\Phi) = 0,\qquad 
\Box := -g^{\mu\nu}\nabla_\mu\nabla_\nu,
\label{Eq:Phi}
\end{equation}
with potential $V(\Phi)$. Later in this article we shall set $V(\Phi)$ to zero, but for the moment we keep $V(\Phi)$ arbitrary for generality. The stress energy-momentum tensor associated with $\Phi$ is
\begin{equation}
T_{\mu\nu} = (\nabla_\mu\Phi)(\nabla_\nu\Phi) - \frac{1}{2} g_{\mu\nu}
\left[ g^{\alpha\beta}(\nabla_\alpha\Phi)(\nabla_\beta\Phi) + 2 V(\Phi) \right].
\label{Eq:SEtensor}
\end{equation}

Under the conformal rescaling
\begin{equation}
g^{\mu\nu} = \Omega^2\tilde{g}^{\mu\nu},\qquad
\Phi = \Omega\tilde{\phi}
\end{equation}
one has the identity\footnote{See, for instance, Appendix D in Ref.~\cite{Wald-Book}; in particular see Eq.~(D.14) with $n=4$.}
\begin{equation}
\Box\Phi + \frac{1}{6} R^{(4)}\Phi 
 = \Omega^3\left[ \tilde{\Box}\tilde{\phi} + \frac{1}{6}\tilde{R}^{(4)}\tilde{\phi} \right],
\end{equation}
with $R^{(4)}$ and $\tilde{R}^{(4)}$ the Ricci scalars belonging to the physical metric $g_{\mu\nu}$ and the conformal metric $\tilde{g}_{\mu\nu}$, respectively. Using this identity, Eq.~(\ref{Eq:Phi}) can be rewritten as
\begin{equation}
\tilde{\Box}\tilde{\phi} + \frac{1}{6}\tilde{R}^{(4)}\tilde{\phi} 
 = \frac{1}{6\Omega^2} R^{(4)}\tilde{\phi}
 - \Omega^{-3}\frac{\partial V}{\partial\Phi}( \Omega\tilde{\phi} ).
\label{Eq:PhiRescaled}
\end{equation}
The left-hand side of this equation is manifestly regular at $\ScriPlus$ since it is independent of $\Omega$. The first term on the RHS is also regular at $\ScriPlus$, since by virtue of Einstein's field equations, $R^{(4)} = -8\pi G T^\mu{}_\mu = 8\pi G\Omega^2(\tilde{\rho} - \tilde{\sigma}^c{}_c)$ which scales as $\Omega^2$. Finally, the second term on the RHS is also regular at $\ScriPlus$ provided $V(\Phi)$ falls off sufficiently fast as $\Phi\to 0$, more specifically if
\begin{equation}
\frac{\partial V}{\partial\Phi}(\Phi) = {\cal O}(\Phi^3)
\label{Eq:VRegularity}
\end{equation}
for small $|\Phi|$.

The rescaled wave equation~(\ref{Eq:PhiRescaled}) can be cast into first-order symmetric hyperbolic form by introducing the quantities $\tilde{\pi} := \tilde{D}_0\tilde{\phi}$ and $\tilde{\chi}_a := \tilde{D}_a\tilde{\phi}$. An evolution equation for the fields $\tilde{\chi}_a$ follows by commutating the derivative operators $\tilde{D}_0$ and $\tilde{D}_a$:
$$
\tilde{D}_0\tilde{\chi}_a = \tilde{D}_0\tilde{D}_a\tilde{\phi}
 = \tilde{D}_a\tilde{\pi} + [\tilde{D}_0,\tilde{D}_a]\tilde{\phi},
$$
and using $[\tilde{D}_0,\tilde{D}_a] = (\tilde{D}_a\log\tilde{\alpha})\tilde{D}_0 + (\tilde{D}_0\tilde{B}_a{}^i)\partial_i$ and the evolution equation~(\ref{Eq:Ba}) in order to eliminate $\tilde{D}_0\tilde{B}_a{}^i$. The evolution equation for $\tilde{\pi}$ follows from the $3+1$ decomposition of the wave operator,
$$
\tilde{\Box}\tilde{\phi} = (\tilde{D}_0 + \tilde{K})\tilde{\pi} 
 - \frac{1}{\tilde{\alpha}}\tilde{D}^a(\tilde{\alpha}\tilde{\chi}_a),
$$
and the decomposition of the rescaled Ricci scalar in the 3D Nester gauge,
\begin{equation}
\tilde{R}^{(4)}
  = 2\tilde{D}_0\tilde{K} + \hat{\tilde K}^{ab}\hat{\tilde K}_{ab} + \frac{4}{3}\tilde{K}^2
 - \hat{\tilde N}^{ab}\hat{\tilde N}_{ab} 
 - \frac{2}{\tilde{\alpha}}\tilde{D}^a\tilde{D}_a\tilde{\alpha}.
\label{Eq:Rtilde}
\end{equation}
With these remarks, Eq.~(\ref{Eq:PhiRescaled}) can be rewritten as
\begin{eqnarray}
\tilde{D}_0\tilde{\phi} &=& \tilde{\pi},
\label{Eq:phi}\\
\tilde{D}_0\tilde{\chi}_a &=& \frac{1}{\tilde{\alpha}}\tilde{D}_a(\tilde{\alpha}\tilde{\pi})
 - \left( \hat{\tilde K}_a{}^b + \varepsilon_a{}^{cb}\tilde{\omega}_c 
 + \frac{\tilde{K}}{3}\delta_a{}^b \right)\tilde{\chi}_b,
\label{Eq:chi}\\
\tilde{D}_0\tilde{\pi} &=& \frac{1}{\tilde{\alpha}}\tilde{D}^a(\tilde{\alpha}\tilde{\chi}_a)
 - \tilde{K}\tilde{\pi}
 - \frac{1}{6}\left( \tilde{R}^{(4)} - \frac{R^{(4)}}{\Omega^2} \right)\tilde{\phi}
 - \frac{1}{\Omega^3}\frac{\partial V}{\partial\Phi}(\Omega\tilde{\phi}),
\label{Eq:pi}
\end{eqnarray}
where $\tilde{R}^{(4)}$ is computed using Eq.~(\ref{Eq:Rtilde}) and where by virtue of Einstein's field equations we may write $R^{(4)}/\Omega^2 = 8\pi G(\tilde{\rho} - \tilde{\sigma}^c{}_c)$. Although Eqs.~(\ref{Eq:phi},\ref{Eq:chi},\ref{Eq:pi}) form a symmetric hyperbolic system for $(\tilde{\phi},\tilde{\chi},\tilde{\pi})$, there is an issue regarding the RHS of the equation for $\tilde{\pi}$, since the expression for $\tilde{R}^{(4)}$ contains the term $\tilde{D}_0\tilde{K}$ which cannot be eliminated since there is no evolution equation for $\tilde{K}$. In order to remedy this problem, we replace $\tilde{\pi}$ with the new variable
\begin{equation}
\hat{\pi} := \tilde{\pi} + \frac{1}{3}\tilde{K}\tilde{\phi}.
\end{equation}
In terms of the variables $(\tilde{\phi},\tilde{\chi}_a,\hat{\pi})$ we obtain the new symmetric hyperbolic system
\begin{eqnarray}
\tilde{D}_0\tilde{\phi} &=& \hat{\pi} - \frac{1}{3}\tilde{K}\tilde{\phi},
\label{Eq:phitilde}\\
\tilde{D}_0\tilde{\chi}_a &=& \frac{1}{\tilde{\alpha}}\tilde{D}_a(\tilde{\alpha}\hat{\pi})
 - \left( \hat{\tilde K}_a{}^b + \varepsilon_a{}^{cb}\tilde{\omega}_c 
 + \frac{2}{3}\tilde{K}\delta_a{}^b \right)\tilde{\chi}_b
 - \frac{1}{3\tilde{\alpha}}\tilde{D}_a(\tilde{\alpha}\tilde{K})\tilde{\phi},
\label{Eq:chitilde}\\
\tilde{D}_0\hat{\pi} &=& \frac{1}{\tilde{\alpha}}\tilde{D}^a(\tilde{\alpha}\tilde{\chi}_a)
 - \frac{2}{3}\tilde{K}\hat{\pi}
 - \frac{1}{6}\left[ \hat{\tilde K}^{ab}\hat{\tilde K}_{ab} - \hat{\tilde N}^{ab}\hat{\tilde N}_{ab} 
 - \frac{2}{\tilde{\alpha}}\tilde{D}^a\tilde{D}_a\tilde{\alpha} 
 + 8\pi G(\tilde{\sigma}^c{}_c - \tilde{\rho}) \right] \tilde{\phi}
 - \frac{1}{\Omega^3}\frac{\partial V}{\partial\Phi}(\Omega\tilde{\phi}),
\label{Eq:pihat}
\end{eqnarray}
which no longer contains any time derivatives of the fields in the RHS. We stress again that these equations are manifestly regular at $\ScriPlus$, as long as the potential $V(\Phi)$ satisfies the condition~(\ref{Eq:VRegularity}).

Noting that $D_0\Phi = \Omega^2\hat{\pi} - C\Omega\tilde{\phi}$ and $D_a\Phi = \Omega^2\tilde{\chi}_a + \Omega(\tilde{D}_a\Omega)\tilde{\phi}$, the explicit expressions for the rescaled energy density, energy flux and stress tensor are
\begin{eqnarray}
\tilde{\rho} &=& \frac{1}{2}\left( \Omega\hat{\pi} - C\tilde{\phi} \right)^2
 + \frac{1}{2}\left( \Omega\tilde{\chi}^a + \tilde{\phi}\tilde{D}^a\Omega \right)
 \left( \Omega\tilde{\chi}_a + \tilde{\phi}\tilde{D}_a\Omega \right)
 + \Omega^{-2} V(\Omega\tilde{\phi}),
\label{Eq:rho}\\
\tilde{j}_b &=& -\left( \Omega\hat{\pi} - C\tilde{\phi} \right)
\left( \Omega\tilde{\chi}_b + \tilde{\phi}\tilde{D}_b\Omega \right),
\label{Eq:jb}\\
\sigma^a{}_a &=& \frac{3}{2}\left( \Omega\hat{\pi} - C\tilde{\phi} \right)^2
 - \frac{1}{2}\left( \Omega\tilde{\chi}^a + \tilde{\phi}\tilde{D}^a\Omega \right)
 \left( \Omega\tilde{\chi}_a + \tilde{\phi}\tilde{D}_a\Omega \right)
 - 3\Omega^{-2} V(\Omega\tilde{\phi}),
\label{Eq:sigmatr}\\
\hat{\tilde \sigma}_{ab} &=&  \left[ 
\left( \Omega\tilde{\chi}_a + \tilde{\phi}\tilde{D}_a\Omega \right)
\left( \Omega\tilde{\chi}_b + \tilde{\phi}\tilde{D}_b\Omega \right)
\right]^{TF},
\label{Eq:sigmahat}
\end{eqnarray}
and we see that these quantities are manifestly regular at $\ScriPlus$, provided $V(\Phi) = {\cal O}(\Phi^2)$.

\section{Self-gravitating, spherically symmetric scalar field}
\label{Sec:SphSym}

In the particular case of a spherically symmetric spacetime and scalar field configuration the quantities $\tilde{\alpha}$, $\hat{\tilde N}_{ab}$, $\hat{\tilde K}_{ab}$, $\tilde{K}$, $\tilde{\omega}_a$, $\tilde{\phi}$, $\tilde{\chi}_a$, $\tilde{\pi}$, etc. are functions of $(t,R)$ only, with $R$ a radial coordinate which will be determined later. The shift vector is radial,
\begin{equation}
\beta^i = b(t,R)\hat{x}^i,
\end{equation}
where here and in the following $(x^i) = (x^1,x^2,x^3)$ are Cartesian spatial coordinates on $\Sigma_t$ such that $\delta_{ij}x^i x^j = R^2$ and $\hat{x}^i = x^i/R$. A natural choice for the spatial legs of the tetrad is\footnote{A different possibility would be to choose $\tilde{\bf e}_1$, say, in the radial direction. However, the remaining two legs $\tilde{\bf e}_2$ and $\tilde{\bf e}_3$ would be tangent to the two-spheres and thus would not be globally well defined. See~\cite{lBjB05b} for a related discussion.}
\begin{equation}
\tilde{\bf e}_a = \tilde{B}_a{}^i\frac{\partial}{\partial x^i}
 = \left[ \tilde{B}_R(t,R)\hat{x}_a\hat{x}^i + \tilde{B}_T(t,R)\hat{\delta}_a{}^i \right]
 \frac{\partial}{\partial x^i},
\label{Eq:eaSphSym}
\end{equation}
with $\hat{\delta}_{ab} := \delta_{ab} - \hat{x}_a\hat{x}_b$. The coordinate components of the conformal inverse spatial metric are given by
$$
\tilde{h}^{ij} = \delta^{ab}\tilde{B}_a{}^i\tilde{B}_b{}^j
 = \tilde{B}_R^2\hat{x}^i\hat{x}^j + \tilde{B}_T^2\hat{\delta}^{ij},
$$
and consequently, the conformal spatial metric is
$$
\tilde{\bf h} = \tilde{B}_R^{-2} dR^2 + \tilde{B}_T^{-2} R^2
\left( d\vartheta^2 + \sin^2\vartheta d\varphi^2 \right)
$$
in standard spherical coordinates $(R,\vartheta,\varphi)$ associated with $(x^i)$. Choosing the ``background'' metric $\hz$ to be the one for which $\tilde{B}_R$ and $\tilde{B}_T$ are equal to one, the spatial harmonic gauge condition~(\ref{Eq:HarmGauge}) yields
\begin{equation}
\partial_R(\log\tilde{B}_R) - 2\partial_R(\log\tilde{B}_T) 
 + \frac{2}{R}\left( 1 - \frac{\tilde{B}_T^2}{\tilde{B}_R^2} \right) = 0.
\label{Eq:SpatialHarmSph}
\end{equation}

By computing the commutators $[\tilde{\bf e}_0,\tilde{\bf e}_b]$ and $[\tilde{\bf e}_a,\tilde{\bf e}_b]$ and using the torsion-free property of the connection one finds the following expressions for the connection coefficients in spherical symmetry:
\begin{eqnarray}
\tilde{K}_{ab} &=& 
 -\left( \tilde{D}_0\log\tilde{B}_R + \frac{\partial_R b}{\tilde{\alpha}} \right)\hat{x}_a\hat{x}_b
  - \left( \tilde{D}_0\log\tilde{B}_T + \frac{b}{\tilde{\alpha}R} \right)\hat{\delta}_{ab},\\
\tilde{N}_{ab} &=& \varepsilon_{ab}{}^c\hat{x}_c
\left[ \tilde{B}_R\partial_R(\log\tilde{B}_T) + \frac{\tilde{B}_T - \tilde{B}_R}{R} \right],\\
\tilde{\omega}_b &=& 0.
\end{eqnarray}
In particular, it follows that our tetrad choice in Eq.~(\ref{Eq:eaSphSym}) automatically satisfies the Nester gauge since $\tilde{N}_{ab}$ is antisymmetric and dual to a purely radial vector field. Furthermore, with our choice for the conformal factor, the antisymmetric part of $\tilde{N}_{ab}$ vanishes identically, and hence $\tilde{N}_{ab} = 0$, implying that the conformal spatial metric $\tilde{h}_{ij}$ is flat. Consequently, all the information about the geometry of the spatial physical metric is encoded in the conformal factor $\Omega$. 

Introducing the quantity $\tilde{\nu}$ which parametrizes the traceless part of the conformal extrinsic curvature according to $\hat{\tilde K}_{ab} = \tilde{\nu}(\hat{x}_a\hat{x}_b - \hat{\delta}_{ab}/2)$, the evolution equations~(\ref{Eq:Ba},\ref{Eq:Nab},\ref{Eq:Kab}) in spherical symmetry simplify to
\begin{eqnarray}
\tilde{D}_0(\log\tilde{B}_R) 
 &=& -\frac{1}{\tilde{\alpha}}\partial_R b - \tilde{\nu} - \frac{\tilde{K}}{3},
\label{Eq:BR}\\
\tilde{D}_0(\log\tilde{B}_T) 
 &=& -\frac{1}{\tilde{\alpha}}\frac{b}{R} + \frac{\tilde{\nu}}{2} - \frac{\tilde{K}}{3},
\label{Eq:BT}\\
\tilde{D}_0\tilde{\nu} &=& -\frac{\tilde{K}}{3}\tilde{\nu} 
 + \frac{2}{3\tilde{\alpha}}\left[ \tilde{\alpha}'' - \frac{\tilde{B}_T}{R}\tilde{\alpha}' \right]
 - \frac{4}{3\Omega}\left[ \Omega'' - \frac{\tilde{B}_T}{R}\Omega' 
 + \frac{K}{2}\tilde{\nu} \right]
 + 8\pi G\tilde{\sigma}_R,
\label{Eq:nu}
\end{eqnarray}
where here and in the following a prime denotes application of the operator $\tilde{B}_R\partial_R$, and where we have expanded $\hat{\tilde\sigma}_{ab} = \tilde{\sigma}_R(\hat{x}_a\hat{x}_b - \hat{\delta}_{ab}/2)$. The Hamiltonian and momentum constraints reduce to
\begin{eqnarray}
\Omega\left( \Omega'' + \frac{2}{R}\tilde{B}_T\Omega' \right)
 &=& \frac{3}{2}(\Omega'^2 - C^2) + \frac{3}{8}\Omega^2\tilde{\nu}^2 
  + 4\pi G\Omega^2\tilde{\rho},
\label{Eq:HamSphSym}\\
\tilde{\nu}' + \frac{3}{R}\tilde{B}_T\tilde{\nu} - \frac{2}{\Omega}\Omega'\tilde{\nu} 
 &=& -8\pi G\tilde{j}_R
\label{Eq:MomSphSym},
\end{eqnarray}
with $\tilde{j}_a = \tilde{j}_R\hat{x}_a$. The remaining constraints are the preservation conditions for the Nester gauge, our choice of the conformal factor, and CMC slicing:
\begin{eqnarray}
(\log\tilde{B}_T)' &=& \frac{1}{R}(\tilde{B}_R - \tilde{B}_T),
\label{Eq:Cons1}\\
\left( \frac{2}{3}\tilde{\alpha}\tilde{K} \right)' &=& (\tilde{\alpha}\tilde{\nu})' 
 + \frac{3}{R}\tilde{B}_T\tilde{\alpha}\tilde{\nu},
\label{Eq:Cons2}\\
\Omega\left[ \tilde{\alpha}'' + \frac{2}{R}\tilde{B}_T\tilde{\alpha}' \right] 
 - 3\Omega'\tilde{\alpha}'
 + \left[ \Omega'' + \frac{2}{R}\tilde{B}_T\Omega' - \frac{9}{4}\Omega\tilde{\nu}^2
 \right]\tilde{\alpha}
 &=& 4\pi G\Omega(3\tilde{\rho} + \tilde{\sigma}^c{}_c)\tilde{\alpha}.
\label{Eq:Cons3}
\end{eqnarray}
For the following, we set $\tilde{B}_T = 1$ which means that $R$ is the areal radius of the conformal metric. The constraint Eq.~(\ref{Eq:Cons1}) then implies that $\tilde{B}_R = 1$, so that the conformal spatial metric is equal to the flat background metric $\hz$ with areal radius $R$. In this gauge, Eqs.~(\ref{Eq:BR},\ref{Eq:BT}) reduce to two equations which relate the radial component of the shift, $b$, with $\tilde{\nu}$ and $\tilde{K}$. It is simple to verify that these two conditions imply the validity of Eq.~(\ref{Eq:Cons2}). Furthermore, the choice $\tilde{B}_T = 1$ leads to the general solution $\tilde{B}_R^2 = 1 + A R^{-4}$ (with $A$ a constant) of the spatial harmonic condition~(\ref{Eq:SpatialHarmSph}), which is compatible with $\tilde{B}_R = 1$.

The evolution and constraint equations in this gauge, which in the following we shall call the ``conformally flat gauge'', are summarized next.

\subsection{Summary of evolution and constraint equations in the conformally flat gauge}
\label{SubSec:Summary}

In the conformally flat gauge, where $\tilde{B}_R = \tilde{B}_T = 1$, our system describing a self-gravitating, spherically symmetric scalar field in the compactified CMC foliation consists of the following hyperbolic-elliptic system. First, we have a set of first-order hyperbolic equations for the scalar field quantities $(\tilde{\phi},\tilde{\chi},\hat{\pi})$ (where $\tilde{\chi}_a = \tilde{\chi}\hat{x}_a$) and the component $\tilde{\nu}$ of the traceless part of the conformal extrinsic curvature, given by
\begin{eqnarray}
\tilde{D}_0\tilde{\phi} &=& \hat{\pi} - \tilde{C}\tilde{\phi},
\label{Eq:phiSphSymBis}\\
\tilde{D}_0\tilde{\chi} &=& \frac{1}{\tilde{\alpha}}(\tilde{\alpha}\hat{\pi})'
 - (\tilde{\nu} + 2\tilde{C})\tilde{\chi}
 - \frac{1}{\tilde{\alpha}}(\tilde{\alpha}\tilde{C})'\tilde{\phi},
\label{Eq:chiSphSymBis}\\
\tilde{D}_0\hat{\pi} &=& \frac{1}{\tilde{\alpha} R^2} (\tilde{\alpha} R^2\tilde{\chi})'
 - 2\tilde{C}\hat{\pi}
 + \left[ \frac{1}{3\tilde{\alpha}}\left( \tilde{\alpha}'' + \frac{2}{R}\tilde{\alpha}' \right)
 - \frac{1}{4}\tilde{\nu}^2
 - \frac{4\pi G}{3}(\tilde{\sigma}^c{}_c - \tilde{\rho}) \right]\tilde{\phi}
 - \Omega^{-3}\frac{dV}{d\Phi}(\Omega\tilde{\phi}),
\label{Eq:piSphSymBis}
\end{eqnarray}
with
$$
\tilde{\sigma}^c{}_c - \tilde{\rho} = (\Omega\hat{\pi} - C\tilde{\phi})^2 
 - (\Omega\tilde{\chi} + \tilde{\phi}\Omega')^2 - 4\Omega^{-2} V(\Omega\tilde{\phi}),
$$
and
\begin{equation}
\tilde{D}_0\tilde{\nu} = -\tilde{C}\tilde{\nu} 
 + \frac{2}{3\tilde{\alpha}}\left( \tilde{\alpha}'' - \frac{1}{R}\tilde{\alpha}' \right)
 - \frac{4}{3\Omega}\left( \Omega'' - \frac{1}{R}\Omega' + \frac{3}{2}C\tilde{\nu} \right)
 + \frac{16\pi G}{3}(\Omega\tilde{\chi} + \tilde{\phi}\partial_R\Omega)^2.
\label{Eq:nuBis}
\end{equation}
Here, we recall the definition $C := K/3$ of the mean extrinsic curvature, and accordingly, $\tilde{C} := \tilde{K}/3$ denotes the conformal mean extrinsic curvature. The radial component of the shift $b$ which appears in the operator $\tilde{D}_0 = \tilde{\alpha}^{-1}(\partial_t - b\partial_R)$ is determined by the algebraic equation
\begin{equation}
\frac{b}{\tilde{\alpha}} = R\left( \frac{\tilde{\nu}}{2} - \tilde{C} \right),
\label{Eq:AlgebraicShift}
\end{equation}
which follows from Eq.~(\ref{Eq:BT}) and the gauge choice $\tilde{B}_T = 1$.

Next, the conformal factor $\Omega$, conformal lapse $\tilde{\alpha}$, and conformal mean extrinsic curvature $\tilde{C}$ are determined by the elliptic equations
\begin{eqnarray}
\Omega\left( \Omega'' + \frac{2}{R}\Omega' \right)
 - \frac{3}{2}(\Omega'^2 - C^2) - \frac{3}{8}\Omega^2\tilde{\nu}^2 
 &=& 4\pi G\Omega^2\tilde{\rho},
\label{Eq:HamSphSymBis}\\
\Omega\left( \tilde{\alpha}'' + \frac{2}{R}\tilde{\alpha}' \right) 
 - 3\Omega'\tilde{\alpha}'
 + \left( \Omega'' + \frac{2}{R}\Omega' - \frac{9}{4}\Omega\tilde{\nu}^2 \right)\tilde{\alpha}
 &=& 4\pi G\Omega(3\tilde{\rho} + \tilde{\sigma}^c{}_c)\tilde{\alpha},
\label{Eq:CMCConstrSphSymBis}
\end{eqnarray}
with
\begin{eqnarray*}
\tilde{\rho} &=& \frac{1}{2}(\Omega\hat{\pi} - C\tilde{\phi})^2 
 + \frac{1}{2}(\Omega\tilde{\chi} + \tilde{\phi}\Omega')^2 + \Omega^{-2} V(\Omega\tilde{\phi}),\\
3\tilde{\rho} + \tilde{\sigma}^c{}_c &=& 3(\Omega\hat{\pi} - C\tilde{\phi})^2 
 + (\Omega\tilde{\chi} + \tilde{\phi}\Omega')^2,
\end{eqnarray*}
and
\begin{equation}
(2\tilde{\alpha}\tilde{C})' = \frac{1}{R^3}(R^3\tilde{\alpha}\tilde{\nu})'.
\label{Eq:Cons2Bis}
\end{equation}
These elliptic equations are subject to the following boundary conditions at $\ScriPlus$ (see~\cite{jBoSlB11}):
\begin{equation}
\left. \Omega \right|_{\ScriPlus} = 0,\qquad
\left. \tilde{\alpha} \right|_{\ScriPlus} = R_+ C,\qquad
\left. 2\tilde{\alpha}\tilde{C} \right|_{\ScriPlus} = 2C,
\label{Eq:ScriPlusBC}
\end{equation}
where $R_+$ is the coordinate radius of $\ScriPlus$ which, in the conformally flat gauge, determines the scalar curvature of the cross sections of $\ScriPlus$ through $2/R_+^2$.

Finally, we have the evolution equation for the conformal factor,
\begin{equation}
\tilde{D}_0\Omega = -(C - \Omega\tilde{C}),
\label{Eq:OmegaEvol}
\end{equation}
and the momentum constraint which can be rewritten as
\begin{equation}
\frac{\Omega^2}{R^3}\left( \frac{R^3}{\Omega^2}\tilde{\nu} \right)'
 = 8\pi G(\Omega\hat{\pi} - C\tilde{\phi})(\Omega\tilde{\chi} + \tilde{\phi}\Omega').
\label{Eq:MomSphSymBis}
\end{equation}

Note that the only evolution equation which is formally singular at $\ScriPlus$ is Eq.~(\ref{Eq:nuBis}), which requires
$$
\Omega'' - \frac{1}{R}\Omega' + \frac{3}{2} C\tilde{\nu} = 0
$$
at $\ScriPlus$. The elliptic equations~(\ref{Eq:HamSphSymBis},\ref{Eq:CMCConstrSphSymBis}) as well as the momentum constraint~(\ref{Eq:MomSphSymBis}) are formally singular at $\ScriPlus$ and require suitable regularity conditions which will be analyzed in detail in the next section and in Apps.~\ref{App:FormalExpansions} and \ref{App:LocSolutionsScriPlus}. If a black hole is present whose interior is excised from the computational domain, further conditions on the inner boundary are required. A rather rudimentary approach for treating such inner boundary conditions based on a similar approach in~\cite{oRvM13} will be discussed in Sec.~\ref{Sec:NumResults}.

\subsection{In- and outgoing expansions, mass function}

Next, we discuss some geometric invariant quantities which will be useful for the interpretation of the numerical results in Sec.~\ref{Sec:NumResults} and for monitoring the behavior of the fields. First, let us consider a sphere $S_{t,r}$ of fixed areal radius $r$ embedded in a constant time slice $\Sigma_t$. The future-directed in- (-) and outgoing (+) null vectors orthogonal to $S_{t,r}$ are given by
$$
{\bf k}^{\pm} := {\bf e}_0 \pm \hat{x}^a {\bf e}_a,
$$
and the corresponding expansions are
$$
\Theta^\pm = {\bf k}^{\pm}[r] = D_0 r \pm \hat{x}^a D_a r.
$$
The two-surface $S_{t,r}$ is called {\it trapped} if both these expansions are negative, and {\it marginally trapped} if
$$
\Theta^+ = 0,\qquad \Theta^-\leq 0.
$$
Using the relation $r = R/\Omega$ between the areal radii $r$ and $R$ of the physical and conformal metrics, respectively, the evolution equation~(\ref{Eq:OmegaEvol}) for the conformal factor, and the expression~(\ref{Eq:AlgebraicShift}) for the shift, we find that in the conformally flat gauge,
\begin{equation}
\Theta^\pm = R\left( \frac{C}{\Omega} - \frac{\tilde{\nu}}{2} \right)
 \pm \left( 1 - R\partial_R\log\Omega \right).
\label{Eq:Expansions}
\end{equation}
The derivatives along the radial direction of these expansions can be computed using the Hamiltonian and momentum constraints, and yield
\begin{equation}
\partial_R\Theta^\pm = \pm \frac{1}{2R}\left\{
 1 + [ 6C r - 3(D_0 r) \mp (\hat{x}^a D_a r) ]\Theta^\pm 
 - 8\pi G R^2 {\bf T}(\tilde{\bf e}_0,\tilde{\bf k}^\pm) \right\},
\label{Eq:RadDerivExpansions}
\end{equation}
with $\tilde{\bf k}^\pm = \Omega^{-1}{\bf k}^\pm = \tilde{\bf e}_0 \pm \hat{x}^a\tilde{\bf e}_a$. Likewise, using the constraints and the evolution equations for $\Omega$ and $\tilde{\nu}$, we find
\begin{equation}
\tilde{D}_0\Theta^\pm = \pm \frac{\tilde{\alpha}'}{\tilde{\alpha}}\Theta^\pm
 - \frac{1}{2R}\left\{ 1 + \Theta^+\Theta ^- \pm 2(1 - \hat{x}^a D_a r)\Theta^\pm
 \pm 8\pi G R^2 {\bf T}(\hat{x}^a\tilde{e}_a,\tilde{\bf k}^\pm) \right\}.
\label{Eq:NormDerivExpansions}
\end{equation}
In particular, at a marginally trapped surface $R = R_{MTS}$,
$$
\left. 2R\partial_R\Theta^+ \right|_{R = R_{MTS}} 
 = \left. 1 - 8\pi G R^2 {\bf T}(\tilde{\bf e}_0,\tilde{\bf k}^+) \right|_{R = R_{MTS}},
$$
which is positive if and only if the positive semi-definite quantity ${\bf T}(\tilde{\bf e}_0,{\bf k}^+)$ is smaller than $1/(8\pi G R_{MTS}^2)$, and
$$
\left. 2R\tilde{D}_0\Theta^+ \right|_{R = R_{MTS}} 
 = \left. -1 - 8\pi G R^2 {\bf T}(\hat{x}^a\tilde{\bf e}_a,\tilde{\bf k}^+) \right|_{R = R_{MTS}}.
$$

The product of the expansions $\Theta^\pm$ determine the Hawking~\cite{sH68} or Misner-Sharp (MS)~\cite{cMdS64} mass function $m$ according to the relation
\begin{equation}
1 - \frac{2m}{r} = {\bf g}(dr,dr) = -\Theta^+\Theta^-, \label{Eq:MisnerSharpMass}
\end{equation}
so that $r < 2m$ at a trapped surface, and $r = 2m$ at a marginally trapped surface. Using Eqs.~(\ref{Eq:RadDerivExpansions}) and~(\ref{Eq:NormDerivExpansions}) one finds the simple equations
\begin{eqnarray}
m' &=& 4\pi G R^2 {\bf T}( \tilde{\bf e}_0,\tilde{\bf X} ),
\label{Eq:RadDerivMass}\\
\tilde{D}_0 m &=& 4\pi G R^2 {\bf T}( \hat{x}^a\tilde{e}_a,\tilde{\bf X} ),
\label{Eq:NormDerivMass}
\end{eqnarray}
for the first derivatives of the mass function, with the vector field $\tilde{\bf X} := r'\tilde{\bf e}_0 - (\tilde{D}_0 r)\hat{x}^a\tilde{e}_a$. Note that $\tilde{\bf X}$ is future-directed timelike for $r > 2m$ so that along the $t = const$ time slices, the mass function $m$ increases monotonically with $R$.

\section{Asymptotic behavior at $\ScriPlus$}
\label{Sec:Asymptotics}

In this section, we derive formal local expansions at $\ScriPlus$ for the metric fields $(\Omega,\tilde{\nu},\tilde{\alpha})$ which are constrained by the singular equations~(\ref{Eq:HamSphSymBis},\ref{Eq:MomSphSymBis},\ref{Eq:CMCConstrSphSymBis}). For general discussions regarding the vacuum case without symmetries, we refer the reader to Refs.~\cite{lApChF92,lApC94,vMoR09,jBoSlB11,jBlB12}. The analysis below, although restricted to spherical symmetry, includes the presence of a non-trivial scalar field which is assumed to be sufficiently regular at $\ScriPlus$.

In order to expand the fields, it is convenient to rewrite them in terms of the dimensionless functions $(u(z),v(z),a(z))$ defined by
$$
\Omega(R) = R_+ C u(z),\qquad
\tilde{\nu}(R) = \frac{v(z)}{R_+},\qquad
\tilde{\alpha}(R) = R_+ C a(z),
$$
where $z = 1 - R/R_+$. With this notation, the Hamiltonian and momentum constraints, as well as the elliptic equation~(\ref{Eq:CMCConstrSphSymBis}) responsible for preserving the CMC gauge, can be rewritten as
\begin{eqnarray}
u\left( u_{zz} - \frac{2}{1-z} u_z \right) - \frac{3}{2}(u_z^2-1) &=& F_1(z,u,u_z,v) u^2,
\label{Eq:HamRescaled}\\
u v_z - 2u_z v - \frac{3u}{1-z} v &=& F_2(z,u,u_z,v) u,
\label{Eq:MomRescaled}\\
u\left( a_{zz} - \frac{2}{1-z} a_z \right) - 3u_z a_z 
 + \left( u_{zz} - \frac{2}{1-z} u_z \right) a &=& F_3(z,u,u_z,v) u a,
\label{Eq:CMCRescaled}
\end{eqnarray}
where here we have introduced the notation $u_z := \partial_z u$, $u_{zz} := \partial_z^2 u$ etc., and have defined the three functions
\begin{eqnarray*}
F_1(z,u,u_z,v) &:=& \frac{3}{8} v^2 + 2\pi G (R_+ C)^2
\left[ (\tilde{\phi} - u R_+\hat{\pi})^2 + (\tilde{\phi} u_z - u R_+\tilde{\chi})^2
\right] + \frac{4\pi G}{C^2 u^2} V(R_+ C u\tilde{\phi}),\\
F_2(z,u,u_z,v) &:=& -8\pi G (R_+ C)^2
(\tilde{\phi} - u R_+\hat{\pi})(\tilde{\phi} u_z - u R_+\tilde{\chi}),
\\
F_3(z,u,u_z,v) &:=& \frac{9}{4} v^2 + 4\pi G (R_+ C)^2\left[
 3(\tilde{\phi} - u R_+\hat{\pi})^2 + (\tilde{\phi} u_z - u R_+\tilde{\chi})^2
\right].
\end{eqnarray*}
For the following, we assume that $\tilde{\phi}$, $\tilde{\chi}$ and $\hat{\pi}$ are {\it a priori} given functions of $z$ which are regular at $z = 0$, and that $V$ decays sufficiently fast at $\Phi = 0$ so that the functions $F_i(z,u,u_z,v)$ are regular at $z = 0$.

We obtain formal expansions near $\ScriPlus$ based on the following considerations. First, evaluation of Eq.~(\ref{Eq:HamRescaled}) at $z = 0$ gives the condition $u_z(0) = 1$, since $u(0) = 0$ and $u(z) > 0$ for $z > 0$. Using this and evaluating Eq.~(\ref{Eq:MomRescaled}) at $z = 0$, we obtain $v(0) = 0$, implying that the trace-free part of the conformal extrinsic curvature vanishes at $\ScriPlus$. Note that this is equivalent to the regularity condition~(\ref{Eq:RegCond}) for the evolution equation for $\hat{\tilde K}_{ab}$ in the spherically symmetric case.

Next, we differentiate Eq.~(\ref{Eq:HamRescaled}) with respect to $z$ and evaluate the result at $z = 0$, obtaining the condition $u_{zz}(0) = -1$. We then use these results to show that (assuming the $V(\Phi)$ decays at least as fast as $\Phi^4$)
\begin{eqnarray}
F_1(z,u,u_z,v) &=& \frac{1}{2} f_0 - \frac{1}{2}(f_0 + f_1) z + {\cal O}(z^2),
\label{Eq:F1Exp}\\
F_2(z,u,u_z,v) &=& -f_0 + (f_0 + f_1) z + {\cal O}(z^2),
\label{Eq:F2Exp}\\
F_3(z,u,u_z,v) &=& 2f_0 - (f_0 + f_1^*) z + {\cal O}(z^2),
\label{Eq:F3Exp}
\end{eqnarray}
with coefficients
$$
f_0 = 8\pi G C^2 R_+^2\left. \tilde{\phi}^2 \right|_\ScriPlus,\qquad
f_1 = 8\pi G C^2 R_+^3\left.\tilde{\phi}(3\tilde{\chi} + \hat{\pi}) \right|_\ScriPlus,\qquad
f_1^* = 8\pi G C^2 R_+^3\left.\tilde{\phi}(5\tilde{\chi} + 3\hat{\pi}) \right|_\ScriPlus,
$$
where we have also used the Taylor expansion
$$
\tilde{\phi} = \left. \tilde{\phi} \right|_\ScriPlus + (\left. \partial_R\tilde{\phi}) \right|_\ScriPlus (R - R_+) + {\cal O}(z^2)
 = \left. \tilde{\phi} \right|_\ScriPlus - (\left. R_+\tilde{\chi})\right|_\ScriPlus z
  + {\cal O}(z^2).
$$

Next, differentiation of Eq.~(\ref{Eq:MomRescaled}) with respect to $z$ gives $v_z(0) = f_0$. Differentiating this equation a second time and evaluating at $z=0$, we get
$$
-3v_z(0) = 3f_0 + 2f_1.
$$
Not only does this condition leave $v_{zz}(0)$ indeterminate, but moreover it is incompatible with the previous result $v_z(0) = f_0$, unless $3f_0 + f_1 = 0$. Under suitable fall-off conditions for the initial data, the vanishing of $3f_0 + f_1$ is only satisfied in the absence of outgoing scalar radiation, see below.

In order to obtain asymptotic expressions which are valid in the presence of radiation, one needs to include a term of the form $z^2\log z$ in the expansion for $v(z)$. This yields:
\begin{equation}
v(z) = f_0 z + (3f_0 + f_1)z^2\log z + v_2 z^2 + 2(3f_0 + f_1)z^3\log z + {\cal O}(z^3),
\label{Eq:vExpansion}
\end{equation}
with $v_2$ a free parameter. Likewise, logarithmic terms appear in the asymptotic expansion for $u(z)$:
\begin{equation}
u(z) = z - \frac{1}{2} z^2 - \frac{1}{6} f_0 z^3 - \frac{1}{8}(3f_0 + f_1) z^4\log z + u_4 z^4
- \frac{1}{8}(3f_0 + f_1) z^5\log z + {\cal O}(z^5),
\label{Eq:uExpansion}
\end{equation}
with $u_4$ a free parameter. From the corresponding expressions in Schwarzschild spacetimes (see Eq.~(\ref{Eq:OmegaSchwExpansion})), we know that $v_2 = 2D C^2$ is related to the residual freedom to choose the CMC slicing, and that $u_4 = -(C m + D C^2)/4$ is related to the MS mass $m$ at $\ScriPlus$. More generally, we find that the expansions~(\ref{Eq:vExpansion},\ref{Eq:uExpansion}) imply that the expansions $\Theta^\pm$ are given by
\begin{eqnarray}
\frac{u}{2}\Theta^+ &=& 1 - z + \frac{1}{4}(1 - 2f_0)z^2 - \frac{1}{2}(3f_0 + f_1) z^3\log z 
 + {\cal O}(z^3),
\label{Eq:ThetaPlusExp}\\
-\frac{2}{u}\Theta^- &=& 1 + \left( 8u_4 + v_2 - \frac{19}{12} f_0 - \frac{1}{4} f_1 + 1 \right) z
 + {\cal O}(z^2).
\label{Eq:ThetaMinusExp}
\end{eqnarray}
This in turn implies the following expression for the MS mass:
\begin{equation}
2Cm = -8u_4 - v_2 + \frac{19}{12} f_0 + \frac{1}{4} f_1 + {\cal O}(z).
\label{Eq:MassExp}
\end{equation}
The absence of the $z\log z$ term in $m$ means that the radial derivative of $m$ is bounded near $\ScriPlus$.

From the expansions~(\ref{Eq:vExpansion},\ref{Eq:uExpansion}), we obtain the following expansion for the lapse from Eq.~(\ref{Eq:CMCRescaled}):
\begin{equation}
a(z) = 1 - z + \frac{1}{4}(2 - 3f_0) z^2 - \frac{1}{2}(3f_0 + f_1)z^3\log z
 + \left( 4u_4 + \frac{1}{8} f_0 - \frac{5}{8} f_1 + \frac{1}{3} f_1^* \right) z^3
 + {\cal O}(z^4\log z).
\label{Eq:aExpansion}
\end{equation}
From these results, it is not difficult to check the consistency of the regularity condition $v(0) = 0$ at $\ScriPlus$ with the evolution equation~(\ref{Eq:nuBis}) at $\ScriPlus$. First, one finds that the apparently singular term
$$
S := \frac{1}{\Omega}\left( \Omega'' - \frac{1}{R}\Omega' + \frac{3}{2} C\tilde{\nu} \right)
 = \frac{1}{R_+^2 u}\left( u_{zz} + \frac{1}{1 - z} u_z + \frac{3}{2} vÊ\right)
$$
has the expansion
$$
R_+^2 S = \frac{1}{2} f_0 
 + \frac{3}{2}\left( 8 u_4 + v_2 - \frac{23}{12} f_0 - \frac{7}{12} f_1 \right) z + {\cal O}(z^2),
$$
which is regular at $\ScriPlus$. Likewise, one finds from Eq.~(\ref{Eq:aExpansion}) the expansion
$$
\frac{R_+^2}{\tilde{\alpha}}\left( \tilde{\alpha}''  - \frac{1}{R}\tilde{\alpha}' \right)
 = -\frac{3}{2} f_0 - 3(3f_0 + f_1)z\log z + 3\left( 8u_4 - \frac{13}{4} f_0 - \frac{25}{12} f_1
 + \frac{2}{3}f_1^* \right) z + {\cal O}(z^2\log z).
$$
Using Eqs.~(\ref{Eq:AlgebraicShift}) and (\ref{Eq:Cons2Bis}) one also finds
\begin{equation}
\frac{b}{\tilde{\alpha}} = -1 + \frac{1}{2} z^2 + {\cal O}(z^3)
\end{equation}
for the ratio between the radial component of the shift and the conformal lapse. Using these expansions, it is simple to check that the RHS of Eq.~(\ref{Eq:nuBis}) yields
$$
\left. \partial_t v \right|_\ScriPlus = 0,
$$
which preserves the regularity condition $v(0) = 0$ at $\ScriPlus$.

We end this section by analyzing the role played by the Newman-Penrose (NP) constant~\cite{eNrP68} in our expansions. Assuming that the physical scalar field $\Phi$ admits a Bondi-type expansion of the form
$$
\Phi = \frac{\phi_0(u)}{r} + \frac{\phi_1(u)}{r^2} + \ldots,
$$
at $\ScriPlus$, where $u$ is the retarded time coordinate at $\ScriPlus$ and $r$ the areal radius along outgoing null geodesics, it can be shown~\cite{eNrP68} that the evolution equations for the scalar field plus regularity assumptions at $\ScriPlus$ imply that $\phi_1$ must be independent of $u$. Knowing $\Phi$, the NP constant $\phi_1$ can be extracted using the formula
$$
\phi_1 = -\lim\limits_{r\to \infty} r^2\left. \frac{\partial}{\partial r} \right|_u [r\Phi].
$$
In terms of the rescaled fields used in this article we obtain
\begin{equation}
\phi_1 = -\frac{R_+}{4C}\left[ \tilde{\phi} + R_+(\hat{\pi} + \tilde{\chi}) \right]_{\ScriPlus}.
\label{Eq:NPConstant}
\end{equation}
Using the expansions above it is not difficult to check that Eqs.~(\ref{Eq:phiSphSymBis},\ref{Eq:chiSphSymBis},\ref{Eq:piSphSymBis}) induce the following evolution equations at $\ScriPlus$:
\begin{eqnarray}
C^{-1}\partial_t\tilde{\phi} &=& -\tilde{\phi} + R_+(\hat{\pi} - \tilde{\chi}),
\label{Eq:ScalarFieldScri1}\\
C^{-1}\partial_t(\hat{\pi} + \tilde{\chi}) &=& -(\hat{\pi} - \tilde{\chi}) + \frac{1}{R_+}\tilde{\phi},
\label{Eq:ScalarFieldScri2}
\end{eqnarray}
which imply that $\phi_1 = const$. For initial data satisfying $\tilde{\phi} = \tilde{\chi} = \hat{\pi} = 0$ at $\ScriPlus$ we have $\phi_1 = 0$, and Eq.~(\ref{Eq:ScalarFieldScri1}) yields
\begin{equation}
C^{-1}\partial_t(\tilde{\phi}^2) = -\frac{1}{4\pi G C^2 R_+^2} (3f_0 + f_1).
\end{equation}
This shows that the $\log$ terms appear in the expansions whenever outgoing scalar radiation is present at $\ScriPlus$. Therefore, the situation is completely analogous to the nonspherical vacuum case for initial data satisfying the Penrose regularity condition, where the presence of outgoing gravitational radiation at $\ScriPlus$ implies a lack of smoothness of the extrinsic curvature~\cite{jBoSlB11}.

In App.~\ref{App:FormalExpansions}, we generalize the expansions~(\ref{Eq:vExpansion},\ref{Eq:uExpansion},\ref{Eq:MassExp},\ref{Eq:aExpansion}) to include higher-order terms. Then, in App.~\ref{App:LocSolutionsScriPlus} we prove that these formal expansions do in fact correspond to a three-parameter family of local solutions of the system of equations~(\ref{Eq:HamRescaled},\ref{Eq:MomRescaled},\ref{Eq:CMCRescaled}) in the vicinity of $z = 0$. The three parameters are the free parameters $v_2$ and $u_4$ in the above expansions for $v(z)$ and $u(z)$ plus an additional free parameter $a_4$ that appears in the expansion for $a(z)$ at order $z^4$.

\section{Numerical implementation and results}
\label{Sec:NumResults}

In this section, we numerically implement the hyperbolic-elliptic system summarized in Section~\ref{SubSec:Summary}. We start with a detailed description of the numerical construction of the initial data which represent a scalar field configuration outside an apparent horizon. Then, we describe the discretization method for the evolution system and run several tests. Finally, we perform long-term evolutions and analyze the tail decay of the scalar field. All the results presented in this section refer to the simple case where the potential $V(\Phi)$ vanishes.

\subsection{Initial data}
\label{SubSec:ID}

We construct the initial data as follows. Instead of specifying the conformally rescaled scalar field $\tilde{\phi}$ and the corresponding momentum $\tilde{\pi}$, we specify directly the physical field $\Phi$ and its corresponding moment $\Pi := D_0\Phi$ as a function of the (unphysical) radial coordinate $R$ at the initial CMC surface $t = 0$. Consequently, the source terms appearing in the Hamiltonian and momentum constraints~(\ref{Eq:HamSphSym},\ref{Eq:MomSphSym}) have the following form:
\begin{eqnarray}
\Omega^2\tilde{\rho} &=& \rho = \frac{1}{2}\left( \Pi^2 + \Omega^2\Phi'^2 \right), \label{Eq:rhoIVP} \\
\Omega^2\tilde{j}_R &=& j_R = -\Omega\Pi\Phi'.
\end{eqnarray}
For the sake of simplicity, we restrict ourselves to initial data satisfying $\Pi = 0$, in which case the momentum constraint equation~(\ref{Eq:MomSphSymBis}) can be integrated explicitly and yields $\tilde{\nu} = 2D\Omega^2/R^3$, with $D$ a constant. We choose a Gaussian pulse for $\Phi$ of the form
$$
\Phi(R) = A e^{-\frac{1}{2}\frac{(R-R_0)^2}{w^2}},
$$
with $A$, $w$ and $R_0$ denoting the amplitude, width and center of the pulse respectively. Then, it only remains to solve the Hamiltonian constraint equation~(\ref{Eq:HamSphSymBis}) for $\Omega$. We solve this equation on a bounded interval of the form $[R_{in},R_+]$, where at the inner boundary $R = R_{in}$ we specify the following conditions with parameter $D$:
$$
\Omega_{in} := \frac{R_{in}}{r_{in}},\qquad
R_{in}\frac{\Omega'_{in}}{\Omega_{in}} =
 R_{in}\left( \frac{C}{\Omega_{in}} - D\frac{\Omega_{in}{}^2}{R_{in}{}^3} \right)
  + 1 - \Theta^+_{in}
$$
on the conformal factor and its first radial derivative, where $r_{in}$ and $\Theta^+_{in} < 0$ are the areal radius and the outgoing expansion at the inner boundary, respectively (see Eq.~(\ref{Eq:Expansions})). Near the outer boundary $R = R_+$, we impose the expansion~(\ref{Eq:uExpansion}) with free parameter $u_4$.\footnote{Since the scalar field decays exponentially in $R$, the matter terms $f_0$ and $f_1$ vanish identically and the expansion reduces to the same one as in the Schwarzschild case, cf. Eq.~(\ref{Eq:OmegaSchwExpansion}). For the results corresponding to the initial data, we truncate the series at the order $z^9$.} Next, we adjust the free parameters $D$ and $u_4$ to obtain a smooth solution of the Hamiltonian constraint. In order to do so, we use a ``shooting to a matching point" algorithm as described in~\cite{Recipes-Book}, rewriting Eq.~(\ref{Eq:HamSphSymBis}) as a first-order coupled system of differential equations which we integrate using a standard fourth-order Runge-Kutta (RK4) algorithm on the grid $R_j = R_{in} + j\Delta R$, $j=0,1,2,\ldots,N$, with spatial resolution $\Delta R = (R_+ - R_{in})/N$ and $N$ an even number. The matching point is chosen simply as $R_m := (R_{in} + R_+)/2$. A Newton-Raphson routine~\cite{Recipes-Book} in two dimensions is implemented to perform the matching of $(\Omega,\Omega')$ at $R = R_m$ as a function of $(D,u_4)$, where the Jacobian matrix is approximated using centered differencing. We choose the following numerical values: $R_{in} = 0.195$, $R_+ = 1$, $r_{in} = 1/C$, $\Theta_{in}^+ = -0.02$. The expansion~(\ref{Eq:uExpansion}) is used to specify boundary data at $R = 1 - \varepsilon$ (with typical values of $\varepsilon = 4\times 2^l\Delta R$) away from the singular point. Runs with $N = 2^l\times 100$ and $l = 0,1,2,3,\ldots$ are performed.

\begin{table}[h]
\center
\begin{tabular}{|c|c|c|c|}\hline
 $A$ & $R_{AH}$ & $m_{AH} C$ & $m_\ScriPlus C$ \\
\hline
0.0 & $0.2035$ & $0.5048$ & $0.5048$ \\
0.1 & $0.2035$ & $0.5052$ & $0.5380$ \\
0.2 & $0.2036$ & $0.5063$ & $0.6366$ \\
0.3 & $0.2036$ & $0.5082$ & $0.7981$ \\
0.4 & $0.2036$ & $0.5110$ & $1.0180$ \\
\hline
\end{tabular}
\caption{MS masses at the apparent horizon and at null infinity for different values of the scalar field amplitude $A$.}
\label{Tab:MS}
\end{table}

\begin{figure}[htp]
   \begin{center}
   \includegraphics[width=8.2cm]{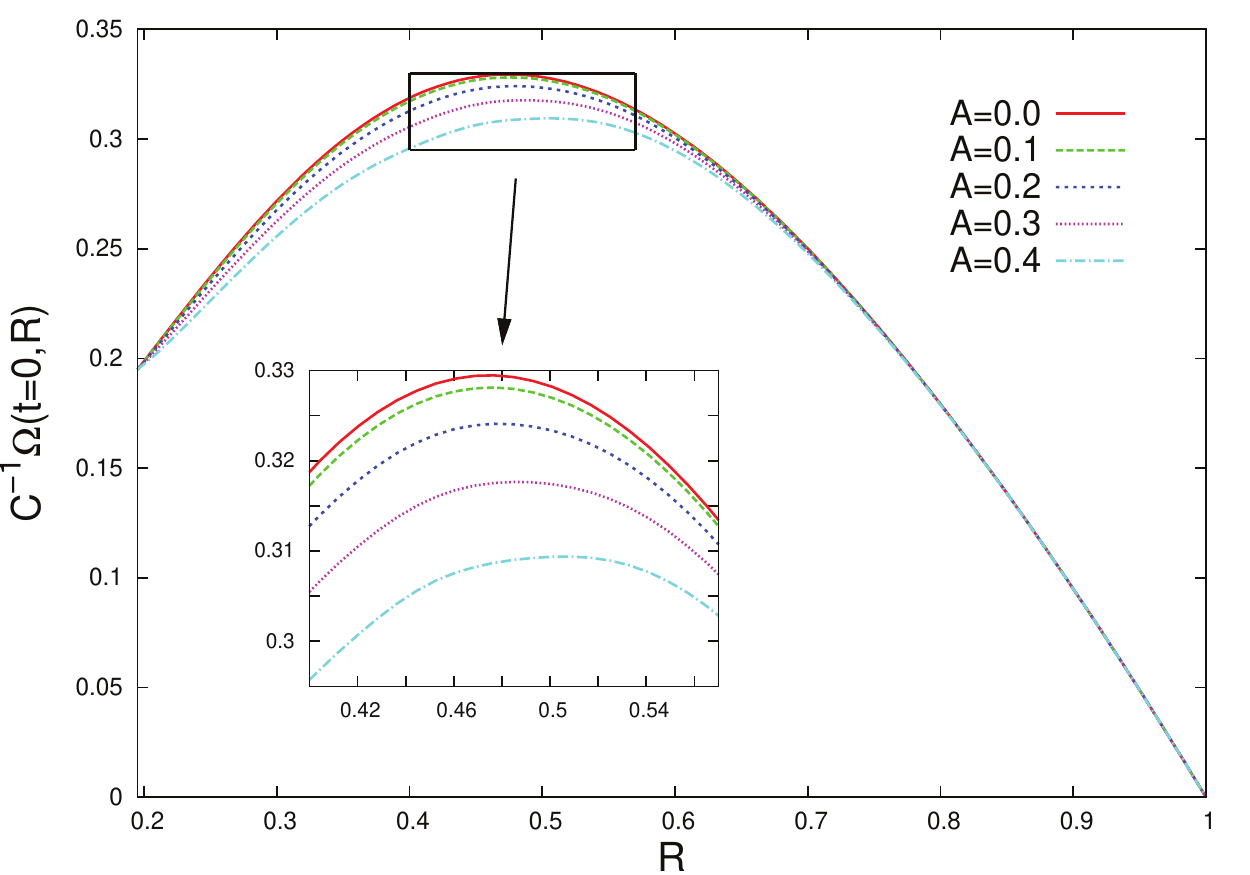}
   \includegraphics[width=8.2cm]{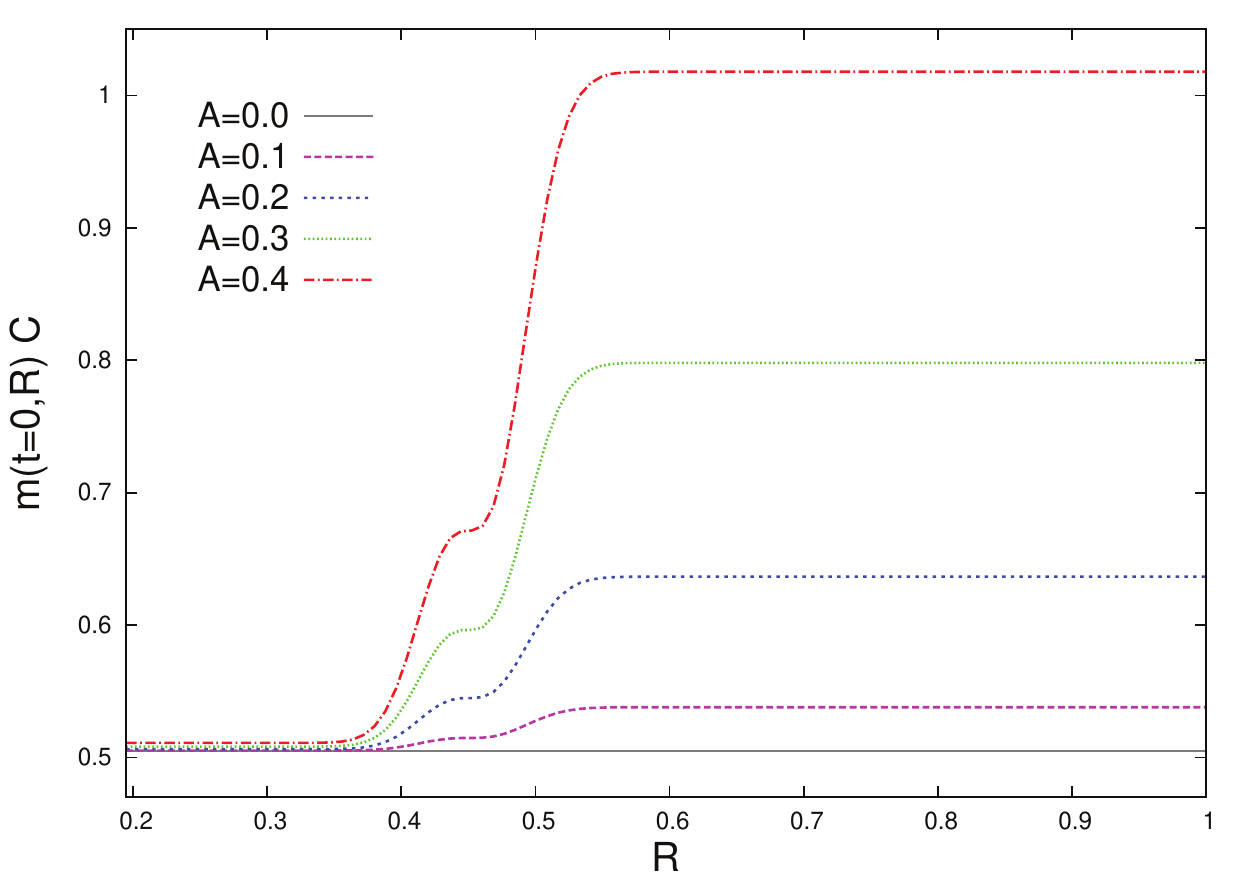}
   \caption{\label{fig:IVP-OmegaMass} The conformal factor (left panel) and the MS mass function (right panel) for the initial configuration. Both quantities are numerically computed using different amplitudes $A$ for the physical scalar field. In these plots the scalar field distribution has width $w = 0.04$ and is centered at $R_0=0.45$. In both plots, our mesh was made up of $1,600$ points.}
   \end{center}
\end{figure}

In Fig.~\ref{fig:IVP-OmegaMass} we show plots for the conformal factor $\Omega(t=0,R)$ and the corresponding MS mass function $m(t=0,R)$, using different values of the amplitude $A$ and fixing $R_0=0.45$ and $w=0.04$. Note that the relative change in the conformal factor for the chosen parameter values of $A$ is small. On the other hand, the ratio between the MS masses at $\ScriPlus$ and at the inner boundary changes by a factor of about $2$ as $A$ increases from $0.0$ to $0.4$. For $A > 0$ the MS mass function is monotonically increasing with steepest gradient in the region where the scalar pulse $\Phi(R)$ is nonzero, as expected. The specific values for the MS mass at the apparent horizon ($m_{AH}$) and at null infinity ($m_\ScriPlus$) are shown in Table~\ref{Tab:MS}.

\begin{figure}[htp]
   \begin{center}
   \includegraphics[width=8.2cm]{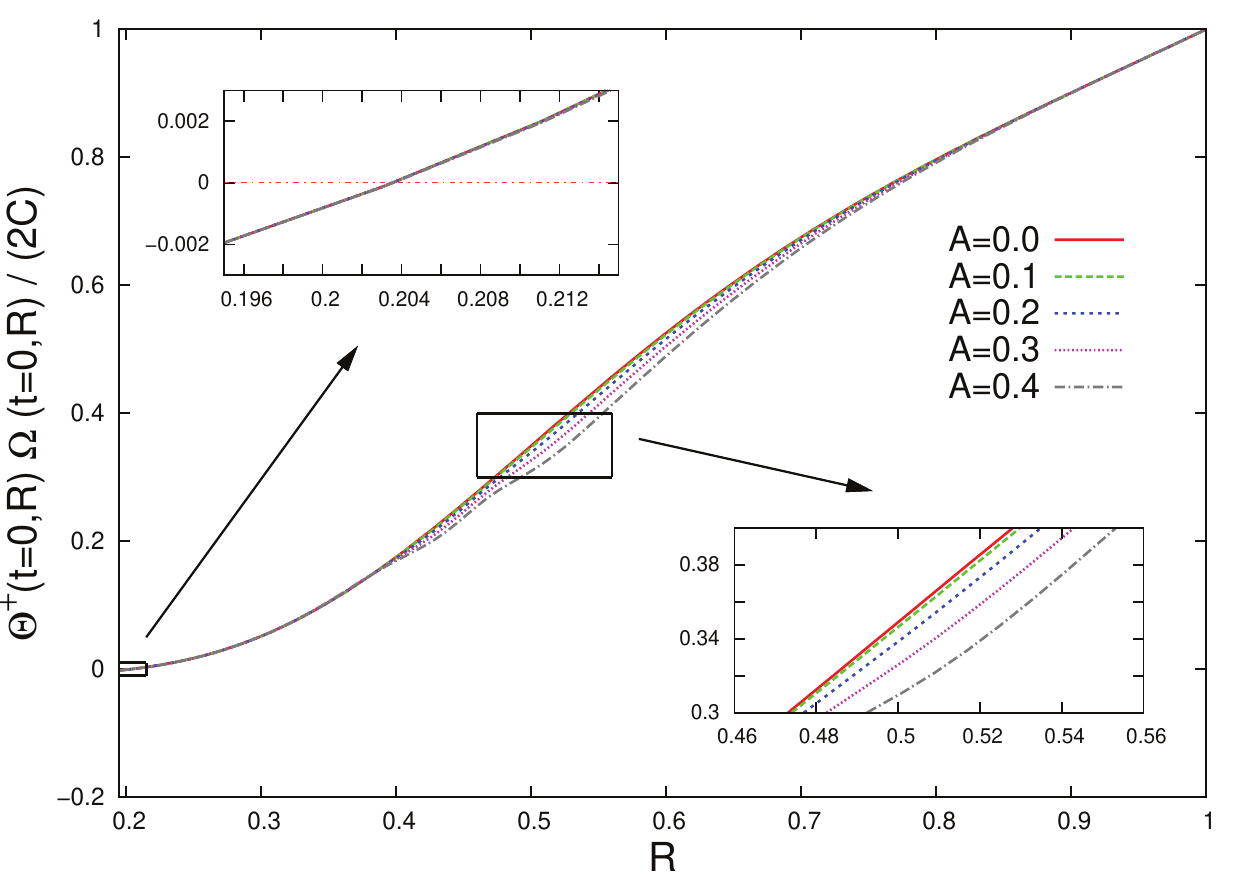}
   \includegraphics[width=8.2cm]{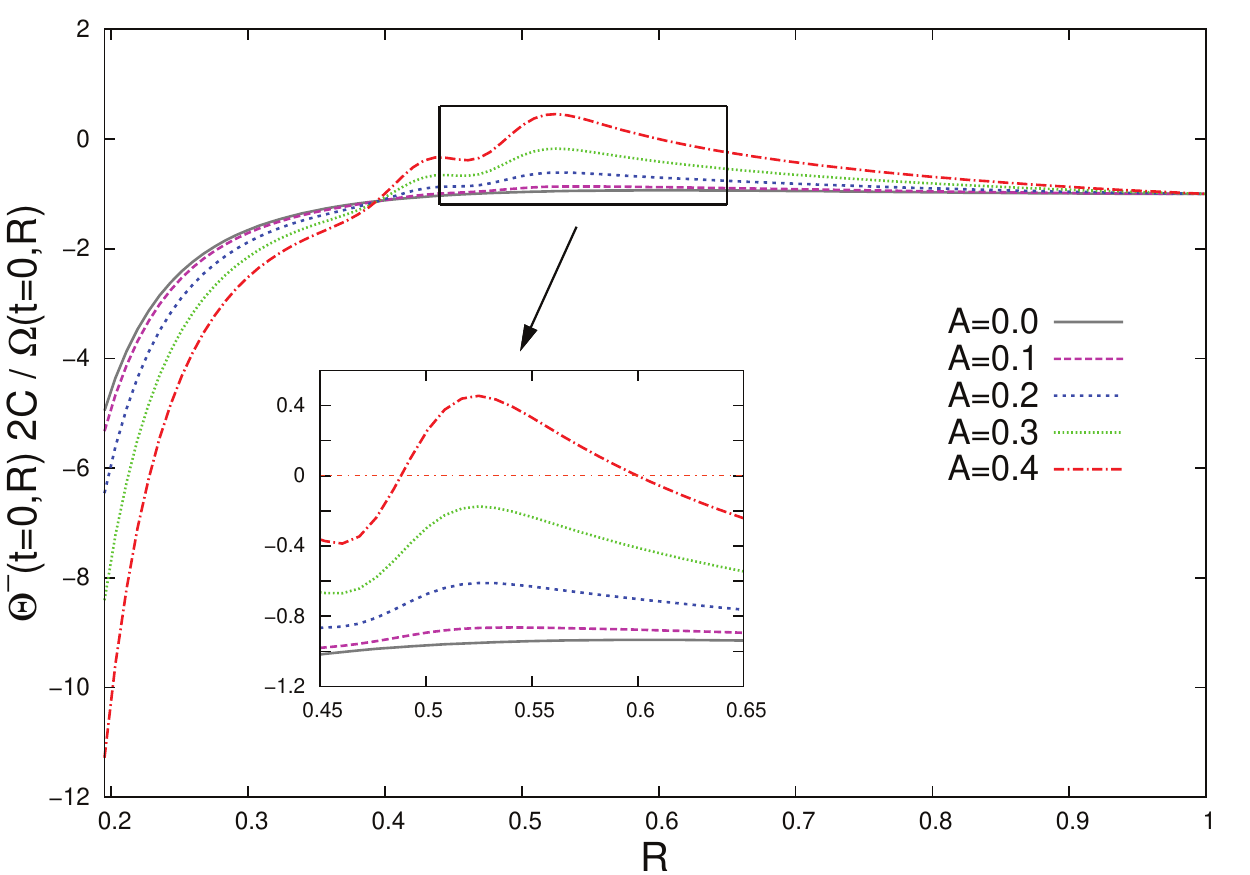}
   \caption{\label{fig:IVP-Expansions} The rescaled outgoing expansion $\frac{\Omega}{2}\Theta^{+}$ (left panel) and the rescaled ingoing expansion $\frac{2}{\Omega}\Theta^{-}$ (right panel) computed  using different amplitudes $A$ for the physical scalar field. The upper left inset in the left panel shows the location where $\Theta^+ = 0$, corresponding to the location of the marginally trapped surface. This location is determined numerically by a linear interpolation, see Table~\ref{Tab:MS} for the resulting values. In the bottom right inset we show the trend for $\frac{\Omega}{2}\Theta^{+}$ as $A$ increases. The inset in the right panel shows that the ingoing expansion is always negative except for the case with amplitude $A = 0.4$, indicating the presence of surfaces which are trapped in the exterior from the point of view of an observer which lies in the interior of these surfaces. As in the previous plot, we  used $1,600$ grid points.}
\end{center}
\end{figure}

Next, we analyze the properties of the in- and outgoing expansions $\frac{2}{\Omega}\Theta^-$ and $\frac{\Omega}{2}\Theta^+$, respectively, corresponding to the initial data configurations shown in Fig.~\ref{fig:IVP-OmegaMass}. Their behavior is shown in Fig.~\ref{fig:IVP-Expansions}. In all cases shown, the rescaled outgoing expansion is monotonously increasing, and has a zero close to the inner boundary, corresponding to the location of the apparent horizon. In contrast to this, the rescaled ingoing expansion is not monotonic. It remains negative for $A = 0.0,0.1,0.2,0.3$. However, for $A = 0.4$ there is a region where $\Theta^-$ becomes positive. This region corresponds to two-spheres along which both expansions $\Theta^\pm$ are positive. Hence, these surfaces are trapped in the exterior region, from the point of view of an observer located in the interior of these surfaces.

\begin{figure}[htp]
   \begin{center}
   \includegraphics[width=8.2cm]{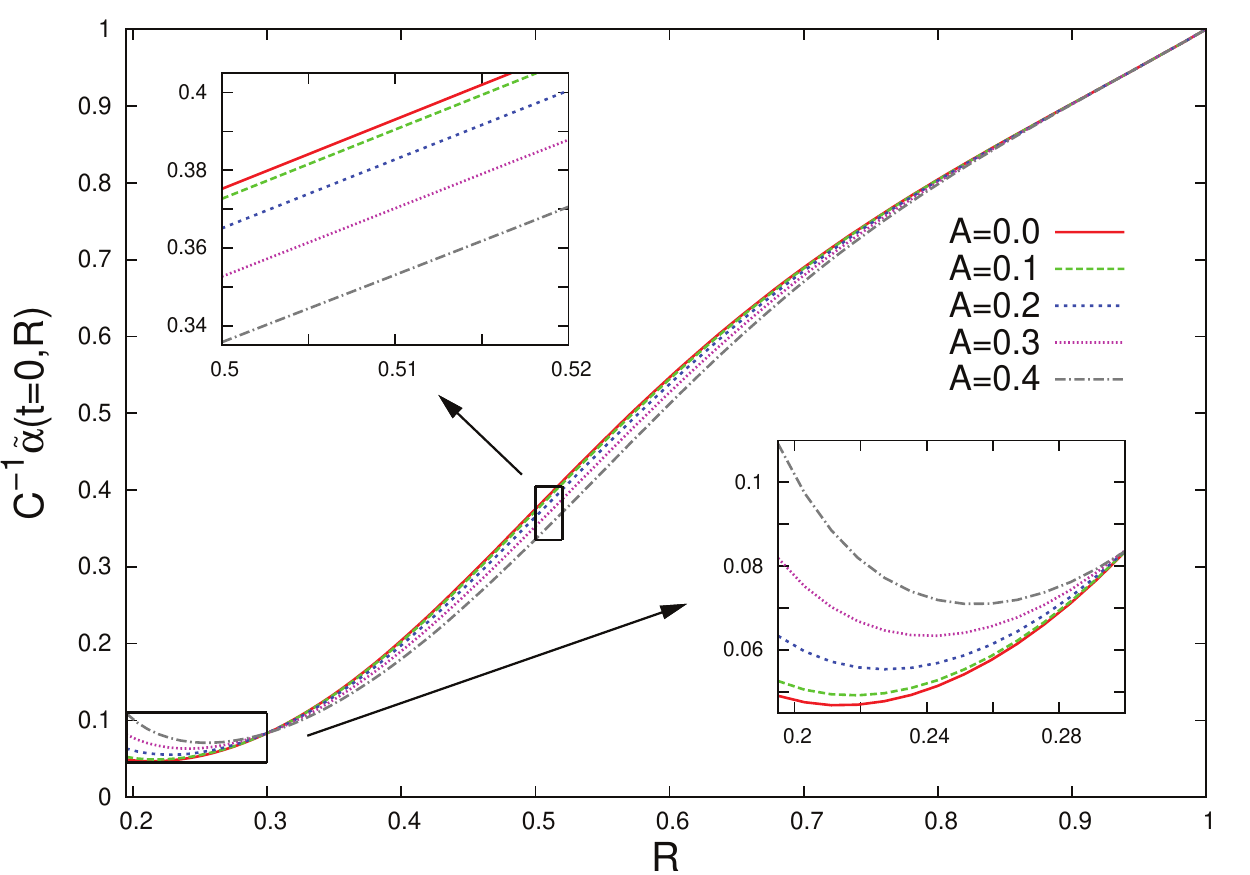}
   \includegraphics[width=8.2cm]{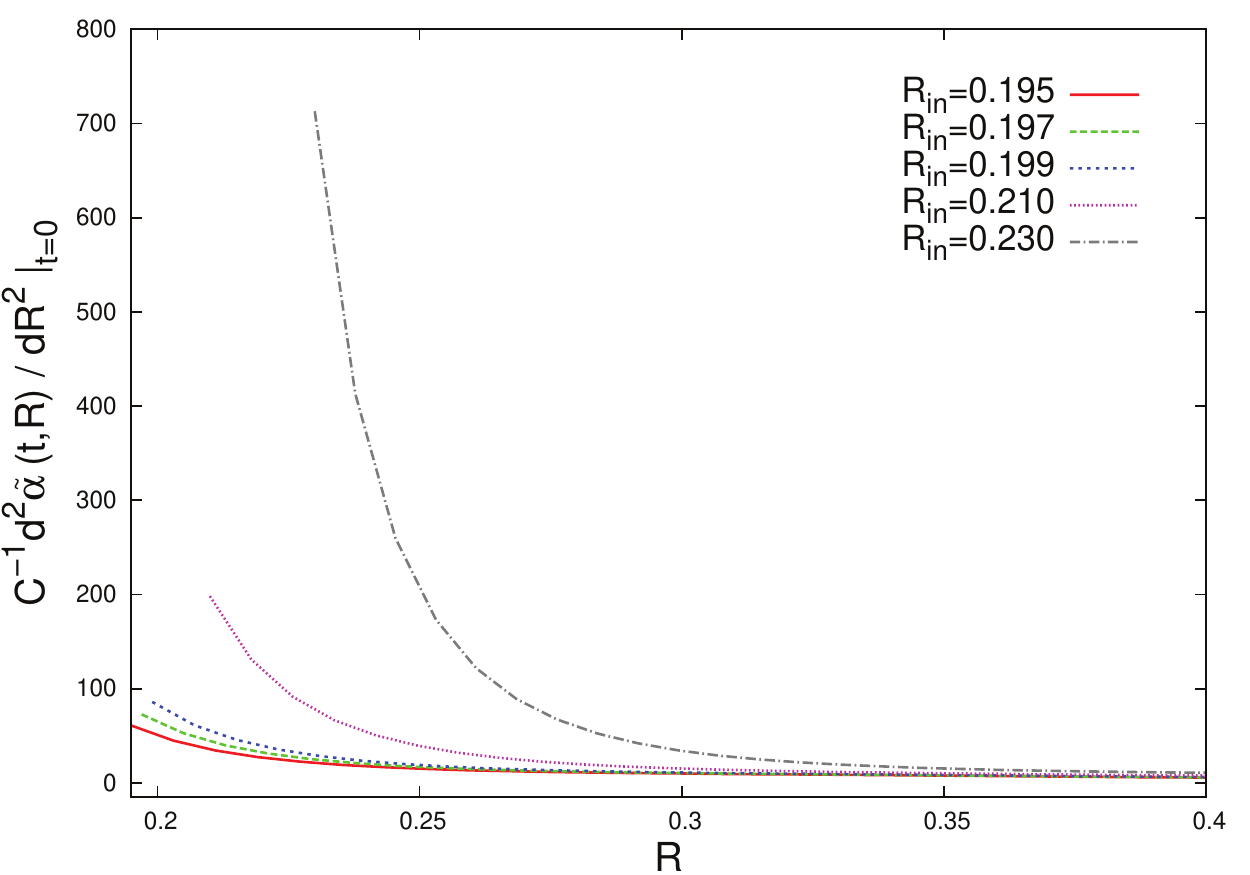}
   \caption{\label{fig:IVP-lapse} Left panel: The conformal lapse $\tilde{\alpha}$ obtained from the CMC constraint. Right panel: The (numerically computed) second derivative of $\tilde{\alpha}$ for $A = 0.4$ and different choices for $R_{in}$ . In both plots we used $1,600$ grid points.}
   \end{center}
\end{figure}

After having solved the Hamiltonian constraint for the conformal factor, we proceed to the numerical solution of the elliptic equations~(\ref{Eq:CMCConstrSphSymBis}) and~(\ref{Eq:Cons2Bis}) for $\tilde{\alpha}$ and $\tilde{C}$, respectively. For the first equation, the matter source term is given by
$$
\Omega^2(3\tilde{\rho} + \tilde{\sigma}^c{}_c) = \Omega^2\Phi'^2 + 3\Pi^2,
$$
and at the inner boundary $R = R_{in}$ we specify the conditions
\begin{eqnarray}
\tilde{\alpha}_{in} &:=& \left| R_{in}-D\frac{\Omega_{in}{}^3}{R_{in}{}^2} \right|  ,\qquad \nonumber\\
\tilde{\alpha}'_{in} &:=& -\frac{\Omega_{in}}{2\tilde{\alpha}_{in}} 
 \left(\Omega'_{in}-\frac{\Omega_{in}}{R_{in}}\right)
 + \frac{1}{\tilde{\alpha}_{in}}\left(R_{in}-D\frac{\Omega_{in}{}^3}{R_{in}{}^2}\right)
 \left(1+2D\frac{\Omega_{in}{}^3}{R_{in}{}^3}-3D\frac{\Omega_{in}{}^2\Omega'_{in}}{R_{in}{}^2}\right) , \nonumber 
\end{eqnarray}
which correspond to the Schwarzschild black hole case with $\zeta = 0$ and $N = 0$, see Eq.~(\ref{Eq:AlphaTildeSchw}). Here, the above value for $\tilde{\alpha}_{in}$ is kept fixed, while the value for $\tilde{\alpha}'_{in}$ and the free parameter $a_4$ arising in the asymptotic expansion~(\ref{Eq:aExpBis}) are adjusted (taking the above expression for $\tilde{\alpha}'_{in}$ as an initial guess) to obtain a smooth solution of the CMC constraint equation~(\ref{Eq:CMCConstrSphSymBis}). As for the Hamiltonian constraint, this is achieved using a ``shooting to a matching point" algorithm. Fig.~\ref{fig:IVP-lapse} shows the result for the conformal lapse $\tilde{\alpha}$. We have found that the behavior of $\tilde{\alpha}$ near the inner boundary depends sensitively on the choice for $R_{in}$. As is shown in the left panel, the second derivative of $\tilde{\alpha}$ can take quite large numerical values (compared to its asymptotic value) if $R_{in}$ is not chosen adequately. In practice, this presents a problem during the evolutions, as the scalar field equation~(\ref{Eq:piSphSymBis}) involves a term which depends on the second derivative of $\tilde{\alpha}$. These considerations motivated the choice $R_{in} = 0.195$ in our simulations. For larger values, we have found that spurious peaks appears in the early evolution of $\tilde{\phi}(t,R)$. Although at each fixed time these peaks converge away when increasing the resolution, their presence impedes the possibility of performing long-term stable evolutions.

\begin{figure}[htp]
   \begin{center}
   \includegraphics[width=8.2cm]{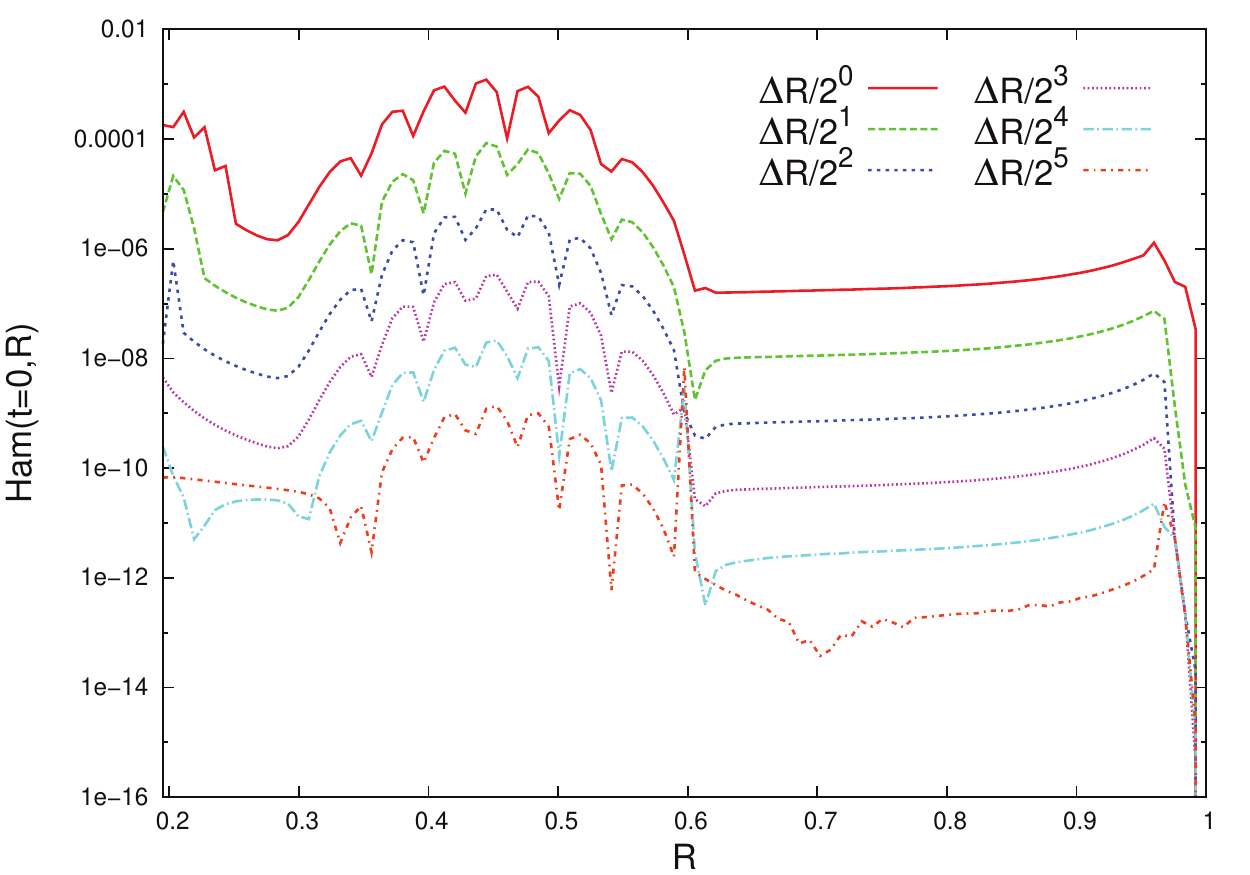}
   \includegraphics[width=8.2cm]{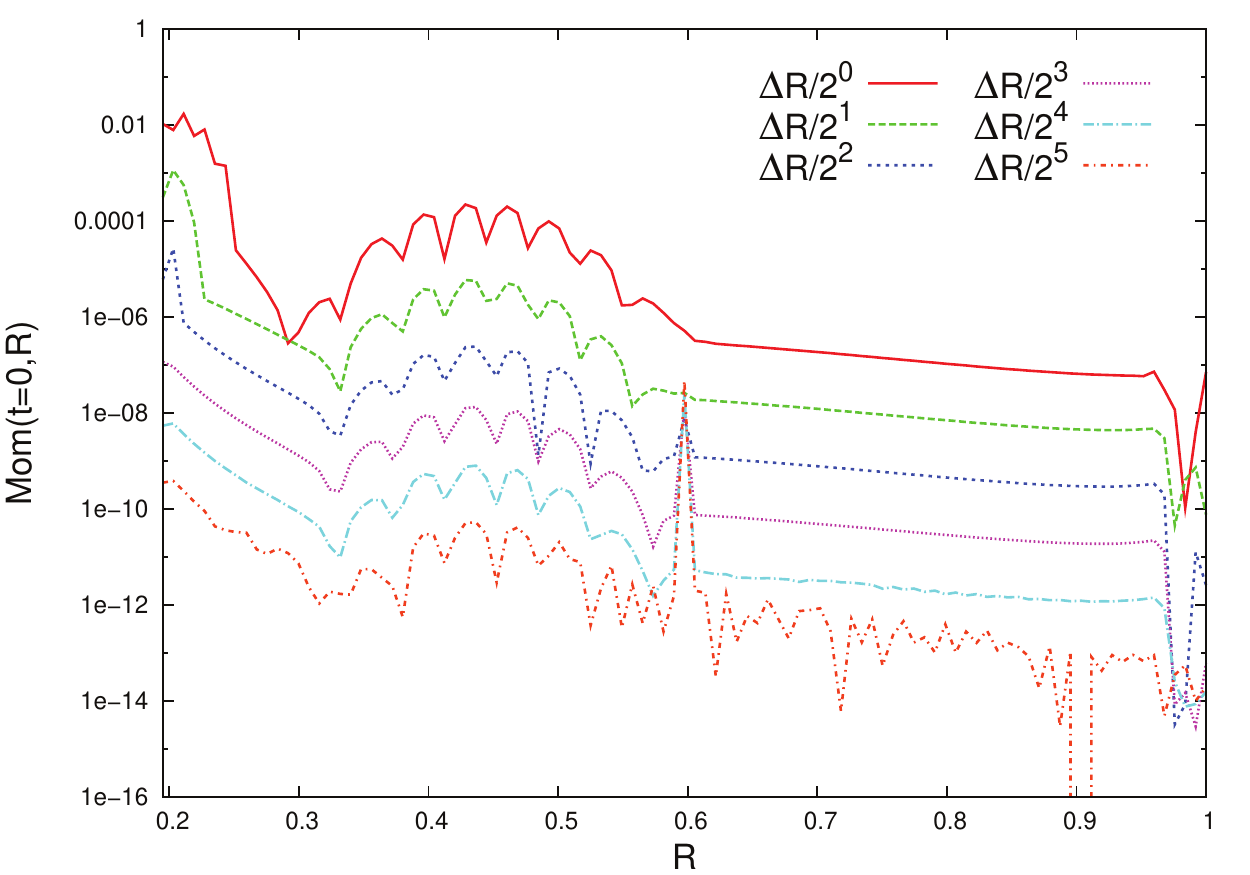}
   \caption{\label{fig:IVP-convergence} Convergence tests showing the numerical residuals of the Hamiltonian constraint (left panel) and the momentum constraint (right panel). In both cases we have checked that the order of convergence lies about the expected order $4$. Here, the coarsest resolution $\Delta R$ corresponds to $N = 100$ points. The scalar field amplitude is $A = 0.3$, and as before the width and center of the Gaussian pulse are $w = 0.04$ and $R_0 = 0.45$, respectively. Note that at the matching point $R = R_m$ a peak appears as we increase the resolution. This behavior is due to the combination of two issues in the Newton-Raphson algorithm: first, the tolerance $tol$ that we chose for the iterations, and second, the length of the infinitesimal steps $\Delta u_4$ and $\Delta D$ that we assign when we evaluate the Jacobian of the parameter vector at the point $R_m$. In practice, we observe that as we diminish the values of $tol$, $\Delta u_4$ and $\Delta D$, the peak at $R_m$ also diminishes. Here and during the evolution, we have found that the choice $tol = \Delta u_4 = \Delta D = 0.2\times 10^{-10}$ seems to work best.}
   \end{center}
\end{figure}

Finally, convergence tests showing the residuals of the Hamiltonian and momentum constraints are shown in Fig.~\ref{fig:IVP-convergence}. In general, we find $4th$ order convergence with residuals reaching $10^{-12}$ lying close to machine precision.

\subsection{Evolution}

Having discussed the construction of our initial data, we now describe the details for the numerical implementation of the evolution scheme. This scheme is directly based on the hyperbolic-elliptic system summarized in Sec.~\ref{SubSec:Summary} with one important difference: at each time step the field $\tilde{\nu}$ parametrizing the trace-free part of the conformal extrinsic curvature is determined from the momentum constraint equation~(\ref{Eq:MomSphSymBis}) instead of the evolution equation~(\ref{Eq:nuBis}). This change was found to be necessary in order to obtain long-term stable numerical evolutions.

The resulting evolution scheme consists of the following iterative procedure:
\begin{itemize}
\item[1.] We evolve the scalar field equations~(\ref{Eq:phiSphSymBis},\ref{Eq:chiSphSymBis},\ref{Eq:piSphSymBis}) and the evolution equations~(\ref{Eq:OmegaEvol}) and (\ref{Eq:nuBis}) for $\Omega$ and $\tilde{\nu}$ one step in time, using a RK4 integrator. The spatial derivatives are discretized using the finite difference operator $D_{6-5}$ (which is sixth order accurate in the inside of the domain and fifth order accurate near the boundaries) satisfying the ``summation by parts" property as implemented and tested in~\cite{pDeDeSmT07}. During this step, the values of the fields $(\Omega,\Omega',\Omega'',\tilde{\alpha},\tilde{\alpha}',\tilde{\alpha}'')$ appearing in the RHS of these equations are kept fixed to their values from the previous time step (with the exception of Eq.~(\ref{Eq:OmegaEvol}) where we use the values of $\Omega$ required by the RK4 algorithm). The second derivatives $\Omega''$ and $\tilde{\alpha}''$ are computed using the $D_{6-5}$ operators. In each sub-iteration of the RK4 algorithm we solve the constraint equation~(\ref{Eq:Cons2Bis}) for $\tilde{C}$, by integrating inwards from $\ScriPlus$ starting from the value given in Eq.~(\ref{Eq:ScriPlusBC}).
 
\item[2.] Next, we solve the Hamiltonian and momentum constraint equations~(\ref{Eq:HamSphSymBis},\ref{Eq:MomSphSymBis}) for $\Omega$ and $\tilde{\nu}$, respectively. This system is solved using a similar algorithm than for the initial data, using the expansions~(\ref{Eq:uExpBis},\ref{Eq:vExpBis}) at $\ScriPlus$ and Dirichlet boundary conditions at the inner boundary where the values for $\Omega$ and $\tilde{\nu}$ at the inner boundary are taken from step 1. However, compared to the initial data case, an extra complication arises because the RK4 algorithm used to perform the spatial integration requires evaluating the source terms at midpoints lying between two successive grid points of our mesh. Whereas for the initial data construction this was not a problem since we specified the scalar field analytically, for the evolution, we need to numerically interpolate the source terms appearing in the constraints using a third-order polynomial.
 
\item[3.] Taking the fields $\Omega$, $\Omega'$ and $\tilde{\nu}$ computed from the previous step, we solve the constraints~(\ref{Eq:CMCConstrSphSymBis},\ref{Eq:Cons2Bis}) for $\tilde{\alpha}$ and $\tilde{C}$, respectively. In order to solve the elliptic equation for $\tilde{\alpha}$, we use an algorithm which is similar to the one described in Sec.~\ref{SubSec:ID}, where the value of the conformal lapse at the inner boundary is frozen to its initial value and the asymptotic expansion~(\ref{Eq:aExpBis}) is used.
\end{itemize}

As in the previous subsection, the numerical grid consists of a uniform partition of $N$ elements of the interval $[R_{in},R_+]$ with $R_{in} = 0.195$ and $R_+ = 1$. The Courant-Friedrichs-Lewy (CFL) factor is chosen to be $\lambda_{CFL} = 0.3$.

\begin{figure}[htp]
   \begin{center}
   \includegraphics[width=8.5cm]{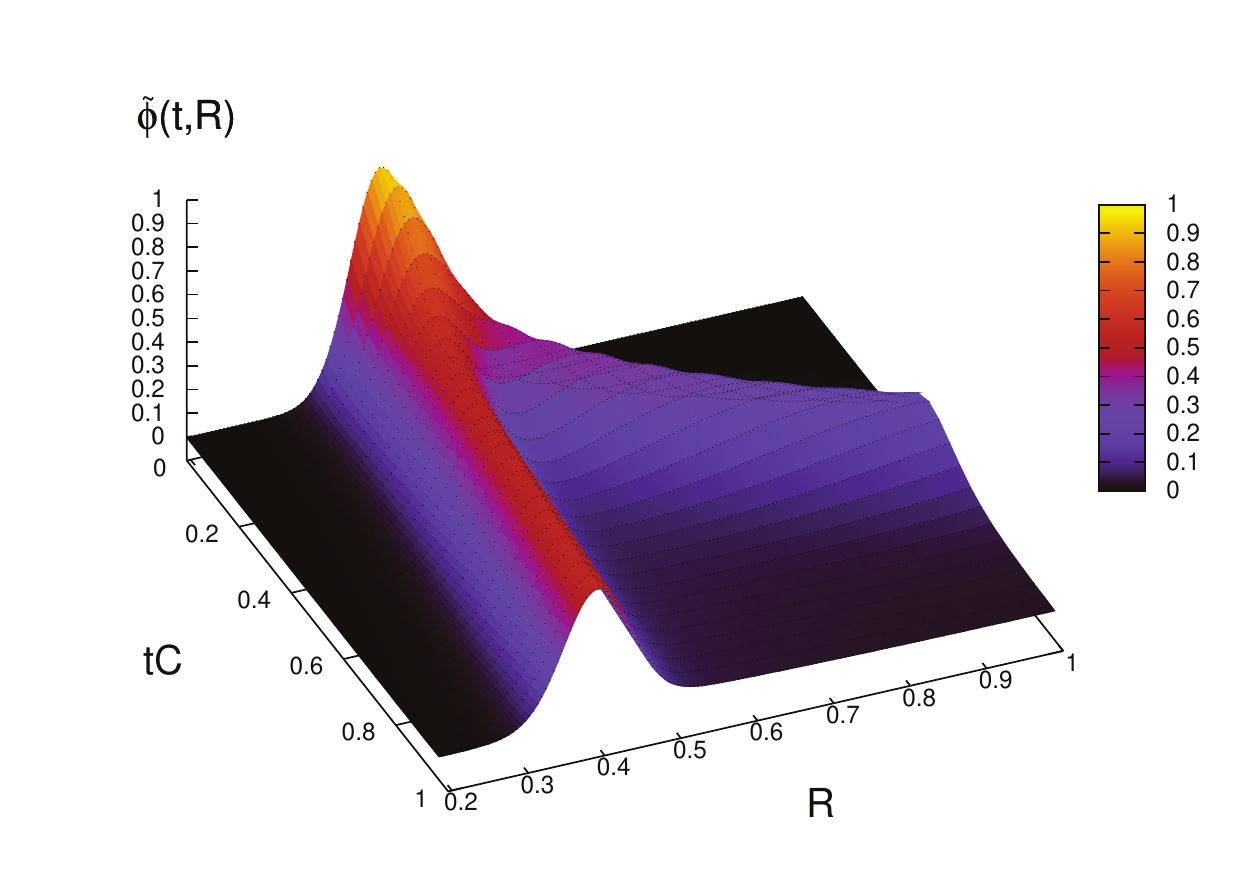}
   \includegraphics[width=8.5cm]{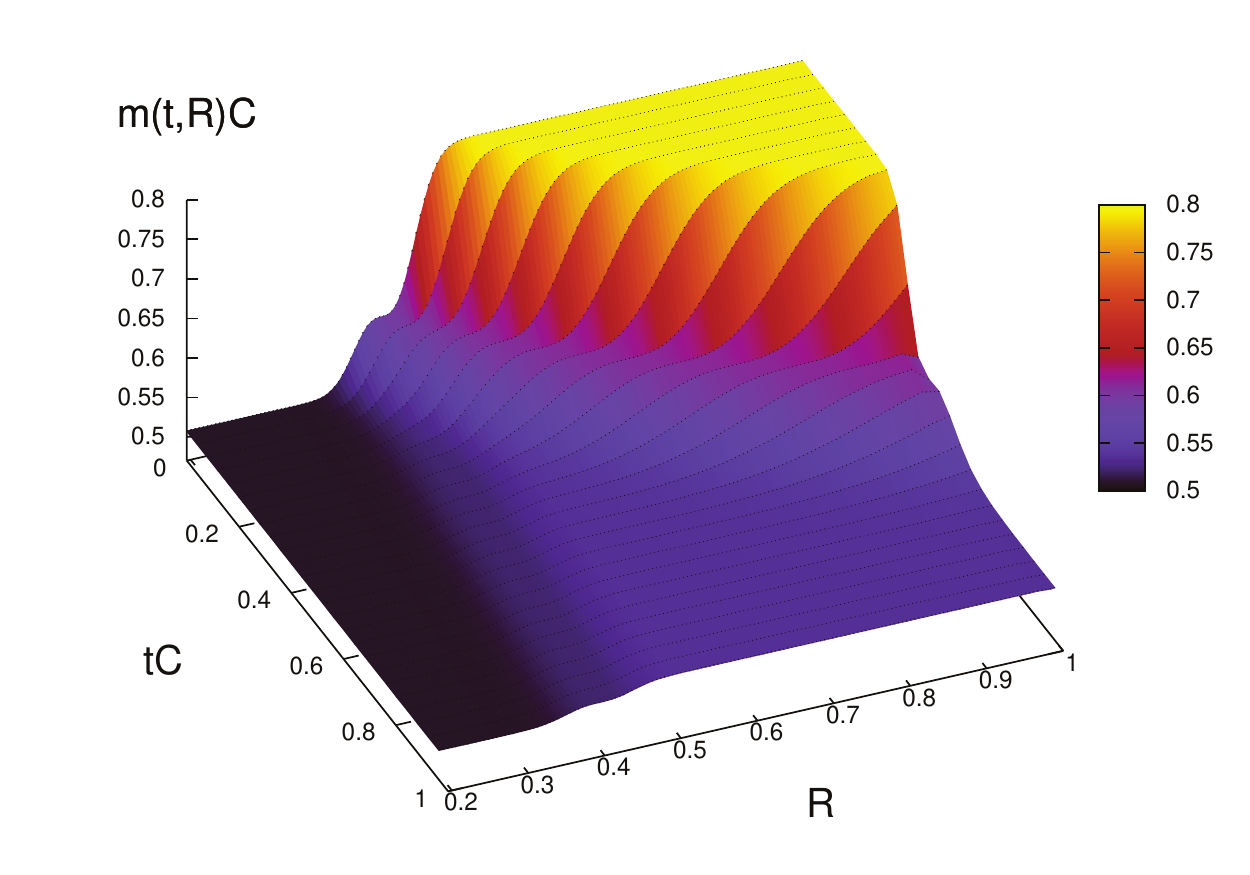}
   \includegraphics[width=8.5cm]{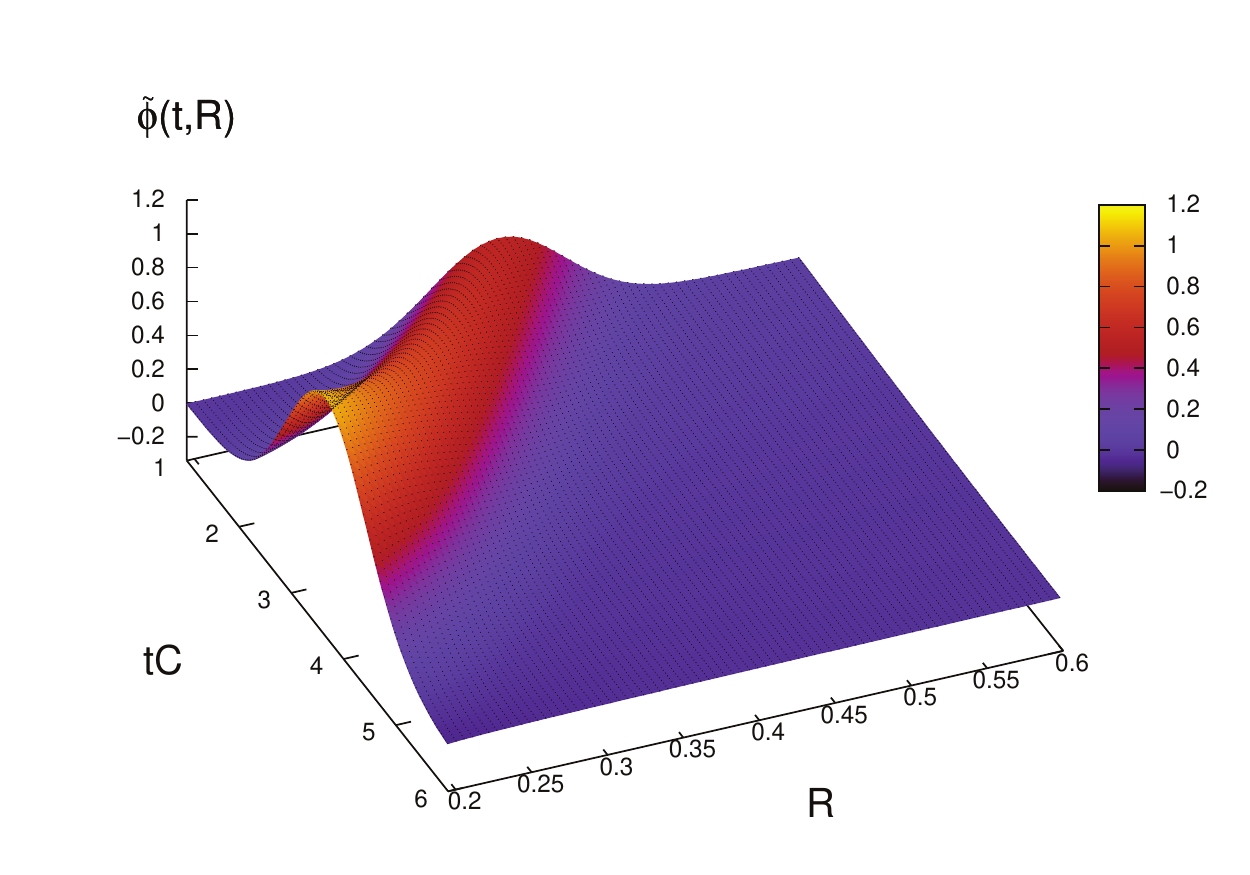}
   \includegraphics[width=8.5cm]{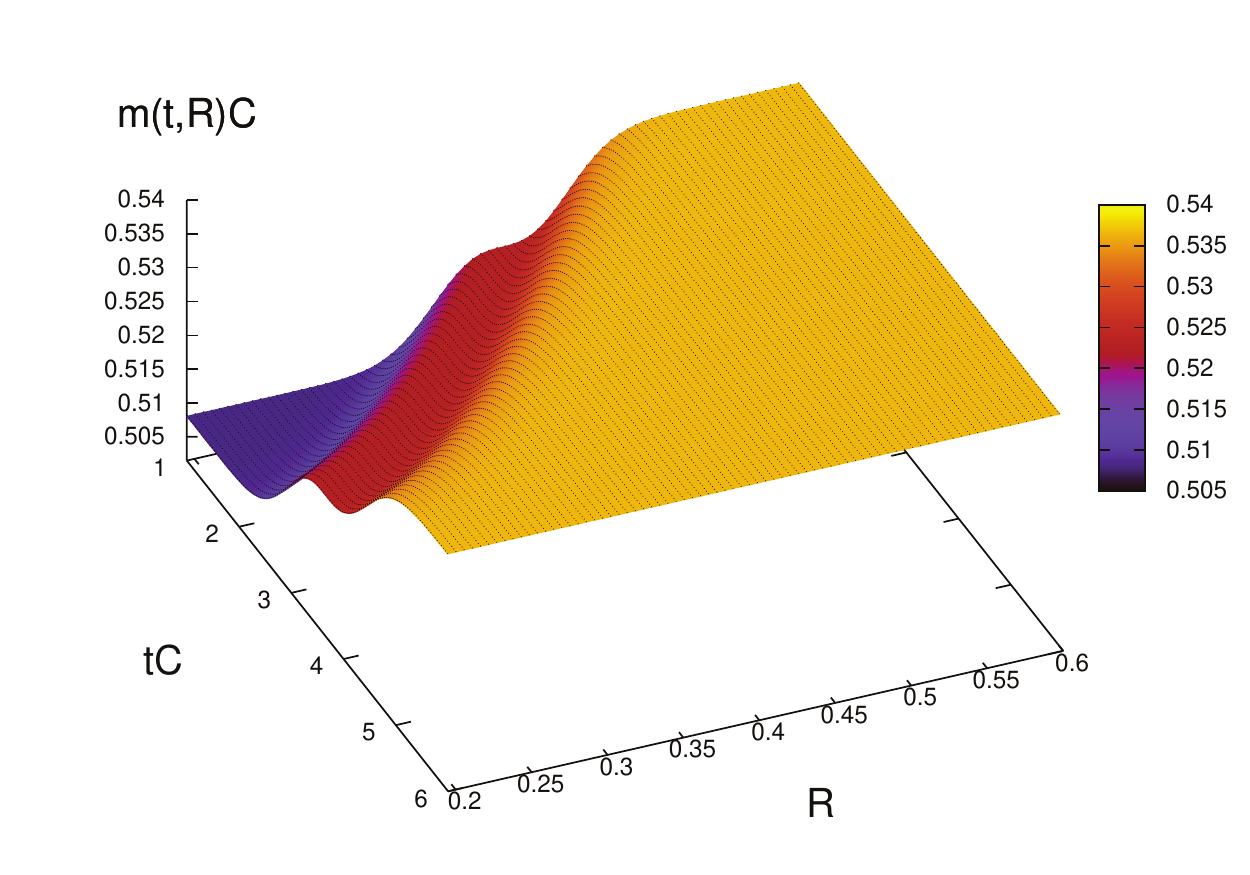}
   \caption{\label{fig:RescaledSF} The evolution of the conformally rescaled scalar field $\tilde{\phi}$ (left panels) and the corresponding evolution of the MS mass function $m$ (right panels) at times $0\leq t\leq 1.0 C^{-1}$ (top panels) and $1.0 C^{-1}\leq t \leq 6.0C^{-1}$ (bottom panels). We used $N = 2,400$ grid points and a CFL factor of $0.3$ for this simulation. The plots show a ``slice" of data each $25$ time steps. The initial data is the one described in Sec.~\ref{SubSec:ID} with amplitude $A = 0.3$, width $w = 0.04$ and centered at $R_0 = 0.45$. Note the radial coordinate only extends from $R\in [0.195,0.6]$ in the bottom panels, while $R$ ranges over the whole computation domain $[0.195,1.0]$ in the top ones.}
\end{center}
\end{figure}

Fig.~\ref{fig:RescaledSF} illustrates the evolution of the conformally rescaled scalar field $\tilde{\phi}$ and the corresponding MS mass function defined in Eq.~(\ref{Eq:MisnerSharpMass}) at early times. As is visible from these plots, part of the scalar field propagates towards $\ScriPlus$ and quickly dissipates (at a time scale smaller than $C^{-1}$) while the remaining part is accreted by the black hole (at a time scale smaller than $6.0 C^{-1}$). However, due to backscattering, the scalar field does not vanish exactly after these time scales, but decays slowly to zero as will be analyzed in more detail in the next subsection. For each fixed value of $t$, we see from these graphs that the mass function $m$ is monotonically increasing in $R$, as expected from Eq.~(\ref{Eq:RadDerivMass}).\footnote{This property is slightly violated during the time span $t\in [0.4,0.5]C^{-1}$ for values of $R\in [0.95,1]$ due to a numerical effect associated with the approximations of the fields we use near $\ScriPlus$ which are based on the truncated expansions and which result in higher errors when the scalar field pulse propagates through $\ScriPlus$. To verify this, we decreased the parameter $\epsilon$ (which determines the point $R = 1 - \epsilon$ at which the expansion is used) and checked that this diminishes this effect.} At $\ScriPlus$, $m$ decreases with time as the scalar field radiates at null infinity, while at the inner boundary (which lies very close to the apparent horizon) $m$ increases until it reaches a comparable value to the one at $\ScriPlus$. The behavior of the scalar field at late times will be analyzed in the next subsection.

\begin{figure}[htp]
   \begin{center}
   \includegraphics[width=8.2cm]{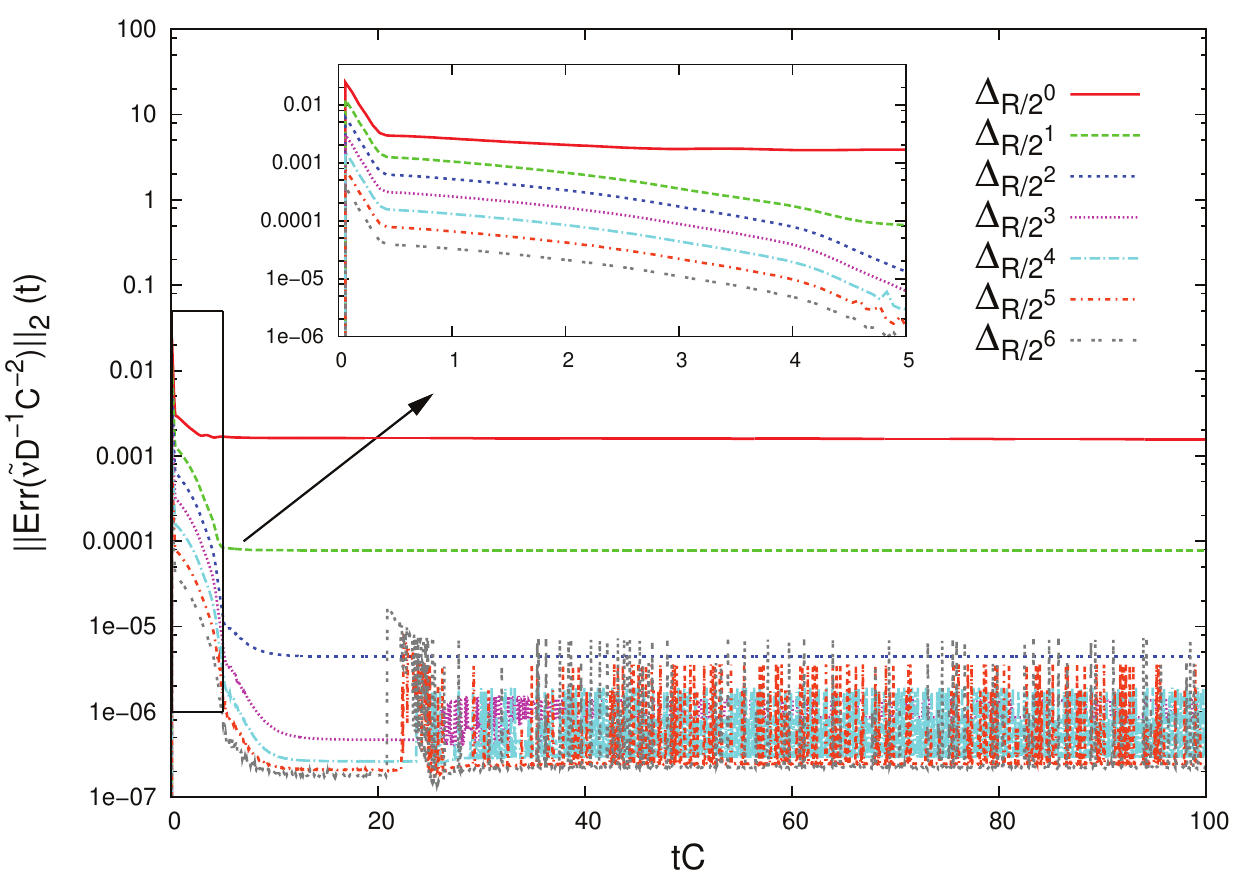}
   \includegraphics[width=8.2cm]{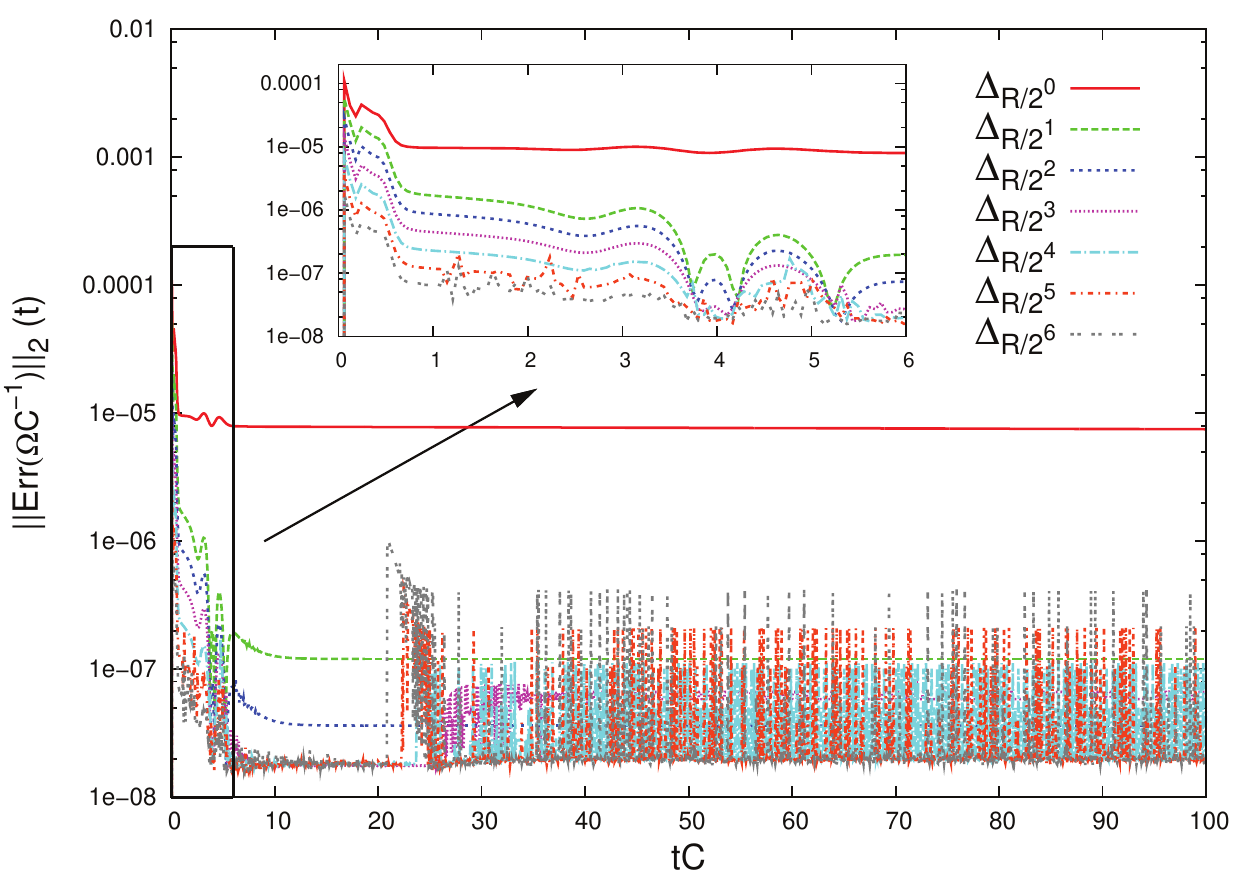}
   \caption{\label{fig:convergence-evolution} $2-$norm of the residual of $\tilde{\nu}$ (left panel) and $2-$norm of the residual of $\Omega$ (right panel) using different resolutions. In both plots we find a good convergence trend for $\Delta R / 2^n$ with $n=0,1,2,3$. However for $n=4,5,6$ we have two numerical effects: first, a saturation when the residual is about $10^{-7}$ in $\|Err(\tilde{\nu})\|$, and about $10^{-8}$ in $\|Err(\Omega)\|$, and second, the occurrence of oscillations starting from different times depending on the resolution. The first effect is associated with the choice of the tolerance value $tol$ used in the Newton-Raphson algorithm while the oscillations depend on the value for $\varepsilon$ that determines the point $R = 1 - \varepsilon$ at which the expansions are applied. We have checked that both these effects diminish as we decrease $tol$ and $\varepsilon$ by running our code with $tol = 10^{-8}$, $10^{-9}$, $10^{-10}$ and $\varepsilon / \Delta R = 5,4,3,2$. From $t = 0C^{-1}$ to $t = 5C^{-1}$ we find $1st$ order convergence in both plots, and for $t > 5C^{-1}$, we find $4th$ order of convergence in $Err(\tilde{\nu})$, and a convergence of order between $2$ and $3$ for $Err(\Omega)$. For these tests we used an initial scalar field pulse with $A=0.3$, $w=0.04$, $R_0=0.45$, and we set $\lambda_{CFL}=0.3$.}
   \end{center}
\end{figure}

In order to validate our numerical evaluation scheme, we perform several convergence tests. In particular, we monitor the $2$-norm of the residuals of the error associated with the evolution Eqs. (\ref{Eq:nuBis}) and (\ref{Eq:OmegaEvol}), computing numerically the derivatives $\partial_t\tilde{\nu}$ and $\partial_t \Omega$ and subtracting from them the corresponding RHS. Namely:
\begin{eqnarray}
Err(\tilde{\nu}) &=& \norm{ \partial_t\tilde{\nu} - \text{rhs}_{\tilde{\nu}} }_2
= \norm{ \partial_t\tilde{\nu} - \tilde{\alpha}
    \left( \tilde{D}_0{\tilde{\nu}} + \frac{b}{\tilde{\alpha}} \partial_R\tilde{\nu}   \right) }_2, \\
Err(\Omega) &=& \| \partial_t\Omega - \text{rhs}_{\tilde{\Omega}} \|_2
 = \norm{ \partial_t\tilde{\nu} - \tilde{\alpha}
    \left( \tilde{D}_0{\Omega} + \frac{b}{\tilde{\alpha}} \partial_R\Omega \right) }_2,
\end{eqnarray}
where the quantities $\tilde{D}_0\tilde{\nu}$, $\tilde{D}_0\Omega$ and $b/\tilde{\alpha}$ are evaluated using the RHSs of Eqs.~(\ref{Eq:nuBis},\ref{Eq:OmegaEvol},\ref{Eq:AlgebraicShift}) respectively. To compute the time derivatives $\partial_t\tilde{\nu}$ and $\partial_t \Omega$, we implemented a fourth order stencil taken from~\cite{fG10} which depends on four previous time steps, so that the actual monitoring begins at $t=5\Delta t=\lambda_{CFL}\Delta x$. For spatial derivatives $\partial_R{\tilde{\nu}}$ and $\partial_R\Omega$, we implemented $D_{6-5}$ ``summation by part'' operators. The results of this test are shown in Fig. \ref{fig:convergence-evolution}, where we have included plots for the errors of $\tilde{\nu}$ and $\Omega$, from $t=0$ to $t=100C^{-1}$. At early times, from $t=0C$ to $t=5C^{-1}$, we find $1st$ order of convergence both in $Err(\tilde{\nu})$ and $Err(\Omega)$. For later times, that is, $t > 5C^{-1}$, we find $4th$ order of convergence in $Err(\tilde{\nu})$, and a convergence of order between $2$ and $3$ for $Err(\Omega)$.

\begin{figure}[htp]
   \begin{center}
   \includegraphics[width=9.7cm]{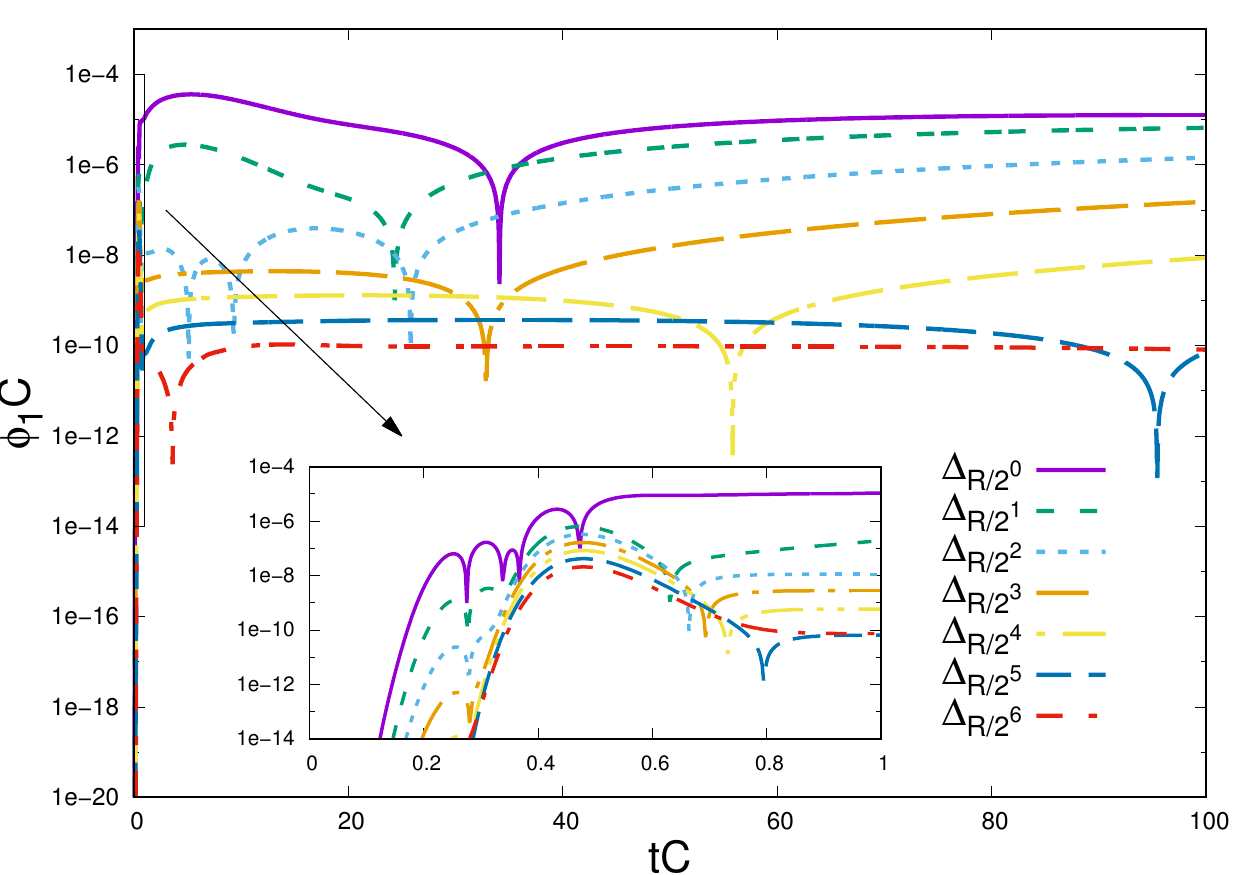}
   \caption{\label{fig:NP-const} The NP quantity $\phi_1$ as a function of time, for the time span $0 \leq tC \leq 100$, using different resolutions. The inset shows a zoom in the time window during which most of the scalar field is radiated at $\ScriPlus$. The convergence order is found to be $1$ for $0\leq tC \leq 0.8$ and about $2$ for larger times. The parameters characterizing the initial scalar field pulse are the same as in the previous figure.}
   \end{center}
\end{figure}

Finally, Fig.~\ref{fig:NP-const} shows the NP quantity $\phi_1$ defined in Eq.~(\ref{Eq:NPConstant}) as a function of time. At $t = 0$ this quantity is practically zero since the initial profile for the scalar field decays exponentially. As can be seen from the plots in Fig.~\ref{fig:NP-const}, $\phi_1$ remains close to zero during the evolution, and its magnitude becomes smaller as resolution is increased. This constitutes a non-trivial test for the constancy of $\phi_1$, which is based on the correct asymptotic values for the metric quantities. For $t \in ]0.0,0.8]C^{-1}$ (corresponding to the time span in which most of the scalar field reaches $\ScriPlus$) the convergence of $\phi_1$ is of $1st$ order, while for larger times the order of convergence becomes $2$.

\begin{figure}[htp]
   \begin{center}
   \includegraphics[width=9.7cm]{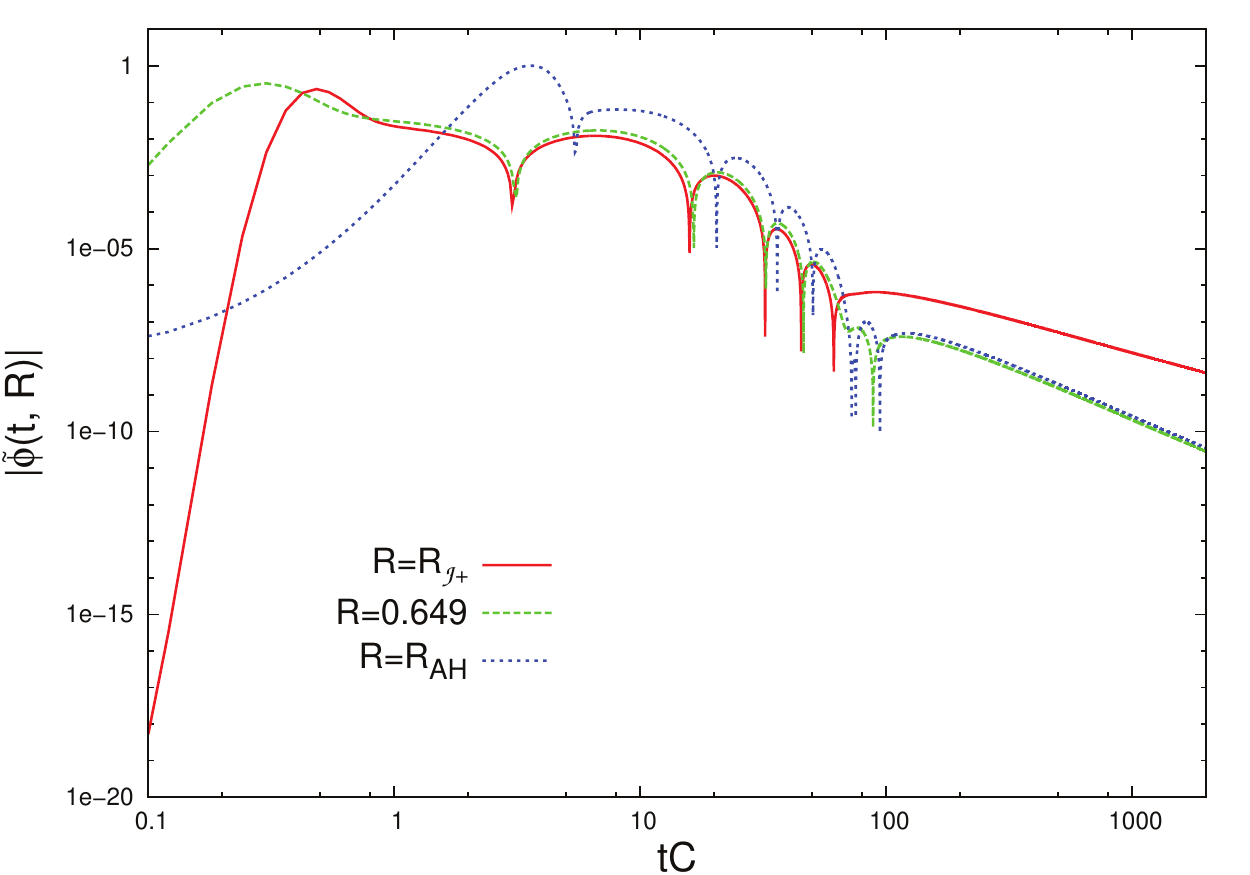}
   \caption{\label{fig:graph-detectors} Conformally rescaled scalar field $\tilde{\phi}$ detected at null infinity ($R=R_+$), along the world line of a timelike observer located at $R=0.649$, and along the apparent horizon of the black hole $R = R_{AH}$. These plots were obtained from an evolution lasting until $t=2000C^{-1}$. Here, we used $1,600$ grid points and a CFL factor of $0.3$. The initial data is the same one as described in the first subsection and has amplitude $A = 0.3$, width $w = 0.04$ and is centered at $R_0 = 0.45$. We see that after an initial period of oscillations, the field starts decaying with an inverse power of $t$ (observe that both axes have a logarithmic scale).}
   \end{center}
\end{figure}

\subsection{Tail decay}

\begin{figure}[htp]
   \begin{center}
   \includegraphics[width=8.5cm]{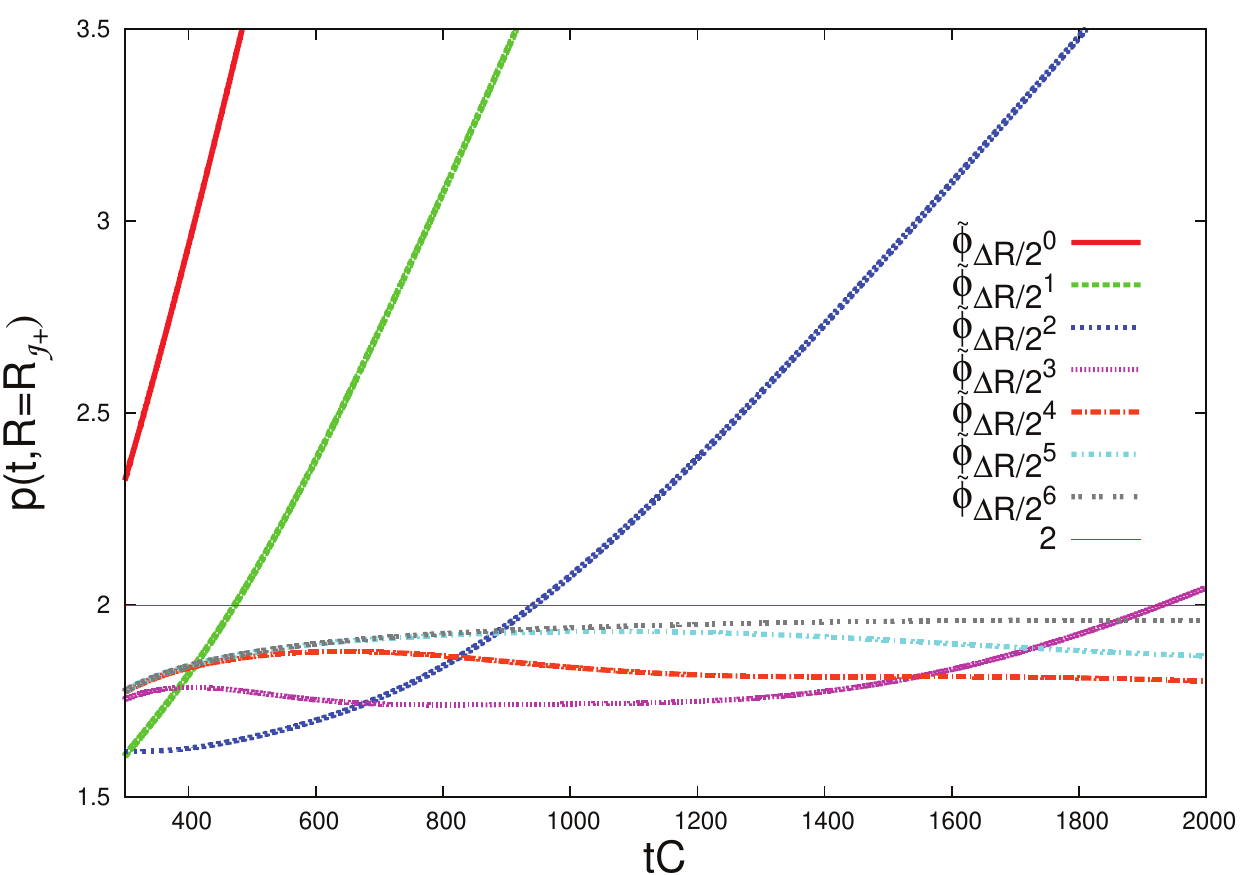}
   \includegraphics[width=8.5cm]{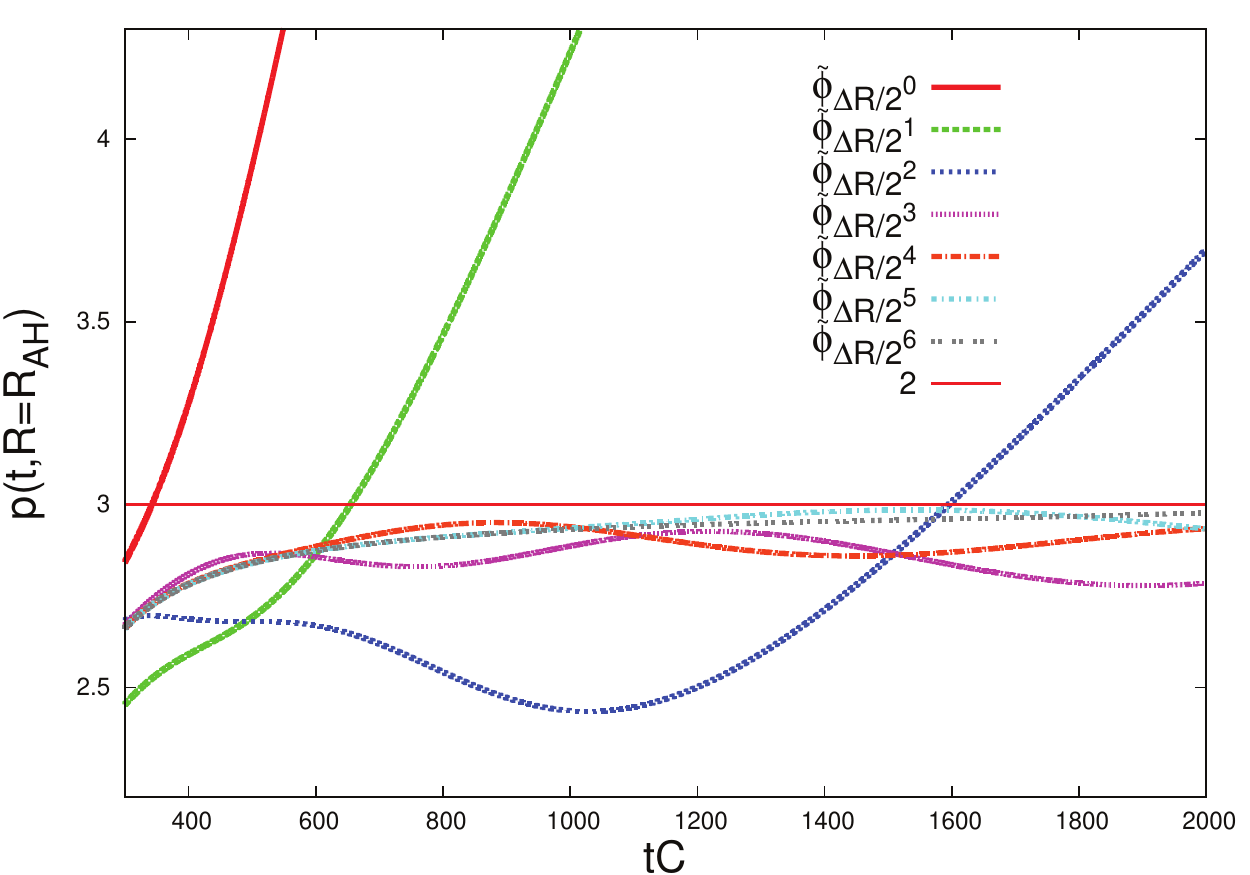}
   \caption{\label{fig:decay-power} Inverse power decay of the conformally rescaled scalar field at late times for different resolutions, measured at null infinity (left panel) and along the apparent horizon (right panel). The parameters for the simulations are the same ones as in the previous figure, except that we use a number of $N = 2^n\times 100$ grid points with $n = 0,1,\ldots 6$. Here, $\Delta R = (R_+ - R_{in})/100$ is the grid space belonging to the coarsest resolution. Note that the case $n = 4$ corresponds to the plots shown in the previous figure. As is clear from this plot, the correct tail decay cannot be measured from simulations using the coarsest resolutions. As will become clear from the next plot, the convergence regime is, in fact, only reached for resolutions $n\geq 4$.}
   \end{center}
\end{figure}

Next, we analyze the decay properties of the scalar field at late times. Fig.~\ref{fig:graph-detectors} shows the behavior of the conformally rescaled scalar field as a function of time along different curves: the first is a radial null geodesic along $\ScriPlus$, the second coincides with the world line of a timelike observer at position $R = 0.649$, and the third is a radial curve along the apparent horizon. In all three cases, we observe an oscillatory behavior until about $t = 100C^{-1}$, after which the field starts decaying as an inverse power in $t$, that is $\tilde{\phi} \sim (t C)^{-p}$ with $p > 0$ a constant. This behavior is known as ``tail decay" in the literature~\cite{rP72}. From the plot, it is also visible that the field decays at about the same rate at the apparent horizon and along the timelike observer, while the decay along $\ScriPlus$ is slower. To compute the inverse power $p$, we monitor the following quantity:
\begin{equation}
 p = -\frac{t}{\tilde{\phi}}\partial_t\tilde{\phi} 
    = \frac{-t}{\tilde{\phi}}\left[\tilde{\alpha}\hat{\pi}
    + \tilde{\alpha}R\left(\frac{\tilde{\nu}}{2}-\tilde{C}\right)\tilde{\chi}
     - \tilde{\alpha}\tilde{C}\tilde{\phi}\right],
\end{equation}
from our numerical data. Fig.~\ref{fig:decay-power} shows the values of $p$ along $\ScriPlus$ and along the apparent horizon, for different resolutions. As the resolution increases, there is a clear trend for $p\to 2$ at $\ScriPlus$ and $p\to 3$ along the apparent horizon, which is consistent with the prediction from linearized theory~\cite{rP72,cGrPjP94}, with numerical studies in the nonlinear case (see for instance~\cite{cGrPjP94b,oRvM13}), and with rigorous results concerning the nonlinear theory~\cite{mDiR05}. Note that our simulations for the tail decay are based on initial data giving rise to a vanishing NP constant. The decay rate in the non-vanishing case has been analyzed in~\cite{rGjWbS94}.

\begin{figure}[htp]
   \begin{center}
   \includegraphics[width=9.5cm]{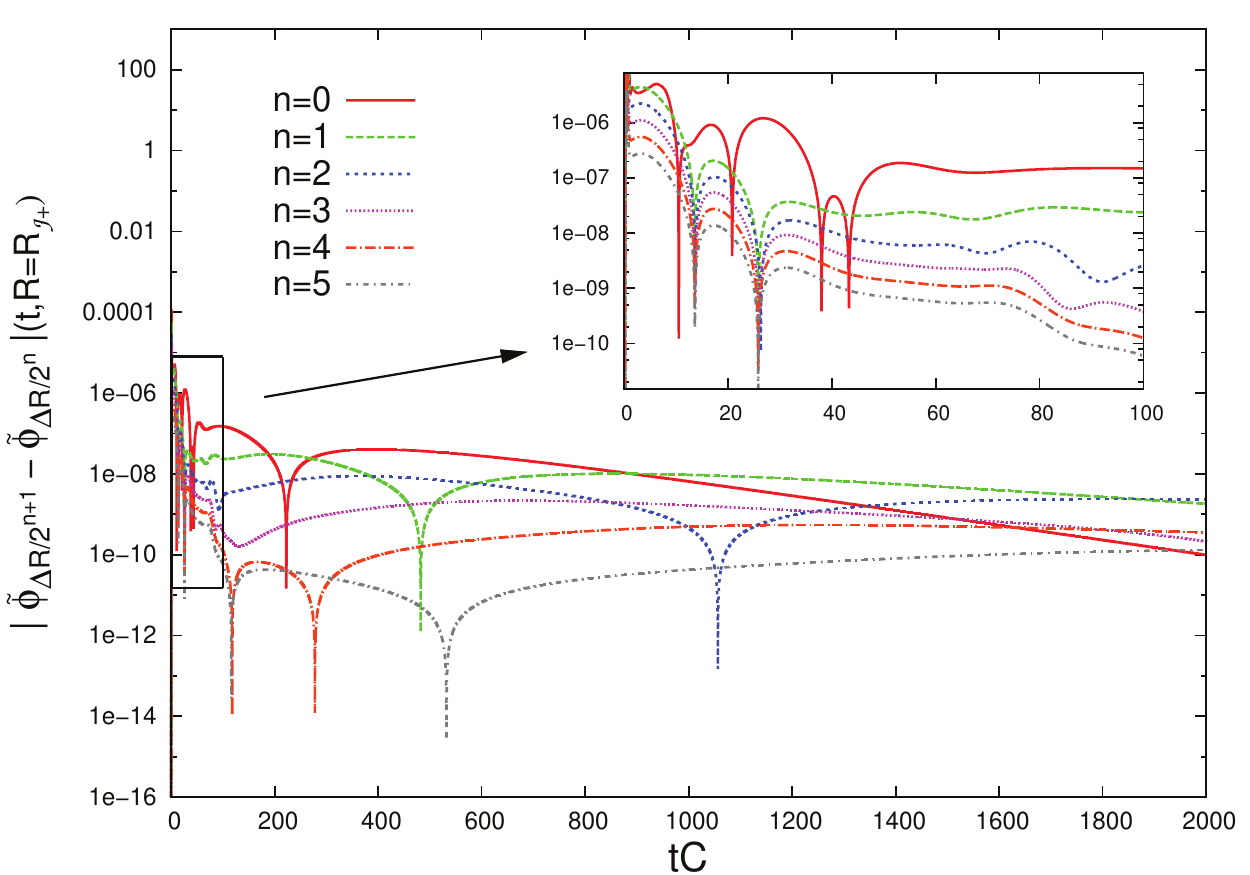}
   \caption{\label{fig:selfconvergence-evolution} \small Self-convergence test for the conformally rescaled scalar field at $\ScriPlus$. The parameters used in the simulations are the same as in the previous plot. As is clear from the plot, at late times, high enough resolutions ($n\geq 4$) are needed in order to be in the convergence regime. The order of convergence has been computed to be between $2$ and $4$.}
   \end{center}
\end{figure}

Finally, Fig.~\ref{fig:selfconvergence-evolution} shows a self-convergence test for the detector located at $\ScriPlus$. Note that for times $t \leq 100C^{-1}$ the error between consecutive resolutions clearly goes down when resolution is increased. The order of self-convergence computed during these times results between $1$ and $2$. However, after times $t\geq 800C^{-1}$ we observe that these errors do not decrease for the lowest resolutions ($\Delta R / 2^n$ with $n=0,1,2,3$), indicating that the convergence regime has not been reached yet. This is consistent with the results from the previous plot (see Fig.~\ref{fig:decay-power}) which shows that one cannot produce the correct tail decay at such coarse resolutions. However, when considering the higher resolutions $n = 4,5,6$, one finds convergence to an order lying between $2$ and $4$ for times $t \geq 800C^{-1}$.

\section{Conclusions}
\label{Sec:Conclusions}

In this work, we presented the first numerical implementation of the tetrad-based formulation of Einstein's field equations on compactified CMC slices proposed in~\cite{jBoSlB11}. For simplicity, we restricted our simulations to spherically symmetric spacetimes, and to obtain non-trivial dynamics, we minimally coupled a scalar field to the gravitational field. We first wrote down the rescaled Einstein-scalar field equations without symmetry assumptions and an arbitrary potential $V(\Phi)$ for the scalar field. Although the wave equation for the scalar field is not conformally covariant, we showed that using the Einstein equations, and assuming that the potential decays sufficiently fast to zero when the scalar field goes to zero, it is possible to rewrite the wave equation as a first-order symmetric-hyperbolic system which is manifestly regular at $\ScriPlus$. In this way, we showed that scalar fields can naturally be incorporated in the formulation of~\cite{jBoSlB11}.

Next, we focused on spherically symmetric configurations and showed that in this case there is a preferred choice for the orientation of the spatial legs of the tetrad vector fields which automatically obeys the 3D Nester gauge on which the formulation in~\cite{jBoSlB11} is based. Further choosing the radial coordinate $R$ to coincide with the areal radius of the conformal three-metric, we obtained a hyperbolic-elliptic system of equations describing the evolution of the gravitational and scalar fields in spherical symmetry. This system is rather similar to the one obtained by Rinne and Moncrief~\cite{oRvM13} from their metric-based formulation which is not surprising since in their formulation the conformal three-metric is also flat and the radial coordinate $R$ is the same as ours. As a result, in both formulations, the geometry of the spatial slices is entirely encoded in the conformal factor. Our equations differ from the ones considered in~\cite{oRvM13} insofar as in our work, the scalar field is minimally coupled to gravity whereas Rinne and Moncrief consider a conformally invariant scalar field.

Our discussion also includes a detailed analysis of the local spherical solutions for the conformal factor, the trace-free part of the conformal extrinsic curvature, and the conformal lapse in the vicinity of $\ScriPlus$. Using the constraint equations and the elliptic equation for the conformal lapse resulting from the CMC slicing condition, we derived formal expansions for these quantities in terms of the radial proper distance $z$ to $\ScriPlus$ and its logarithm. Similar to what occurs in the vacuum case without symmetries~\cite{jBoSlB11}, we found that  even when the Newman-Penrose constant is zero, the $\log$ terms appear whenever outgoing scalar radiation is present at $\ScriPlus$. Explicit expressions for the expansion coefficients were discussed in Sec.~\ref{Sec:Asymptotics} and in App.~\ref{App:FormalExpansions} and a novel, rigorous method for the existence of the corresponding local solutions near $\ScriPlus$ is given in App.~\ref{App:LocSolutionsScriPlus}, based on standard tools from the theory of dynamical systems.

The symmetric hyperbolic system describing the evolution of the rescaled scalar field was numerically implemented using the method of lines, with standard difference operators satisfying the summation by parts property for the discretization of the spatial operators and a fourth-order Runge-Kutta algorithm for the time discretization. At each time step, we numerically solved the (singular) elliptic equations for the conformal factor, the trace-free part of the conformal extrinsic curvature and the conformal lapse. These three variables encode all the information required to determine the gravitational field. The elliptic system is solved using a ``shooting to a matching point" algorithm, placing the inner boundary at a trapped surface and the exterior boundary at $\ScriPlus$. After constructing a family of initial data representing a scalar field configuration outside a marginally trapped surface, we ran several tests for our code, including convergence tests. Next, we performed long-term evolutions and measured the decay of the scalar field at the apparent horizon, at $\ScriPlus$, and along the world lines of a timelike observer. For the initial data used in this article we found that the scalar field decays as $t^{-p}$ with $p = 3$ for timelike observers and along the apparent horizon, and with $p=2$ along $\ScriPlus$. This is fully consistent with known results in the literature~\cite{rP72,cGrPjP94,cGrPjP94b,mDiR05}.

One lesson learned from this work is that to achieve long-term stability, it was necessary to solve the momentum constraint at each time step. This constitutes a slight modification of the proposed scheme put forward in~\cite{jBoSlB11}, where the trace-free part of the conformal extrinsic curvature $\hat{\tilde K}_{ab}$ is evolved freely using Eq.~(\ref{Eq:Kab}) which is singular at $\ScriPlus$. Solving the momentum constraint instead offers the advantage of being able to impose the correct regularity conditions for $\hat{\tilde K}_{ab}$ at $\ScriPlus$. Thus, in our spherically symmetric scheme, one ends up using all the constraints to determine the gravitational variables from the matter ones. This indicates that achieving long-term stable full 3D evolutions based on the proposal~\cite{jBoSlB11} might require a constraint-projection method as in~\cite{mHlLrOhPmSlK04}.

There are several improvements and possible extensions of our work that are worth pursuing. First, it would be interesting to obtain geometrically motivated inner boundary conditions, for example by demanding that the inner sphere correspond to a marginally trapped surface. This would likely require relaxing the gauge condition $\tilde{B}_R = \tilde{B}_T = 1$ which forces the radial coordinate $R$ to measure proper distances in the conformal geometry.

A further interesting extension is to consider non-trivial potentials for the scalar field. Although we have included the corresponding terms in our equations, the presence of a scalar field potential yields additional singular terms, unless $V(\Phi)$ decays at least as fast as $\Phi^4$ as $\Phi$ goes to zero. However, most interesting potentials studied in the literature only decay quadratically with $\Phi$ in which case additional regularity conditions need to be imposed.

Finally, it should be interesting to generalize this work to anti-de-Sitter type spacetimes. A negative cosmological constant can easily be incorporated in our equations by taking the scalar field potential $V(\Phi)$ to be a negative constant (see App.~\ref{App:ExplicitExpressions}). However, this gives rise to additional terms in the equations which are singular at the outer boundary of the numerical domain, where $\Omega$ vanishes. In this case, the outer boundary corresponds to spacelike and null infinity of spacetime, and regularity conditions which are different than the ones considered in the present work need to be imposed.


\acknowledgments

During this work, we benefited from fruitful and stimulating discussions with Luis Lehner, Manuel Tiglio, and Thomas Zannias. We wish to thank Luisa Buchman for reading a previous version of the manuscript and suggesting improvements. This work was supported in part by CONACyT Grants No. 271904 and No. 236810, and by a CIC Grant to Universidad Michoacana. We also thank the Perimeter Institute for Theoretical Physics, where part of this work was performed, for hospitality. Research at Perimeter Institute is supported through Industry Canada and by the Province of Ontario through the Ministry of Research and Innovation.

\appendix
\section{Explicit expressions and expansions of the metric fields for vanishing scalar field}
\label{App:ExplicitExpressions}

For the particular case of a constant scalar field, the stress energy-momentum tensor~(\ref{Eq:SEtensor}) reduces to
$$
T_{\mu\nu} = -\frac{\Lambda}{8\pi G} g_{\mu\nu},\qquad
\Lambda := 8\pi G V(0),
$$
where the effective cosmological constant is related to the value of the potential $V(\Phi)$ at one of its stationary points, say $\Phi = 0$. In this case, Eqs.~(\ref{Eq:RadDerivMass},\ref{Eq:NormDerivMass}) can be integrated explicitly, yielding
$$
N(r) := 1 - \frac{2m}{r} = 1 - \frac{2M}{r} - \frac{\Lambda}{3} r^2
$$
with a constant $M$ representing the total mass. Likewise, when the scalar field vanishes, the momentum constraint equation~(\ref{Eq:MomSphSymBis}) can be solved for explicitly with the result
\begin{equation}
\tilde{\nu} = 2D\frac{\Omega^2}{R^3},
\label{Eq:MomSphSymIntSch}
\end{equation}
with an integration constant $D$ whose meaning will be clarified later. From this one finds
$$
\Theta^\pm = C r - \frac{D}{r^2} \pm R\frac{r'}{r}
$$
for the in- and outgoing expansions. Combining this result with $N(r) = -\Theta^+\Theta^-$ and assuming that $r' > 0$ one finds the equation
$$
R\frac{r'}{r} = \sqrt{N(r) + \left( C r - \frac{D}{r^2} \right)^2},
$$
which yields a relation between the physical areal radial coordinate $r$ and the areal radial coordinate $R$ of the conformal metric. This relation can be written as
\begin{equation}
\frac{R}{R_+} = \exp\left[ -\int\limits_{r(R)}^\infty
\frac{1}{\sqrt{N(r) + \left( Cr - \frac{D}{r^2} \right)^2}}\frac{dr}{r} \right],
\label{Eq:OmegaSchw}
\end{equation}
where we have assumed that $C^2 > \Lambda/3$ such that the integral converges at $r = \infty$. The conformal factor is obtained from this and the relation $\Omega = R/r$.

For the following, we focus on the case of vanishing cosmological constant $\Lambda = 0$, although a similar line of reasoning could be used to deduce the corresponding results for the case $\Lambda < 0$. To obtain the conformal lapse, one needs to integrate the elliptic equation~(\ref{Eq:CMCConstrSphSymBis}) which guarantees the preservation of CMC slicing. For the purpose of explicit integration it is convenient to rewrite this equation in terms of the physical lapse $\alpha = \tilde{\alpha}/\Omega$. Using Eq.~(\ref{Eq:OmegaSchw}) and $\Omega = R/r$ we obtain
\begin{equation}
\alpha_0(r) \frac{d}{dr}Ê\left[ \alpha_0(r) \frac{d\alpha}{dr} \right]
 + \frac{2}{r}\alpha_0(r)^2 \frac{d\alpha}{dr} - 3\left( C^2 + \frac{2D^2}{r^6} \right)\alpha = 0,
\end{equation}
which has the particular solution $\alpha(r) = \alpha_0(r) = \sqrt{N(r) + (Cr - D/r^2)^2}$. Another, independent solution is obtained using the ansatz $\alpha_1(r) = F(r)\alpha_0(r)$ for some function $F(r)$. This yields the following one-parameter family of solutions of Eq.~(\ref{Eq:CMCConstrSphSymBis}) fulfilling the correct boundary conditions at $\ScriPlus$:
\begin{equation}
\tilde{\alpha} = \frac{R}{r}\alpha_0(r) 
 \left[ 1 - \frac{\zeta}{C} \int\limits_r^\infty \frac{ds}{s^2\alpha_0(s)^3} \right],
\label{Eq:AlphaTildeSchw}
\end{equation}
with a dimensionless constant $\zeta$. Note that $\tilde{\alpha}\to R_+ C$ as $r\to \infty$, as required from Eq.~(\ref{Eq:ScriPlusBC}). Next, integrating Eq.~(\ref{Eq:Cons2Bis}) and taking into account the boundary conditions~(\ref{Eq:ScriPlusBC}) we obtain
\begin{equation}
\tilde{C} = \frac{1}{R}\left[ \frac{D}{r^2}
 + \frac{Cr - \frac{D}{r^2} + \frac{D\zeta}{C} r\int\limits_r^\infty
 \left( \frac{1}{r^3} - \frac{1}{s^3} \right) \frac{ds}{s^2\alpha_0(s)^3}}
 {\alpha_0(r)\left[ 1 - C\zeta \int\limits_r^\infty \frac{ds}{s^2\alpha_0(s)^3} \right]} \right],
\end{equation}
from which the radial component of the shift can be computed using Eq.~(\ref{Eq:AlgebraicShift}):
\begin{equation}
b = -R\left[ C - \frac{D}{r^3} 
+ \frac{D\zeta}{C}\int\limits_r^\infty \left( \frac{1}{r^3} - \frac{1}{s^3} \right)\frac{ds}{s^2\alpha_0(s)^3} \right].
\end{equation}
Using all this in the evolution equation for the conformal factor, Eq.~(\ref{Eq:OmegaEvol}), we obtain
$$
\partial_t\Omega = \zeta \frac{R}{r}\alpha_0(r)\int\limits_r^\infty \left( 1 - \frac{D}{C s^3} \right)
\frac{ds}{s^2\alpha_0(s)^3},
$$
which shows that $\zeta = 0$ is characterized by the requirement of $\partial_t$ coinciding with the timelike Killing vector field.

In order to shed some light on the formal expansions discussed in Sec.~\ref{Sec:Asymptotics} it is illustrative to expand the integral in Eq.~(\ref{Eq:OmegaSchw}) in powers of $y := 1/(Cr)$. In a first step we get
$$
\log\left( \frac{R}{R_+} \right) 
 = -y\left[ 1 - \frac{1}{6} y^2 + \frac{\overline{C} + \delta}{4} y^3 + \frac{3}{40} y^4 + {\cal O}(y^5) \right],
$$
where we have set $\overline{C} := C m$ and $\delta := D C^2$. Inverting the power series and expressing the result in terms of the dimensionless quantity $z := 1 - R/R_+$, in terms of which $-\log(R/R_+) = -\log(1 - z) = z + z^2/2 + z^3/3 + \ldots$, we obtain
$$
y = \frac{1}{Cr} = z\left[ 1 + \frac{1}{2}z + \frac{1}{2}z^2 
 + \left( \frac{1}{2} - \frac{\overline{C} + \delta}{4} \right)z^3 
 + \left( \frac{1}{2} - \frac{\overline{C} + \delta}{2} \right)z^4
+ {\cal O}(z^5) \right],
$$
from which the expansion of the conformal factor $\Omega = C R y = C R_+(1-z) y$ can be found:
\begin{equation}
\Omega = R_+ C z\left[ 1 - \frac{1}{2}z - \frac{\overline{C} + \delta}{4} z^3
 - \frac{\overline{C} + \delta}{4} z^4 + {\cal O}(z^5) \right],\qquad
z = 1 - \frac{R}{R_+}.
\label{Eq:OmegaSchwExpansion}
\end{equation}
We notice that the mass only appears at the order $z^4$ in the expansion of the inverse areal radius $1/r$ and the conformal factor $\Omega$. From Eq.~(\ref{Eq:AlphaTildeSchw}) we obtain
\begin{equation}
\tilde{\alpha} = R_+ C\left[ 1 - z + \frac{1}{2}z^2 - (\overline{C} + \delta)z^3
 - \left( \frac{\overline{C} + \delta}{2} + \frac{\zeta}{4} \right) z^4 
 - \left( \frac{\overline{C} + \delta}{2} - \frac{\zeta}{20}  \right) z^5
 +  {\cal O}(z^6) \right].
\end{equation}

The in- and outgoing expansions are
$$
\Theta^\pm = C r - \frac{D}{r^2} \pm \sqrt{N(r) + \left( C r - \frac{D}{r^2} \right)^2}
 = \frac{1}{y}\left[  
 1 - \delta y^3 \pm \sqrt{ 1 + y^2 - 2(\overline{C} + \delta)y^3 + \delta^2 y^6} \right],
$$
and we see that asymptotically, $\Theta^+$ diverges as $2/y$ while $\Theta^-$ falls off as $-y/2$ when $y = 1/(Cr)\to 0$, cf. Eqs.~(\ref{Eq:ThetaPlusExp},\ref{Eq:ThetaMinusExp}).

\section{Formal, polyhomogeneous expansions at $\ScriPlus$}
\label{App:FormalExpansions}

In this appendix, we generalize the expansions~(\ref{Eq:vExpansion},\ref{Eq:uExpansion},\ref{Eq:aExpansion}) to include higher-order terms. To this purpose we first expand the functions $F_1$, $F_2$ and $F_3$ appearing in Eqs.~(\ref{Eq:HamRescaled},\ref{Eq:MomRescaled},\ref{Eq:CMCRescaled}) as follows:
\begin{eqnarray}
F_1 &=& \frac{3}{8}v^2 + 4\pi G R_+^2\tilde{\rho}
 = g_0 + g_1 z + g_2 z^2 + {\cal O}(z^2\log z)
\label{Eq:MatterEvolRscri-gbar},\\
F_2 &=& \tilde{j}_R =  -e_0 - e_1 z - e_2 z^2 + {\cal O}(z^2\log z),\\
F_3 &=& \frac{9}{4}v^2 + 4\pi G R_+^2 \left(3\tilde{\rho} + \tilde{\sigma}^c{}_c\right) 
 = d_0 + d_1 z + d_2 z^2 + {\cal O}(z^2\log z)
\label{Eq:MatterEvolRscri-hbar}.
\end{eqnarray} 
As explained in Sec.~\ref{Sec:Asymptotics}, the expansions for the fields $(u,v,a)$ in terms of $z$ involve logarithmic terms. Denoting by $f$ any of these fields, the expansion has the form
\begin{equation}
f(z) = \sum_{k=0}^{\infty} \sum_{l=0}^{\infty} f_{kl} z^k\log^l(z),
\label{eq:polihomogeneous-exp}
\end{equation}
with coefficients $f_{kl}$ to be determined. Introducing this expansion into the Hamiltonian constraint equation~(\ref{Eq:HamRescaled}) we obtain, for the solution which vanishes at $\ScriPlus$, the following leading-order coefficients:
\begin{eqnarray}
 u_{10} &=& 1,\qquad u_{20} = -\frac{1}{2},\qquad u_{30} = -\frac{1}{3}g_0, \nonumber\\
 u_{41} &=& -\frac{1}{2}g_0 + \frac{1}{4}g_1,\qquad
 u_{40} = u_4 = \text{free parameter associated with the Bondi mass},
\label{Eq:uExpBis}\\
 u_{51} &=& -\frac{1}{2}g_0 + \frac{1}{4}g_1,\qquad
 u_{50} = u_4 + \frac{1}{5}g_0 + \frac{1}{30}g_0{}^2 - \frac{3}{8}g_1 + \frac{1}{5}g_2.
\nonumber
\end{eqnarray}
Next, we solve Eq.~(\ref{Eq:CMCRescaled}) for the rescaled lapse function $a$ and obtain, to leading order,
\begin{eqnarray}
 a_{00} &=& 1,\qquad a_{10} = -1,\qquad a_{20} =\frac{1}{2}-\frac{1}{2}g_0-\frac{1}{4}d_0,
\nonumber\\
 a_{31} &=& - 2g_0 + g_1,\qquad  
 a_{30} = -\frac{11}{6}g_0 + \frac{5}{4}g_1 + \frac{2}{3}d_0 + 4u_4 - \frac{1}{3}d_1,
\label{Eq:aExpBis}\\
 a_{41} &=& -\frac{13}{4}g_0 + 3g_1 - g_2 + \frac{5}{8}d_0 - \frac{5}{8}d_1 + \frac{1}{4}d_2
             +\frac{1}{4}g_0{}^2 - \frac{1}{16}d_0{}^2,\qquad
 a_{40} = a_4 = \text{free parameter}.
\nonumber
\end{eqnarray}
Finally, solving Eq.~(\ref{Eq:MomRescaled}) for $v$ we obtain to leading order
\begin{eqnarray}
v_{10} &=& e_0,\nonumber\\
v_{21} &=& 2e_0 - e_1,\qquad v_{20} = v_2 = \text{free parameter},
\label{Eq:vExpBis}\\
v_{31} &=& 4e_0 - 2e_1,\qquad
v_{30} = 2v_2 - \frac{3}{2}e_0 + 2e_1 -\frac{4}{3}g_0e_0 - e_2, 
\nonumber\\
v_{41} &=& \frac{13}{2}e_0 - \frac{13}{4}e_1 - \frac{17}{6}g_0e_0 + \frac{2}{3}e_1g_0
             + \frac{3}{4}e_0g_1.
\nonumber
\end{eqnarray}

From this one obtains the following expansion coefficients for the MS mass function (setting $8\pi G C^2 R_+^2 = 1$):
\begin{eqnarray}
C m_{00}
 &=& -\frac{2}{3}g_0 + e_0 - \frac{1}{2}v_2 - 4u_4 - \frac{1}{4}g_1 
\nonumber\\
 &=& - 4u_4 - \frac{1}{2}v_2  +\frac{19}{24}\tilde{\phi}_0{}^2 + \frac{1}{8}\tilde{\phi}_0\hat{\pi}_0
   + \frac{3}{8}\tilde{\phi}_0\tilde{\chi}_0 , \nonumber\\
C m_{10}
 &=& \frac{3}{8}e_0 - e_1 - \frac{5}{4}g_0 + \frac{1}{2}e_2 + 2g_1 - g_2 + \frac{1}{3}g_0e_0
            +\frac{1}{3}g_0{}^2 + \frac{1}{8}e_0{}^2 \nonumber\\
 &=& -\frac{1}{2}\tilde{\phi}_0{}^2 
  + \frac{1}{2}\tilde{\phi}_0\hat{\pi}_0 - \frac{1}{2}\tilde{\phi}_0\tilde{\chi}_0 
  - \frac{1}{4}\hat{\pi}_0{}^2 - \frac{1}{4}\tilde{\chi}_0{}^2 
  + \frac{1}{2}\hat{\pi}_0\tilde{\chi}_0,
\end{eqnarray}
while we found that the logarithmic coefficients $m_{0k}$ and $m_{1k}$ with $k > 0$ vanish.

\section{Existence of local solutions of the constraint equations near $\ScriPlus$}
\label{App:LocSolutionsScriPlus}

In this appendix, we prove that the system of equations~(\ref{Eq:HamRescaled},\ref{Eq:MomRescaled},\ref{Eq:CMCRescaled}) possesses a three-parameter family of solutions $(u(z),v(z),a(z))$ which are defined for small enough $z > 0$ and which satisfy $u(z)\to 0$, $v(z)\to 0$ and $a(z)\to 1$ as $z\to 0$. We do this by transforming this system to the ``nicer'' form
\begin{equation}
z\frac{d}{dz} y + B y = z g(z,y),
\label{Eq:NiceForm}
\end{equation}
where here the components of the vector-valued function $y(z)$ are related to $u$, $v$, $a$ and the first derivatives of $u$ and $a$, $B$ is a constant $5\times 5$ matrix with the property that all of its eigenvalues have real parts different than zero, and where $g(z,y)$ is a non-linear function which is $C^k$-differentiable for some $k\geq 1$. The key property of system~(\ref{Eq:NiceForm}) is that the singular part of the equation is entirely contained in the linear operator $z\frac{d}{dz} + B$ on the left-hand side, while the non-linear part on the RHS is regular at $z = 0$. This allows a treatment of the problem based on  standard arguments from the theory of dynamical systems, see Theorem~\ref{Thm:NiceForm} below.

In the next subsection, we first discuss general results which describe the space of local solutions $y(z)$ of Eq.~(\ref{Eq:NiceForm}) satisfying $y(z)\to 0$ for $z\to 0$ and a method to construct them via an iteration scheme. In fact, this result has its own interests, since many other problems in physics can be cast into the form~(\ref{Eq:NiceForm}), including the radial equation describing (relativistic or non-relativistic) spherically symmetric perfect fluid static stars~\cite{aRbS91} and the equations describing null geodesics emanating from singularities~\cite{dC84,nOoS11}. In the subsequent subsections, we apply these general results to the system~(\ref{Eq:HamRescaled},\ref{Eq:MomRescaled},\ref{Eq:CMCRescaled}) of interest to this article.

\subsection{General results describing the regular solutions of Eq.~(\ref{Eq:NiceForm}) near $z=0$}

\begin{theorem}[cf. Theorem~3 in~\cite{nOoS11}]
\label{Thm:NiceForm}
Let $m,k$ be natural numbers, and let $B$ be a real, $m\times m$ matrix with the property that all its eigenvalues $\lambda$ satisfy $\re(\lambda)\neq 0$. Further, let $g: D\subset \Real^{m+1}\to \Real^m$ be a $C^k$ differentiable function defined on an open neighborhood $D$ of the origin in $\Real^{m+1}$.

Let $r$ denote the number of eigenvalues of $B$ with negative real parts. Then, the system~(\ref{Eq:NiceForm}) admits an $r$-parameter family of $C^k$-differentiable local solutions $y: (0,\delta)\to \Real^m$ such that $\lim\limits_{z\to 0} y(z) = 0$.
\end{theorem}

\proof The proof uses standard results from the theory of dynamical systems, see for instance~\cite{Hartman-Book,BrauerNohel-Book}. Let $z_0 > 0$. We first regularize the system~(\ref{Eq:NiceForm}) by introducing the (fictitious) time parameter $\tau = -\log(z/z_0)$. Next, we define $\alpha := g(0,0)\in \Real^m$ and set
$$
U:= \left( \begin{array}{c} z \\ y \end{array} \right).
$$
Then, the system~(\ref{Eq:NiceForm}) is equivalent to the autonomous dynamical system
\begin{equation}
\frac{d}{d\tau} U = A U + G(U),\qquad
A := \left( \begin{array}{cc} -1 & 0 \\ -\alpha & B \end{array} \right),\quad
G(U) := \left( \begin{array}{c} 0 \\ z[g(0,0) - g(z,y)] \end{array} \right).
\label{Eq:DynSystem}
\end{equation}
By construction, $G: D\subset\Real^{m+1}\to \Real^{m+1}$ is a $C^k$ function which vanishes at $U = 0$ and whose differential at $U = 0$ is zero. Therefore, the $(m+1)\times (m+1)$ matrix $A$ describes the linearization of the evolution vector field at the fixed point $U = 0$. Its eigenvalues consist of $-1$ and those of $B$, which by hypothesis have real parts different than zero. Consequently, $U = 0$ is a hyperbolic critical point of Eq.~(\ref{Eq:DynSystem}). Denoting by $\varphi^\tau$ the flow associated with the system~(\ref{Eq:DynSystem}), the local stable manifold associated with $U = 0$ is defined as the following set:
$$
W^+(0) := \{ U\in D : \varphi^\tau(U)\in D \hbox{ for all $\tau\geq 0$ and $\varphi^\tau(U)\to 0$ as $\tau\to\infty$} \}.
$$
According to the standard theory, $W^+(0)$ is an $(r+1)$-dimensional $C^k$-manifold through $U = 0$ which is generated by local solutions $U: (0,\infty)\to D$ of Eq.~(\ref{Eq:DynSystem}) satisfying $U(\tau)\to 0$ as $\tau\to \infty$. Since the first component of the system~(\ref{Eq:DynSystem}) decouples from the remaining ones, $U(\tau)$ has the form
$$
U(\tau) = \left( \begin{array}{c} z_0 e^{-\tau} \\ Y(\tau) \end{array} \right),\qquad
\tau > 0,
$$
for some constant $z_0 > 0$, and hence the corresponding function $y: (0,z_0)\to \Real^m$ defined by $y(z) := Y\left( -\log(z/z_0)\right)$ for $0 < z < z_0$, is a $C^k$-solution of Eq.~(\ref{Eq:NiceForm}) satisfying $y(z)\to 0$ for $z\to 0$. The reason why this family is $r$-parametric and not $(r+1)$-parametric, $r+1$ being the dimension of $W^+(0)$, will become clear from the considerations that follow.
\qed

The $(r+1)$-dimensional manifold $W^+(0)$ can be constructed in the following way: first, using a suitable invertible linear transformation $Q: \Real^{m+1}\to \Real^{m+1}$ we can bring $A$ into the following block-diagonal form:
\begin{equation}
Q^{-1} A Q = \left( \begin{array}{cc} A_+ & 0 \\ 0 & A_- \end{array} \right),
\label{Eq:ABlockDiag}
\end{equation}
where $A_+$ is an $(r+1)\times (r+1)$ matrix whose eigenvalues have negative real parts and where $A_-$ is an $(m-r)\times (m-r)$ matrix whose eigenvalues have positive real parts. For example, this can be achieved by bringing $A$ into its Jordan normal form. Defining $\tilde{U} := Q^{-1} U$ the dynamical system~(\ref{Eq:DynSystem}) is transformed into
$$
\frac{d}{d\tau} \tilde{U} =  \left( \begin{array}{cc} A_+ & 0 \\ 0 & A_- \end{array} \right)\tilde{U}
 + \left( \begin{array}{c} \tilde{G}_+(\tilde{U}) \\ \tilde{G}_-(\tilde{U}) \end{array} \right),\qquad
\left( \begin{array}{c} \tilde{G}_+(\tilde{U}) \\ \tilde{G}_-(\tilde{U}) \end{array} \right)
 = Q^{-1} G(Q\tilde{U}).
$$
Next, one introduces the integral operator $F_a$ defined by
\begin{equation}
(F_a\tilde{U})(\tau) := \left( \begin{array}{c} e^{\tau A_+} a \\ 0 \end{array} \right)
 + \left( \begin{array}{c} 
 \int\limits_0^\tau  e^{(\tau-\tau') A_+} \tilde{G}_+(\tilde{U}(\tau')) d\tau' \\
 -\int\limits_\tau^\infty e^{(\tau-\tau') A_-}\tilde{G}_-(\tilde{U}(\tau')) d\tau'
\end{array} \right),\qquad \tau \geq 0,
\label{Eq:DefFa}
\end{equation}
for $a\in \Real^{r+1}$ and bounded, continuous functions $\tilde{U}: (0,\infty)\to Q^{-1}(D)$. By the spectral properties of $A_+$ and $A_-$, one has $(F_a\tilde{U})(\tau)\to 0$ as $\tau\to \infty$ for such $\tilde{U}$. For small enough $|a|$ one can show that $F_a$ possesses a unique fixed point $\tilde{U}_a$ on an appropriate function space, and it follows that the corresponding function $U_a(\tau) = Q\tilde{U}_a(\tau)$, $\tau > 0$, is a solution of Eq.~(\ref{Eq:DynSystem}) satisfying $U_a(\tau)\to 0$ when $\tau\to \infty$. One then shows that for small enough $|a|$ the map $\Real^{r+1}\to W^+(0)$, $a\mapsto U_a(0)$ is $C^k$ -differentiable and invertible, proving that $W^+(0)$ is an $(r+1)$-dimensional manifold.

The fixed point $\tilde{U}_a$ of $F_a$ can be constructed via the iteration scheme:
$$
\tilde{U}^{(0)}(\tau) := 0,\qquad
\tilde{U}^{(1)}(\tau) = (F_a\tilde{U}^{(0)})(\tau)
 = \left( \begin{array}{c} e^{\tau A_+} a \\ 0 \end{array} \right),\qquad
\tilde{U}^{(2)} = F_a\tilde{U}^{(1)},\qquad
\tilde{U}^{(3)} = F_a\tilde{U}^{(2)},\ldots
$$
The scheme converges for small enough $|a|$.

We can transform this iteration scheme back to the original problem~(\ref{Eq:NiceForm}) and obtain:

\begin{theorem}
\label{Thm:Iteration}
Consider the system~(\ref{Eq:NiceForm}) with the additional hypothesis that $\alpha = g(0,0)$ lies in the image of $I + B$, where here $I$ denotes the identity matrix. (Note that this assumption is automatically satisfied if $-1$ is not an eigenvalue of $B$.) Let $w\in \Real^m$ be a vector such that $(I + B)w = \alpha$. Furthermore, let $T: \Real^m\to \Real^m$ be an invertible linear transformation such that
$$
T^{-1} B T = \left( \begin{array}{cc} B_+ & 0 \\ 0 & B_- \end{array} \right),\qquad
T^{-1} g(z,y) = \left( \begin{array}{c} g_+(z,y) \\ g_-(z,y) \end{array} \right),
$$
where $B_+$ is an $r\times r$ matrix whose eigenvalues have negative real parts and $B_-$ a $(m-r)\times (m-r)$ matrix whose eigenvalues have positive real parts.

Then, for small enough $z_0 > 0$ and $b\in \Real^r$ the integral operator $\hat{F}_b$ defined by\footnote{Here, the notation $x^A = e^{(\log x) A}$ for $x > 0$ and any $n\times n$ matrix $A$ is understood.}
\begin{equation}
(\hat{F}_b y)(z) = z w + T\left( \begin{array}{r} \left( \frac{z}{z_0} \right)^{-B+} b +
  \int\limits_z^{z_0} \left( \frac{z}{z'} \right)^{-B+} [ g_+(0,0) - g_+(z',y(z')) ] dz' \\ 
 -\int\limits_0^z \left( \frac{z'}{z} \right)^{B-} [ g_-(0,0) - g_-(z',y(z')) ] dz'
\end{array} \right),\qquad 0 < z\leq z_0,
\label{Eq:IntegralOp}
\end{equation}
on the space of continuous and bounded functions $y: (0,z_0]\to \Real^m$ possesses a unique fixed point $y_b$ which describes a solution of Eq.~(\ref{Eq:NiceForm}) satisfying $y_b(z)\to 0$ for $z\to 0$. This solution $y_b$ can be obtained from the iteration scheme $y^{(1)}, y^{(2)} := \hat{F}_b y^{(1)}, y^{(3)} := \hat{F}_b y^{(2)},\ldots$ starting with
\begin{equation}
y^{(1)}(z) := z w + T\left( \begin{array}{r} \left( \frac{z}{z_0} \right)^{-B+} b \\ 0 \end{array} \right),
\qquad
0 < z\leq z_0.
\label{Eq:FirstIt}
\end{equation}
All solutions $y(z)$ of Eq.~(\ref{Eq:NiceForm}) satisfying $y(z)\to 0$ for $z\to 0$ can be obtained in this way.
\end{theorem}

\proof It is simple to verify that the invertible matrix
$$
Q := \left( \begin{array}{cc} 1 & 0 \\ w & T \end{array} \right)
$$
satisfies Eq.~(\ref{Eq:ABlockDiag}) with
\begin{equation}
A_+ = \left( \begin{array}{cc} -1 & 0 \\ 0 & B_+ \end{array} \right),\qquad
A_- = B_-.
\label{Eq:APlusBlockDiag}
\end{equation}
Furthermore,
$$
Q^{-1} G(U) = \left( \begin{array}{c} G_+(U) \\ G_-(U) \end{array} \right),
$$
with
$$
G_+(U) = \left( \begin{array}{c} 0 \\ z[ g_+(0,0) - g_+(z,y) ] \end{array} \right),\qquad
G_-(U) = z[ g_-(0,0) - g_-(z,y) ].
$$

Next, one notes that the first component of Eq.~(\ref{Eq:DefFa}) yields
$$
(F_a\tilde{U})_1(\tau) = e^{-\tau} a_1,\qquad \tau\geq 0.
$$
Setting $z_0 := a_1$, $\tau := -\log(z/z_0)$ it follows that $(F_a\tilde{U})_1(\tau) = z$ so that we can identify the first component of $\tilde{U}(\tau)$ with $z$. Substituting $\tau' = - \log(z'/z_0)$ in the integrals appearing in the remaining components of Eq.~(\ref{Eq:DefFa}) yields the desired result with $b = (a_2,a_3,\ldots,a_{r+1})$.
\qed

{\bf Remark}: If $\alpha$ does not lie in the image of $I + B$ one can still apply the iteration scheme defined by Eq.~(\ref{Eq:DefFa}). However, in this case the matrix $A_+$ does not possess the simple block-diagonal structure as in Eq.~(\ref{Eq:APlusBlockDiag}).

\subsection{Application to the system~(\ref{Eq:HamRescaled},\ref{Eq:MomRescaled},\ref{Eq:CMCRescaled})}

To apply the general results described in the previous subsection to our system of equations~(\ref{Eq:HamRescaled},\ref{Eq:MomRescaled},\ref{Eq:CMCRescaled}) we write
$$
u(z) = z - \frac{1}{2}z^2[1 + \eta(z)],\qquad
a(z) = 1 - z[ 1 + \pi(z)],
$$
and set
$$
y := \left( \begin{array}{c} \eta \\ z\eta_z \\ v \\ \pi \\ z \pi_z \end{array} \right).
$$
The system~(\ref{Eq:HamRescaled},\ref{Eq:MomRescaled},\ref{Eq:CMCRescaled}) is then transformed into the form of Eq.~(\ref{Eq:NiceForm}) with $B$ and $g(z,y)$ given by
$$
B := \left( \begin{array}{rrrrr}
  0 & -1 & 0 & 0 & 0 \\ 
 -4 & 0 & 0 & 0 & 0 \\
  0 & 0 & -2 & 0 & 0 \\
  0 & 0 & 0 & 0 & -1Ê\\
  3 & \frac{3}{2} & 0 & -3 & -2  
\end{array} \right),\qquad
g(z,y) = \left( \begin{array}{r}
 0 \\ g_1(z,y) \\ g_2(z,y) \\ 0 \\ g_3(z,y)
\end{array} \right),
$$
where
\begin{eqnarray*}
g_1(z,y) &:=& \frac{2}{1-z}(2\eta + z\eta_z) 
 - \frac{3}{2}\frac{1 + \eta + \frac{1}{2}z\eta_z}{1 - \frac{1}{2} zh} z\eta_z
 - 2\left( 1 - \frac{1}{2}z h \right) F_1(z,u,u_z,v),\\
g_2(z,y) &:=& \left[ \frac{3}{1-z} - \frac{1 + \eta + z\eta_z}{1 - \frac{1}{2} z h} \right] v 
 + F_2(z,u,u_z,v),\\
g_3(z,y) &:=& 3\left( 1 + \eta + \frac{1}{2}z\eta_z \right)\left[ 1 + \pi
 -\frac{1}{2}\frac{1}{1 - \frac{1}{2} zh}\left(
 1 + \eta - \frac{\frac{1}{2} z\eta_z}{1 - \frac{1}{2} z h} \right) a \right]\\
 &+& \left[ \frac{2}{1-z} - \frac{3}{2}\frac{1 + \eta + \frac{1}{2}z\eta_z}{1 - \frac{1}{2} zh} \right]
 ( 1 + \pi + z\pi_z) + [F_1(z,u,u_z,v) - F_3(z,u,u_z,v)] a,
\end{eqnarray*}
where it is understood that in the expressions above, $u = z - \frac{1}{2} z^2 h$, $u_z = 1 - zh - \frac{1}{2} z^2\eta_z$, $a = 1 - z(1 + \pi)$ and $h = 1 + \eta$ should be substituted. We note that $g_1(0,0) = g_2(0,0) = -f_0$ and $g_3(0,0) = 2 - 3f_0/2$, where the coefficient $f_0$ has been defined below Eqs.~(\ref{Eq:F1Exp},\ref{Eq:F2Exp},\ref{Eq:F3Exp}).

The matrix $B$ is diagonalizable with eigenvalues $-3,-2,-2,2,1$ and hence Theorem~\ref{Thm:Iteration} is applicable. The relevant quantities needed to apply it are:
$$
B_+ = \mbox{diag}(-2,-2,-3),\qquad B_- = \mbox{diag}(2,1),
$$
and
$$
T = \left( \begin{array}{rrrrr}
  1 & 0 & 0 & -1 & 0 \\ 
  2 & 0 & 0 & 2 & 0 \\
  0 & 1 & 0 & 0 & 0 \\
  2 & 0 & 1 & 0 & 1Ê\\
  4 & 0 & 3 & 0 & -1  
\end{array} \right),\qquad
T^{-1} = \left( \begin{array}{rrrrr}
  \frac{1}{2} & \frac{1}{4} & 0 & 0 & 0 \\ 
  0 & 0 & 1 & 0 & 0 \\
  -\frac{3}{4} & -\frac{3}{8} & 0 & \frac{1}{4} & \frac{1}{4} \\
  -\frac{1}{2} & \frac{1}{4} & 0 & 0 & 0Ê\\
  -\frac{1}{4} & -\frac{1}{8} & 0 & \frac{3}{4} & -\frac{1}{4}  
\end{array} \right),\qquad
w = \left( \begin{array}{r} \frac{1}{3} f_0 \\ \frac{1}{3} f_0 \\ f_0 \\ 
 -\frac{1}{2} + \frac{3}{4} f_0 \\ -\frac{1}{2} + \frac{3}{4} f_0 \end{array} \right),
$$
and
\begin{eqnarray*}
g_+(0,0) - g_+(z,y) &=& \left( \begin{array}{r}
  -\frac{1}{4}[g_1(z,y) - g_1(0,0)] \\
   -[g_2(z,y) - g_2(0,0)] \\
  \frac{3}{8}[g_1(y,z) - g_1(0,0)] - \frac{1}{4}[g_3(y,z) - g_3(0,0)]
\end{array} \right),\\
g_-(0,0) - g_-(z,y) &=& \left( \begin{array}{r}
  -\frac{1}{4}[g_1(z,y) - g_1(0,0)] \\
  \frac{1}{8}[g_1(y,z) - g_1(0,0)] + \frac{1}{4}[g_3(y,z) - g_3(0,0)]
\end{array} \right).
\end{eqnarray*}

The first iterate, Eq.~(\ref{Eq:FirstIt}), yields
$$
y^{(1)}(z)
 = \left( \begin{array}{r}
\frac{1}{3} f_0 z + \beta_1 z^2 \\
\frac{1}{3} f_0 z + 2\beta_1 z^2 \\
f_0 z + \beta_2 z^2 \\
-\frac{1}{4}\left( 2 - 3f_0 \right) z + 2\beta_1 z^2 + \beta_3 z^3 \\
-\frac{1}{4}\left( 2 - 3f_0 \right) z + 4\beta_1 z^2 + 3\beta_3 z^3
\end{array} \right),
$$
where for simplicity we have introduced the rescaled constants $\beta_1 := b_1/z_0^2$, $\beta_2:= b_2/z_0^2$ and $\beta_3 := b_3/z_0^3$. The corresponding expressions for $u$, $v$ and $a$ are
\begin{eqnarray*}
u^{(1)}(z) &=& z - \frac{1}{2} z^2 - \frac{1}{6} f_0 z^3 - \frac{1}{2}\beta_1 z^4,\\
v^{(1)}(z) &=& f_0 z + \beta_2 z^2,\\
a^{(1)}(z) &=& 1 - z + \frac{1}{4}(2 - 3f_0) z^2 - 2\beta_1 z^3 - \beta_3 z^4,
\end{eqnarray*}
which already coincides with the first few terms in the expansions~(\ref{Eq:uExpansion},\ref{Eq:vExpansion},\ref{Eq:aExpansion}). The logarithmic terms in these expansions appear when computing the higher-order iterates $y^{(2)}$, $y^{(3)}$, etc. In order to compute them, we use
\begin{eqnarray*}
g_1(z,y) - g_1(0,0) &=& 4\eta + \frac{1}{2} z\eta_z + \frac{1}{2}(3f_0 + 2f_1) z 
 + {\cal O}(z^2 + y^2),\\
g_2(z,y) - g_2(0,0) &=& 2v + (f_0 + f_1)z + {\cal O}(z^2 + y^2),\\
g_3(z,y) - g_3(0,0) &=& -\frac{3}{2}\eta + \frac{7}{2}\pi + \frac{1}{2} z\pi_z
 + \left[ 2 + 2f_0 - \frac{1}{2} f_1 + f_1^* \right] z + {\cal O}(z^2 + y^2),
\end{eqnarray*}
where the coefficients $f_0$, $f_1$ and $f_1^*$ are defined below Eqs.~(\ref{Eq:F1Exp},\ref{Eq:F2Exp},\ref{Eq:F3Exp}). This yields
\begin{eqnarray}
g_1(z,y^{(1)}(z)) - g_1(0,0) &=& (3f_0 + f_1)z + {\cal O}(z^2),
\label{Eq:g1Exp}\\
g_2(z,y^{(1)}(z)) - g_2(0,0) &=& (3f_0 + f_1)z +  {\cal O}(z^2),
\label{Eq:g2Exp}\\
g_3(z,y^{(1)}(z)) - g_3(0,0) &=& \left[ \frac{9}{2} f_0 - \frac{1}{2} f_1 + f_1^* \right] z
 + {\cal O}(z^2).
\label{Eq:g3Exp}
\end{eqnarray}
Applying the integral operator defined in Eq.~(\ref{Eq:IntegralOp}) to $y = y^{(1)}$ one obtains
$$
y^{(2)}(z) = \left( \begin{array}{l}
\frac{1}{3} f_0 z + 2\gamma z^2\log z + \beta_1(z_0) z^2 + {\cal O}(z^3) \\
\frac{1}{3} f_0 z + 4\gamma z^2\log z + 2[\beta_1(z_0) + \gamma] z^2 + {\cal O}(z^3) \\
f_0 z + 8\gamma z^2\log z + \beta_2(z_0) z^2 + {\cal O}(z^3) \\
-\frac{1}{4}\left( 2 - 3f_0 \right) z + 4\gamma z^2\log z 
+ 2\left[ \beta_1(z_0) - \frac{1}{16}(f_0 - 5f_1) - \frac{1}{6} f_1^* \right] z^2 + {\cal O}(z^3\log z)\\
-\frac{1}{4}\left( 2 - 3f_0 \right) z + 8\gamma z^2\log z 
+ 4\left[ \beta_1(z_0) + \frac{1}{16}(5f_0 + 7f_1) - \frac{1}{6} f_1^* \right] z^2 
 + {\cal O}(z^3\log z)
\end{array} \right),
$$
where we have introduced the shorthand notation $\gamma := (3f_0 + f_1)/8$ and $\beta_1(z_0)$ and $\beta_2(z_0)$ are rescaled constants depending on $z_0$. Note the logarithmic terms that appear at order $z^2\log z$.

When replacing $y^{(1)}$ with the second iterate $y^{(2)}$ in Eqs.~(\ref{Eq:g1Exp},\ref{Eq:g2Exp},\ref{Eq:g3Exp}) these logarithmic terms give additional contributions:
\begin{eqnarray*}
g_1(z,y^{(2)}(z)) - g_1(0,0) &=& (3f_0 + f_1)z + \frac{5}{4}(3f_0 + f_1)z^2\log z + {\cal O}(z^2),
\\
g_2(z,y^{(2)}(z)) - g_2(0,0) &=& (3f_0 + f_1)z + 2(3f_0 + f_1)z^2\log z +  {\cal O}(z^2),
\\
g_3(z,y^{(2)}(z)) - g_3(0,0) &=& \left[ \frac{9}{2} f_0 - \frac{1}{2} f_1 + f_1^* \right] z
 + \frac{15}{8}(3f_0 + f_1)z^2\log z + {\cal O}(z^2),
\end{eqnarray*}
which in turn yield the third iterate
$$
y^{(3)}(z) = \left( \begin{array}{l}
\frac{1}{3} f_0 z + 2\gamma z^2\log z + \tilde{\beta}_1(z_0) z^2
 + 2\gamma z^3\log z + {\cal O}(z^3) \\
\frac{1}{3} f_0 z + 4\gamma z^2\log z + 2[\tilde{\beta}_1(z_0) + \gamma] z^2 
 + 6\gamma z^3\log z + {\cal O}(z^3) \\
f_0 z + 8\gamma z^2\log z + \tilde{\beta}_2(z_0) z^2 
 + 16\gamma z^3\log z + {\cal O}(z^3) \\
-\frac{1}{4}\left( 2 - 3f_0 \right) z + 4\gamma z^2\log z 
+ 2\left[ \tilde{\beta}_1(z_0) - \frac{1}{16}(f_0 - 5f_1) - \frac{1}{6} f_1^* \right] z^2 
 + {\cal O}(z^3\log z)\\
-\frac{1}{4}\left( 2 - 3f_0 \right) z + 8\gamma z^2\log z 
+ 4\left[ \tilde{\beta}_1(z_0) + \frac{1}{16}(5f_0 + 7f_1) - \frac{1}{6} f_1^* \right] z^2 
 + {\cal O}(z^3\log z)
\end{array} \right),
$$
with new constants $\tilde{\beta}_1(z_0)$ and $\tilde{\beta}_2(z_0)$. The next iterates have exactly the same form (to the order $z^3$ respectively $z^3\log z$ considered here), and they yield the expansions~(\ref{Eq:uExpansion},\ref{Eq:vExpansion},\ref{Eq:aExpansion}) with $u_4 = -\tilde{\beta}_1(z_0)/2$ and $v_2 = \tilde{\beta}_2(z_0)$.

\bibliographystyle{unsrt}
\bibliography{../references/refs}

\end{document}